\documentclass[11pt]{article}
\usepackage{graphicx}
\usepackage[T1]{fontenc}
\usepackage[latin1]{inputenc}
\usepackage[english]{babel}
\usepackage{moreverb}
\usepackage[hang]{subfigure}
\usepackage{amsfonts, amssymb}
\usepackage{amsmath}
\usepackage{amsthm}
\usepackage{a4}
\usepackage{float}
\usepackage{epsfig}
\usepackage{array}
\usepackage{url}
\usepackage{natbib}
\usepackage{mathabx}
\usepackage[usenames,dvipsnames]{color}
\usepackage{algorithmic}

\bibpunct{(}{)}{;}{a}{,}{,}

\usepackage{amsfonts,amssymb}
\usepackage{amsmath}

\newtheorem{theorem}{Theorem}

\begin{document}
\title{Approximations and bounds for binary Markov random fields}

\author{Haakon Michael Austad and H\aa kon Tjelmeland\\
Department of Mathematical Sciences\\
Norwegian University of Science and Technology, Norway}

\maketitle

\begin{abstract}
Discrete Markov random fields form a natural class of models to 
represent images and spatial data sets. The use of such models is,
however, hampered by a computationally intractable normalising
constant. This makes parameter estimation and a fully 
Bayesian treatment of discrete Markov random fields difficult.
We apply approximation theory for pseudo-Boolean functions to
binary Markov random fields and construct approximations and 
upper and lower bounds
for the associated computationally intractable normalising constant.
As a by-product of this process we also get a partially ordered 
Markov model approximation of the binary Markov random field.
We present numerical examples with both the pairwise interaction 
Ising model and with higher-order interaction models, showing
the quality of our approximations and bounds. We also present
simulation examples and one real data example 
demonstrating how the approximations and bounds
can be applied for parameter estimation and to handle 
a fully-Bayesian model computationally.
\end{abstract}


\section{Introduction}
In statistics in general and perhaps especially in spatial statistics we 
often find ourselves with distributions known only up to an 
unknown normalising constant. Calculating this
normalising constant typically involves high dimensional 
summation or integration. This is the case for the class of 
discrete Markov random fields (MRF).
A common situation in spatial statistics is that we have some 
unobserved latent field $x$
for which we have noisy observations $y$. We model $x$ as an MRF with 
unknown parameters
$\theta$ for which we want to do inference of some kind. If we are 
Bayesians we could
imagine adding a prior for our parameters $\theta$ and 
studying the posterior distribution
$p(\theta |y)$. A frequentist approach could involve finding a maximum 
likelihood estimator for
our parameters. 
Without
the normalising constant, these become non-trivial tasks.

There are a number of techniques that have been proposed to overcome this 
problem. The normalising constant can be estimated by running a
Markov chain Monte Carlo (MCMC) algorithm,
which can then be combined with various techniques to produce maximum 
likelihood estimates, see for instance 
\citet{art52}, \citet{art111} and \citet{art134}. 
Other approaches take advantage of the fact that exact sampling
can be done, see \citet{art120}. In the present report however, we focus 
on the class of deterministic methods, where deterministic in this 
setting is referring to that repeating the estimation process 
yields the same estimate. 
In \citet{art100} the authors devise
a computationally efficient algorithm for handling so called 
general factorisable models of
which MRFs are a common example. This algorithm, which we refer to from 
here on as
the variable elimination algorithm, grants a large computational saving 
in calculating the
normalising constant by exploiting the factorisable structure of 
the models. For MRFs
defined on a lattice this allows for calculation of the normalising 
constant on lattices with
up to around 20 rows for models with first order neighbourhoods. 
In \citet{art105} and \citet{art117} the authors construct 
approximations for larger lattices by doing
computations for a number of sub-lattices using the algorithm 
in \citet{art100}.

The energy function of a binary MRF is an example of a so called
pseudo-Boolean function. In general, a pseudo-Boolean function is a 
function of the following type,
$f:\{0,1\}^n\rightarrow \mathbb{R}$. 
A full representation of a 
pseudo-Boolean function requires $2^n$ terms. Finding approximate
representations of pseudo-Boolean functions that require fewer 
coefficients is a well studied field, see
\citet{art137} and \citet{art138}. In \citet{book35} the authors show
how any pseudo-Boolean function can be expressed as a 
binary polynomial in $n$ variables. \citet{art130} expressed the energy
function of MRFs in this manner and by dropping small terms during the
variable elimination algorithm constructed an approximate MRF. 

Our approach and the main contribution of this paper is to apply and
extend approximation theory for pseudo-Boolean function to design
an approximate variable elimination algorithm. By approximating the
binary polynomial representing the distribution before summing
out each variable we get an algorithm less restricted by the
dependence structure of the model, thus capable of handling MRFs
defined on large lattices and MRFs with larger neighbourhood structures. 
For the MRF application this
approximation defines an approximation to the normalising constant, and 
as a by-product we also get a partially ordered Markov model (POMM) 
approximation to the MRF. For the POMM approximation
we can calculate the normalising
constant and evaluate the likelihood, as well as generate
realizations. We also discuss how to modify our approximation strategy
to instead get upper and lower bounds for the normalising constant, 
and how this in turn can be used to construct an interval in which 
the maximum likelihood estimate must lie. We also discuss another
variant of the approach which produces an approximate version of the Viterbi
algorithm \citep{col5}.

The article has the following layout. In Section \ref{sec:pbf} we define 
pseudo-Boolean functions and give a number of approximation theorems
for this function class. Thereafter, in Section \ref{sec:mrf} we 
introduce binary MRFs and the variable elimination algorithm,
and in Section \ref{sec:mrfApprox} we apply
the approximation theorems for pseudo-Boolean functions to define our
approximative variable elimination algorithm for binary MRFs. 
In Section \ref{sec:marg} we define a modified variant of the approximate
variable elimination algorithm through which we obtain upper and lower bounds for 
the normalising constant of a binary MRF, and we discuss how to modify the
approximation algorithm to obtain an 
approximate version of the Viterbi algorithm. We briefly discuss some 
implementational issues in Section \ref{sec:implementation}, and 
in Section \ref{sec:examples} we present simulation 
and data examples. Finally,
in Section \ref{sec:cr} we provide closing remarks.

\section{\label{sec:pbf}Pseudo-Boolean functions}
In this section we introduce the class of pseudo-Boolean functions and discuss
various aspects of approximating pseudo-Boolean functions partly based on results of
\citet{art137} and \citet{art138}.

\subsection{\label{sec:pbfdef}Definition and notation}
Let $x=(x_1,\hdots,x_n) \in \Omega = \{0,1\}^n$ be a vector of
binary variables and let $N=\{1,\hdots,n\}$ be the corresponding list of indices. Then for any subset $\Lambda
\subseteq N$ we associate an incidence vector $x$ of length $n$ whose $k$th
element is $1$ if $k\in \Lambda$ and $0$ otherwise. We refer to
an element of $x$, $x_k$, as being "on" if it has value $1$ and "off" if it is $0$. A pseudo-Boolean
function $f$, of dimension $\mbox{dim}(f) = n$, is a function that
associates a real value to each vector, $x \in \{0,1\}^n$, i.e $f:\{0,1\}^n\rightarrow \mathbb{R}$. 
\cite{book35} showed that any
pseudo-Boolean function can be expressed uniquely as a binary polynomial,
\begin{equation}
f(x) = \sum_{\Lambda \subseteq N} \beta^{\Lambda} \prod_{k \in
  \Lambda}x_k,
\label{eq:pb1}
\end{equation}
where $\beta^{\Lambda}$ are real
coefficients which we refer to as interactions. We define the degree of 
$f$, $\mbox{deg}(f)$ as the degree of the
polynomial.
In general the
representation of a function in this manner requires $2^n$
coefficients. In some cases one or more
$\beta^{\Lambda}$ might be zero and in this case a reduced
representation of the pseudo-Boolean function can be defined by
excluding some or all the terms in the sum in \eqref{eq:pb1} where
$\beta^\Lambda = 0$. Thus we get,
 \begin{equation}\label{eq:canonical}
f(x) = \sum_{\Lambda \in S} \beta^{\Lambda} \prod_{k \in
  \Lambda}x_k,
\end{equation}
where $S$ is a set of subsets of $N$ at least containing all $\Lambda
\subseteq N$ for which $\beta^\Lambda \neq 0$. We then say that $f(x)$ is
represented on $S$. Moreover, we say that our
representation of $f$ is dense if for all $\Lambda \in S$, all subsets
of $\Lambda$ are also included in $S$. The minimal dense representation of
$f$ is thereby \eqref{eq:canonical} with,
\begin{equation}
S = \{\lambda \subseteq N:\beta^\Lambda \neq 0 \text{ for some }
\Lambda \supseteq \lambda \}.
\label{eq:Sdef}
\end{equation}
Throughout this report we restrict the attention to dense
representations of pseudo-Boolean functions.

We also need notation for some subsets of $S$ and $\Omega$.
For $\lambda\in S$ we define $S_\lambda = \{\Lambda
\in S: \lambda \subseteq \Lambda\}$, the set of all
interactions that include $\lambda$. For example if
$n=3$ and $S=\{ \emptyset,\{ 1\},\{ 2\},\{ 3\},\{ 1,2\}, \{ 1,3\},
\{ 2,3\},\{ 1,2,3\}\}$, we have 
$S_{\{1,2\}} = \{\{1,2\},\{1,2,3\}\}$. 
Equivalently for the set $\Omega$, for $\lambda\in S$ we define
$\Omega_\lambda = \{x \in \Omega: x_k=1
,\text{ }\forall k \in \lambda\}$, the set of all states $x$ where
$x_k$ are on for all $k\in\lambda$. For the same $S$ as above we have 
for instance $\Omega_{\{1,2\}} = \{(1,1,0),(1,1,1)\}$. We
will use also the complements of these two
subsets, $S_\lambda^c = S \setminus S_\lambda$ and $\Omega_\lambda^c = \Omega \setminus
\Omega_\lambda$. 
Lastly we define $S_\lambda^0 = \{\Lambda \in S: \lambda
\cap \Lambda = \emptyset\}$ and $\Omega_\lambda^0 = \{x \in \Omega:x_k=0,\text{ }
\forall k \in \lambda\}$. If we think of the sets $S_\lambda$ and
$\Omega_\lambda$ as the sets where $\lambda$ is on, then $S_\lambda^0$
and $\Omega_\lambda^0$ are the sets where $\lambda$ is off. Again, using the same example $S$
as above we have for
instance $S_{\{1,2\}}^0 = \{\emptyset,3\}$ and $\Omega_{\{1,2\}}^0 =
\{\{0,0,0\},\{0,0,1\}\}$. Note that in general $S_\lambda^c \neq
S_\lambda^0$ and equivalently $\Omega_\lambda^c \neq \Omega_\lambda^0$.

\subsection{\label{sec:pbfapprox}Approximating pseudo-Boolean functions}
For a general pseudo-Boolean function, the number of possible interactions in our representation grows
exponentially with the dimension $n$. It is therefore natural to ask if we can
find an approximate representation which requires less memory. We could choose 
some set $\tilde{S} \subseteq S$ to define our approximation, thus choosing
which interactions to retain, $\tilde{S}$, and which to remove, $S
\setminus \tilde{S}$. For a given $\tilde{S}$ our interest lies in the best such approximation
according to some criteria. We define 
$A_{\tilde{S}}\{f(x)\} = \tilde{f}(x) = \sum_{\Lambda \in \tilde{S}} \tilde{\beta}^{\Lambda} \prod_{k \in \Lambda}x_k$ as the
operator which returns the approximation that, for some given approximation set $\tilde{S}$, minimises the
error sum of squares ($\mbox{SSE}$), 
\begin{equation}\label{eq:SSE}
\mbox{SSE}(f,\tilde{f}) = \sum_{x \in \Omega} \left\{f(x)-\tilde{f}(x) \right\}^2.
\end{equation}
We find the best approximation by
taking partial derivatives with respect to $\tilde{\beta}^\lambda$ for all
$\lambda \in \tilde{S}$ and setting these expressions equal to zero. This gives us a system of linear equations,
\begin{equation}
\sum_{x \in \Omega_\lambda}\tilde{f}(x) = \sum_{x \in \Omega_\lambda}\left(\sum_{\Lambda \in \tilde{S}} \tilde{\beta}^{\Lambda} \prod_{k \in
  \Lambda}x_k\right) = \sum_{x \in \Omega_\lambda}f(x), \text{ } \forall \text{ } \lambda
\in \tilde{S}.
\label{eq:equations}
\end{equation}
Existence and uniqueness of a solution is assured since we have
$|\tilde{S}|$ unknown variables and the same number of linearly 
independent equations. If $\tilde{S} \supseteq S$ the best approximation is clearly
the function itself, $\tilde{f}(x) = f(x)$. 

It is common practice in statistics and approximation theory in
general to approximate higher order terms by lower order terms. A natural way to design an 
approximation would be to let $\tilde{S}$
include all interactions of degree less than or equal to some value
$k$. In \citet{art137} the authors focus on approximations of this type
and proceed to show how the resulting system of linear equations through clever
reorganisation can be transformed into a lower triangular system. They solve this 
for $k=1$ and $k=2$ as well as proving a
number of useful properties. \citet{art138} solve this for general $k$. 

In the present article we consider the situation where $\tilde{S}$
is a dense subset of $S$. So if $\lambda \in
\tilde{S}$, then all $\Lambda \subset \lambda$ must also be included
in $\tilde{S}$. 
Clearly the approximation using all interactions up to degree $k$ is a special
case of our class of approximations. Our motivation for studying this
particular design of $\tilde{S}$ will become clear as we study the
variable elimination algorithm in Section \ref{sec:mrf}. 
We now give some useful properties of this
approximation. The two first theorems are from \citet{art137}.
\begin{theorem}
\label{th:1}
The above approximation $A_{\tilde{S}}\{f(x)\}$ is a
linear operator, i.e. for any constants $a,b \in \mathbb{R}$ and
pseudo-Boolean functions $g(x)$ and $h(x)$ represented on $S$,
we have that $A_{\tilde{S}}\{a g(x) + b h(x)\} =
a A_{\tilde{S}}\{g(x)\} + b A_{\tilde{S}}\{h(x)\}$.
\end{theorem}
\begin{theorem}
\label{th:2}
Assume we have two approximations of $f(x)$, $A_{\tilde{S}}\{f(x)\}$ and $A_{\tilde{\tilde{S}}}\{f(x)\}$, 
such that $\tilde{\tilde{S}} \subseteq \tilde{S}
\subseteq S$. Then $A_{\tilde{\tilde{S}}}[A_{\tilde{S}}\{f(x)\}] =  A_{\tilde{\tilde{S}}}\{f(x)\}$.
\end{theorem} 
Proofs can be found in \citet{art137} and in \citet{phd6}.
Since each interaction term in a pseudo-Boolean function is a pseudo-Boolean function in itself, 
Theorem \ref{th:1} is important because it means that we can approximate a pseudo-Boolean function
by approximating each of the interaction terms involved in the function
individually. Also, since the best approximation of a pseudo-Boolean
function is itself, we only need to worry about how to approximate the
interaction terms we want to remove. Theorem \ref{th:2} shows that a sequential scheme for calculating the
approximation is possible. 

Next we give two theorems characterising the properties of the error introduced by the approximation.
\begin{theorem}\label{th:3}
Assume again that we have two approximations of $f(x)$, $A_{\tilde{S}}\{f(x)\}$ 
and $A_{\tilde{\tilde{S}}}\{f(x)\}$, such that $\tilde{\tilde{S}} \subseteq \tilde{S}
\subseteq S$. Letting $\tilde{f}(x) = A_{\tilde{S}}\{f(x)\}$ and
$\tilde{\tilde{f}}(x) = A_{\tilde{\tilde{S}}}\{f(x)\}$, we 
then have $\mbox{SSE}(f,\tilde{\tilde{f}}) = \mbox{SSE}(f,\tilde{f}) +
\mbox{SSE}(\tilde{f},\tilde{\tilde{f}})$. 
\end{theorem} 
\begin{theorem}
\label{th:4}
Given a pseudo-Boolean function $f(x)$ and an approximation $\tilde{f}(x)$
constructed as described, the error sum of squares can be written as,
\begin{equation}
\sum_{x \in \Omega} \left\{f(x) - \tilde{f}(x) \right\}^2 = \sum_{\Lambda \in
  S\setminus \tilde{S}}\left[ \beta^{\Lambda} \sum_{x \in \Omega_{\Lambda}}
  \{f(x)-\tilde{f}(x)\} \right]
\end{equation}
\end{theorem}
Proofs of Theorems \ref{th:3} and \ref{th:4} are in Appendices \ref{app:proof3} and 
\ref{app:proof4}, respectively. Note that Theorem \ref{th:4} tells us that the error can be 
expressed as a sum over the $\beta$'s that we remove when constructing our
approximation. Note also the special case where $\tilde{S} = S \setminus \lambda$
for some $\lambda\in S$, i.e we remove only one interaction $\beta^{\lambda}$. Then,
\begin{equation}\label{eq:errorRemoveOne}
\sum_{x \in \Omega} \left\{f(x) - \tilde{f}(x) \right\}^2 = \beta^{\lambda} \left[
  \sum_{x \in \Omega_{\lambda}}
  \{f(x)-\tilde{f}(x)\} \right].
\end{equation}
With these theorems in hand we can go from $S$ to $\tilde{S}$ by removing all
nodes in $S \setminus \tilde{S}$. Theorems \ref{th:1} and \ref{th:2}
allow us to remove these interactions sequentially one
at a time. We start by removing the interaction (or one of, in the case
of several) $\beta^\lambda,\lambda\in S\setminus \widetilde{S}$ 
with highest degree and approximate it
by the set containing all $\Lambda \subset \lambda$. In other words,
if the interaction has degree $k=|\lambda|$ we design the $k-1$ order approximation of
that interaction term. \citet{art138} gives us the expression for
this,
\begin{equation}
\tilde{\beta}^{\Lambda} = \left\{
  \begin{array}{ll}
    \beta^{\Lambda}  +
(-1)^{|\lambda|-1-|\Lambda|} \left(\frac{1}{2}\right)^{|\lambda| -
  |\Lambda|} \beta^{\lambda} & \quad \text{if~~}\forall \Lambda \subset \lambda, \\
    \beta^{\Lambda} & \quad \text{otherwise.} \\
  \end{array} \right.
\label{eq:sol}
\end{equation}
We then proceed by removing one interaction at a time until we reach the set of interest,
$\tilde{S}$. The approximation error in one step of this procedure
is given by (\ref{eq:errorRemoveOne}) and the total approximation 
error is given as the sum of the errors in each of the approximation steps.

\subsection{\label{sec:pairwise}Second order interaction removal}
In this section we discuss pseudo-Boolean function approximation
for a specific choice of $\tilde{S}$, which is of particular interest
for the variable elimination algorithm. We show how we can construct a new
way of solving the resulting system of equations and term this approximation the
second order interaction removal (SOIR) approximation.

For $i,j\in N$, $i\neq j$ and $\{ i,j\} \in S$, assume we 
have $\tilde{S} = S_{\{i,j\}}^c$. In other words we want to remove all interactions
involving both $i$ and $j$ and approximate these by lower order
interactions. 
To find this approximation we could of course proceed as in the 
previous section, sequentially
removing one interaction at the time until we reach our desired
approximation. However, the following theorem gives explicit
expressions for both the approximation and 
the associated error. 
\begin{theorem}\label{th:6}
Assume we have a pseudo-Boolean function $f(x)$ represented on 
a dense set $S$. 
For $i,j\in N$, $i\neq j$ and $\{ i,j\}\in S$, the least squares approximation of $f(x)$
on $\tilde{S} = S_{\{ i,j\}}^c$ is given by  
\begin{equation}\label{eq:ftilde}
\tilde{f}(x)=A_{\tilde{S}}\{f(x)\} = \sum_{\Lambda\in\tilde{S}} \tilde{\beta}^\Lambda \prod_{k\in\Lambda}x_k,
\end{equation}
where $\tilde{\beta}^\Lambda$ for $\Lambda\in \tilde{S}$ is given as
\begin{equation}\label{eq:soir}
\tilde{\beta}^\Lambda = \left\{ \begin{array}{ll}
\beta^\Lambda - \frac{1}{4}\beta^{\Lambda\cup\{ i,j\}} & \mbox{~~if $\Lambda \cup \{i,j\} \in S$}, \\
\beta^\Lambda + \frac{1}{2}\beta^{\Lambda\cup\{ i\}} & \mbox{~~if $\Lambda\cup\{ i\}\in S$ 
  and $\Lambda\cup\{ j\}\not\in S$}, \\
\beta^\Lambda + \frac{1}{2}\beta^{\Lambda\cup \{ j\}} & \mbox{~~if $\Lambda\cup\{ i\}\not\in S$ 
  and $\Lambda\cup\{ j\}\in S$}, \\
\beta^\Lambda & \mbox{~~otherwise.}
\end{array}\right.
\end{equation}
The associated approximation error is
\begin{equation}\label{eq:soirerror}
f(x)-\tilde{f}(x) = \left(x_ix_j + \frac{1}{4} - \frac{1}{2}x_i -
  \frac{1}{2}x_j\right)\sum_{\Lambda \in
    S_{\{i,j\}}}\left(\beta^{\Lambda}\prod_{k \in \Lambda \setminus \{i,j\}}x_k \right).
\end{equation}
\end{theorem}
A proof is given in Appendix \ref{app:proof6}.
Clearly, the approximation solution of this theorem 
corresponds to the solution we would get using the sequential scheme discussed above, 
but (\ref{eq:soir}) is much faster to calculate and the above theorem has the advantage
of giving us a nice explicit expression for the error. 
Note that the absolute value of the parenthesis outside the sum in (\ref{eq:soirerror}) is
always $\frac{1}{4}$ and thus the absolute value of $f(x)-\tilde{f}(x)$ does not depend on $x_i$ or $x_j$.
We thereby get the following expression for the
error sum of squares,
\begin{align}
\mbox{SSE}(f,\tilde{f}) = \sum_{x \in \Omega} \{f(x)-\tilde{f}(x)\}^2 =  \frac{1}{4}\sum_{x \in
  \Omega_{\{i,j\}}}\left(\sum_{\Lambda \in
    S_{\{i,j\}}}\beta^{\Lambda}\prod_{k \in \Lambda \setminus \{i,j\}}x_k\right)^2.
\end{align}

\subsection{\label{sec:maxmin}Upper and lower bounds for pseudo-Boolean functions}
In this section we construct upper and lower bounds for
pseudo-Boolean functions.
We denote the upper and lower bounds
by $f_U(x)$ and $f_L(x)$, respectively, i.e. we require
$f_L(x)\leq f(x)\leq f_U(x)$ for all $x\in\Omega$. Just like for the 
SOIR approximation $\tilde{f}(x)$, we require that all interactions
involving both $i$ and $j$ are removed from $f_U(x)$ and 
$f_L(x)$. Using the expression for the SOIR approximation error
in (\ref{eq:soirerror}) we have
\begin{equation}
f(x) = \tilde{f}(x) + \left(x_ix_j+\frac{1}{4} -\frac{1}{2}x_i-\frac{1}{2}x_j\right)
\sum_{\Lambda\in S_{\{ i,j\}}}\left( \beta^\Lambda \prod_{k\in\Lambda\setminus\{ i,j\}}x_k\right).
\end{equation}
The terms that are constant or linear in $x_i$ and $x_j$ can be kept unchanged in $f_U(x)$
and $f_L(x)$, whereas we need to find bounds for the interaction part
\begin{equation}\label{eq:g}
g(x) = x_ix_j\sum_{\Lambda\in S_{\{ i,j\}}} \left( \beta^\Lambda\prod_{k\in\Lambda\setminus\{ i,j\}}x_k\right).
\end{equation}
As $x_j\in \{ 0,1\}$, an upper bound for $g(x)$ 
which is linear in $x_i$ and constant
as a function of $x_j$ is
\begin{equation}\label{eq:upper1}
g(x) \leq x_i\max\left\{ 0,
  \sum_{\Lambda\in S_{\{ i,j\}}}\left( \beta^\Lambda \prod_{k\in\Lambda\setminus\{ i,j\}}x_k\right)\right\}.
\end{equation}
An upper bound which is linear $x_j$ and constant as
a function of $x_i$ is correspondingly found by using that $x_i\in \{ 0,1\}$. 
Moreover, any convex linear combination of these two bounds
is also a valid upper bound for $g(x)$. 
Similar reasoning for a lower bound produces the same type of expressions,
except that the $\max$ operators are replaced by
$\min$ operators.

The bounds defined above are clearly valid upper and lower bounds,
and by construction they have no interactions involving both $i$ and $j$.
However, in the next section we need to find the canonical forms, given by 
(\ref{eq:canonical}), of the bounds and then the computational complexity of
constructing these representations is important. 
Focusing on the bound in (\ref{eq:upper1}),
the $\max$ function is a function of
$d_{ij}=|\{ \Lambda\in S_{\{ i,j\}} : |\Lambda|=3\}$ variables, so 
we need to compute the values of $2^{d_{ij}}$ interaction coefficients. 
This is
computationally feasible only if $d_{ij}$ is small enough. If $d_{ij}$
is too large we need to consider computationally cheaper, and coarser, 
bounds. To see how this can be done, first 
note that for any $r\in N\setminus \{ i,j\}$ 
and $\{ i,j,r\}\in S_{\{ i,j\}}$ we have
$S_{\{ i,j,r\}}\subset S_{\{ i,j\}}$ and thereby the $g(x)$ defined
in (\ref{eq:g}) can alternatively be expressed as
\begin{equation}\label{galt}
g(x) = x_i x_j x_r \sum_{\Lambda\in S_{\{ i,j,r\}}} \left(
\beta^\Lambda \prod_{k\in \Lambda\setminus\{ i,j,r\}} x_k\right) +
x_i x_j \sum_{\Lambda\in S_{\{ i,j\}} \setminus S_{\{ i,j,r\}}} \left(
\beta^\Lambda \prod_{k\in\Lambda\setminus\{ i,j\}}\right). 
\end{equation}
An alternative upper bound can then be defined by following a similar strategy
as above, but for each of the two terms in (\ref{galt}) separately. This gives
the upper bound
\begin{equation}\label{boundalt}
f(x) \leq \sum_{\Lambda\in S_{\{ i,j\}}^c} \beta^\Lambda \prod_{k\in \Lambda} x_k + 
x_i\max\left\{ 0,\sum_{\Lambda\in S_{\{ i,j,r\}}}\left(
\beta^\Lambda \prod_{k\in \Lambda\setminus\{ i,j,r\}}x_k\right)\right\}
\end{equation}
\[
+ x_i\max \left\{0,\sum_{\Lambda\in S_{\{ i,j\}}\setminus S_{\{ i,j,r\}}}\left(
\beta^\Lambda\prod_{k\in \Lambda\setminus\{ i,j\}} x_k\right)\right\},
\]
and the corresponding lower bound is again given by the same type of 
expression except that the $\max$ operators are
replaced by $\min$ operators. One should note the canonical form of the bound in 
(\ref{boundalt}) can be done for each $\max$ term separately. Moreover,
as both $\max$ terms in (\ref{boundalt}) are functions of strictly less than $d_{ij}$ variables
the computational complexity of finding the canonical representation of the bound in 
(\ref{boundalt}) is smaller than the complexity of the 
corresponding operation for the bound in (\ref{eq:upper1}). 
However, for one or both of the $\max$ functions in (\ref{boundalt}), the task of transforming it 
into the canonical form may still be too computationally expensive. If so, the process of splitting
a sum into a sum of two sums must be repeated. For example, if the first $\max$ function is
problematic, one needs to locate an $s\in N\setminus\{ i,j,r\}$ so that $\{ i,j,r,s\}\in S_{\{ i,j,r\}}$ and the sum 
over $\Lambda\in S_{\{ i,j,r\}}$ can be split into a sum over $\Lambda \in S_{\{  i,j,r,s\}}$ and 
a sum over $\Lambda \in S_{\{ i,j,r\}}\setminus S_{\{ i,j,r,s\}}$ and finding bounds as before. 
By repeating this process
sufficiently many times one will eventually end up with a bound consisting of a sum of $\max$ 
terms that can be transformed to the canonical form in a reasonable computation time.

\section{\label{sec:mrf}MRFs and the variable elimination algorithm}
In this section we give a short introduction to binary MRFs. In particular we explain how the
variable elimination algorithm can be applied to this class of models and
point out its computational limitation. For a general
introduction to MRFs see \citet{art3} or \citet{book10} and for more
on the connection between binary MRFs and pseudo-Boolean functions see
\citet{art130}. For more on the variable elimination algorithm and
applications to MRFs see \citet{art100} and \citet{art105}.

\subsection{Binary Markov random fields}
Assume we have a vector of $n$ binary variables $x = \{x_1,\hdots,x_n\} \in \Omega =
\{0,1\}^n$, $N=\{1,\hdots,n\}$. Let $\mathcal{N} =
\{\mathcal{N}_1,\hdots,\mathcal{N}_n\}$ denote a neighbourhood system where
$\mathcal{N}_k$ denotes the set of indices of nodes that are neighbours of node
$k$. As usual we require a symmetrical neighbourhood system, so if $i
\in \mathcal{N}_j$ then $j \in \mathcal{N}_i$, and by convention a node
is not a neighbour of itself. Then $x$ is a binary MRF with respect to a neighbourhood
system $\mathcal{N}$ if $p(x) > 0$ for all $x \in \Omega$ and the full conditionals $p(x_k|x_{-k})$ have
the Markov property,
\begin{equation}
p(x_k|x_{-k}) = p(x_k|x_{\mathcal{N}_k}) \text{ } \forall \text{ }x
\in \Omega \mbox{~and~}k\in N,
\label{fullconditionals}
\end{equation}
where $x_{\mathcal{N}_k} = (x_i:i \in \mathcal{N}_k)$. A clique $\Lambda$ is 
a set $\Lambda \subseteq N$ such
that for all pairs $i,j\in \Lambda$ we have $i \in \mathcal{N}_j$. 
A clique is a maximal clique if it is not a subset of another clique. The set of all
maximal cliques we denote by $\mathcal{C}$. The Hammersley-Clifford theorem
\citep{art3,col7} tells us
that we can express the distribution of $x$ either through the full
conditionals in \eqref{fullconditionals} or through clique potential functions,  
\begin{equation}\label{eq:p}
p(x) = \frac{1}{c}  \exp\{U(x)\} = \frac{1}{c} \exp\left\{\sum_{\Lambda \in \mathcal{C}}U_\Lambda(x_\Lambda)\right\},
\end{equation}
where $c$ is a normalising constant, $U_\Lambda(x_\Lambda)$ is a 
potential function for a clique $\Lambda$ and $x_\Lambda=(x_i:i\in\Lambda)$. $U(x)$ is commonly referred to as the
energy function. From the previous sections we know that $U(x)$ is a
pseudo-Boolean function and can be expressed as,
\begin{equation}
U(x) = \sum_{\Lambda \subseteq N} \beta^{\Lambda} \prod_{k \in
  \Lambda}x_k =\sum_{\Lambda \in S} \beta^{\Lambda} \prod_{k \in
  \Lambda}x_k,
\label{eq:ux}
\end{equation}
where $S$ is defined as in \eqref{eq:Sdef}. For a given energy
function $U(x)$, \citet{art130} show how the interactions $\beta^\Lambda$ can be calculated 
recursively by evaluating $U(x)$. Moreover, \citet{art130} show that $\beta^\Lambda=0$
whenever $\Lambda$ is not a clique. From this we understand that 
it is important that we represent $U(x)$ on $S$ as defined in (\ref{eq:Sdef}), and not
use the full representation.

\subsection{\label{sec:exactVEA}The variable elimination algorithm}
As always the problem when evaluating the likelihood or generating
samples from MRFs is that $c$ is a function of the model parameters and in
general unknown. Calculation involves a sum over $2^n$
terms,
\begin{equation}
c = \sum_{x \in \Omega} \exp\left\{U(x)\right\} = \sum_{x \in \Omega} \exp\left(\sum_{\Lambda
    \in S} \beta^{\Lambda} \prod_{k \in \Lambda}x_k\right).
\label{eq:normsum1}
\end{equation}
The variable elimination algorithm \citep{art100,art105}
calculates the sum in \eqref{eq:normsum1} by
taking advantage of the fact that we
can calculate this sum more efficiently by factorising the
un-normalised distribution. We now cover this
recursive procedure. 

Clearly we can always split the set $S$ into two parts, $S_{\{i\}}$ and
$S_{\{i\}}^c$ where $i\in N$. Thus we can
split the energy function in \eqref{eq:ux} into a sum of two sums,
\begin{equation}
U(x) = \sum_{\Lambda \in S_{\{i\}}^c} \beta^{\Lambda} \prod_{k \in
  \Lambda}x_k + \sum_{\Lambda \in S_{\{i\}}} \beta^{\Lambda} \prod_{k \in
  \Lambda}x_k.
\label{uxsplit}
\end{equation}
Note that the first sum contains no interaction terms involving
$x_i$. Letting $x_{-i} = (x_1,\hdots,$ $x_{i-1},x_{i+1},\hdots,x_n)$, we
note that this is essentially equivalent to factorising $p(x) =
p(x_i|x_{-i})$ $p(x_{-i})$, since
\begin{equation}
p(x_i|x_{-i}) \propto \exp\left(\sum_{\Lambda \in S_{\{i\}}} \beta^{\Lambda} \prod_{k \in
  \Lambda}x_k \right).
\end{equation}
By summing out $x_i$ from $p(x)$ we get
the distribution of $p(x_{-i})$. Taking advantage of the split in
\eqref{uxsplit} we can write this as,
\begin{equation}\label{eq:marginal}
p(x_{-i}) = \sum_{x_i}p(x) = \frac{1}{c}\exp\left(\sum_{\Lambda \in S_{\{i\}}^c}
    \beta^{\Lambda} \prod_{k \in \Lambda}x_k \right) \sum_{x_i} \exp\left(\sum_{\Lambda \in S_{\{i\}}}
    \beta^{\Lambda} \prod_{k \in \Lambda}x_k \right).
\end{equation}
The last sum over $x_i$ can be expressed as the exponential of a new
binary polynomial, i.e.
\begin{equation}
\exp\left(\sum_{\Lambda \subseteq \mathcal{N}_i} \check{\beta}^{\Lambda} \prod_{k \in
  \Lambda}x_k\right) = \sum_{x_i} \exp\left(\sum_{\Lambda \in S_{\{i\}}}
    \beta^{\Lambda} \prod_{k \in \Lambda}x_k \right),
\label{eq:newpbf}
\end{equation}
where the interactions $\check{\beta}^\Lambda$ can be sequentially
calculated by evaluating the sum over $x_i$ in \eqref{eq:newpbf} as 
described in \citet{art130}. 
Note that this new function is a pseudo-Boolean
function potentially of full degree. The number of non-zero interactions in this
representation could be up to $2^{|\mathcal{N}_i|}$. Summing out $x_i$
leaves us with a new MRF with a new neighbourhood system. This is the first step in a sequential
procedure for calculating the normalising constant $c$. In each step
we sum over one of the remaining variables by splitting the energy
function as above. Repeating this procedure until we have summed out
all the variables naturally yields the normalising constant.

The computational bottleneck for this algorithm occurs when
representing the sum in \eqref{eq:newpbf}. Assume we have summed out
variables $x_{1:i-1} = (x_1,\hdots,x_{i-1})$, have an MRF with a neighbourhood system
$\check{\mathcal{N}} = \{\check{\mathcal{N}}_i,\hdots,\check{\mathcal{N}}_n\}$ and want 
to sum out $x_i$. If
$\check{\mathcal{N}}_i$ is too large we run into trouble 
with the sum corresponding to
\eqref{eq:newpbf} since this requires us to compute and store up to
$2^{\check{\mathcal{N}}_i}$ interaction terms. In models where
$\check{\mathcal{N}}_i$ increases as we sum out variables the
exponential growth causes us to run into problems very quickly. As a practical example of
this consider the Ising model defined on a lattice. Assuming we sum out variables in the
lexicographical order, the size of the 
neighbourhood will grow to the number of rows in our lattice. 
This thus restricts the number of rows in the lattice to less or 
equal to $20$ for practical purposes.

\section{\label{sec:mrfApprox}Construction of an approximate variable 
elimination algorithm}
In this section we include the approximation results of Section
\ref{sec:pbf} in the variable elimination algorithm described in the previous
section to obtain an approximate, but computationally more efficient variant of
the variable elimination algorithm.
To create an algorithm that is computationally viable we must seek to
control $|\check{\mathcal{N}}_i| = \eta_i$ as we sum out variables. If this
neighbourhood becomes too large, we run into problems both with
memory and computation time. Our idea is to construct an
approximate representation of the MRF before summing out each
variable. The approximation is chosen so that
$\eta_i \leq \nu$, where $\nu$ is an input to our
algorithm. Given a design for the approximation we then want to
minimise the error sum of squares of our energy function.

Assume we have an MRF and have (approximately) summed out variables $x_{1:i-1}$, so we
currently have an MRF with a neighbourhood structure
$\check{\mathcal{N}}$ and energy function $\check{U}(x_{i:n}) = \sum_{\Lambda \in \check{S}}
\check{\beta}^{\Lambda}\prod_{k \in \Lambda}x_k $, so,
\begin{equation}
c = \sum_{x_{i:n}} \exp\left\{\check{U}(x_{i:n}) \right\}.
\end{equation}
If $\eta_i$ is too large we run into problems when summing over
$x_i$. Our strategy for overcoming this problem is first to create an
approximation of the energy function
$\check{U}(x_{i:n})$,
\begin{equation}
\check{U}(x_{i:n}) = \sum_{\Lambda \in \check{S}}
\check{\beta}^{\Lambda}\prod_{k \in \Lambda}x_k \approx
\tilde{U}(x_{i:n}) = \sum_{\Lambda \in \tilde{S}}
\tilde{\beta}^{\Lambda}\prod_{k \in \Lambda}x_k. 
\end{equation}
We control the size of $\eta_i$, by designing our approximation set
$\tilde{S}$ and thus the new approximate neighbourhood $\tilde{\mathcal{N}}$ in such a
way that $|\tilde{\mathcal{N}}_i| = \tilde{\eta}_i \leq
\nu$. Assuming we can do this, we could construct an approximate
variable elimination algorithm where we check the size of the neighbourhood
$\eta_i$ before summing out each variable. If this is greater than
the given $\nu$ we approximate the energy function
before summation. This leaves two questions; how do we choose the set
$\tilde{S}$ and how do we define the approximation?
The two questions are obviously linked, however we start by looking
more closely at how we may choose the set $\tilde{S}$. Our
tactic is to reduce $\eta_i$ by one at the time. 
To do this we need to design $\tilde{S}$ in such
a way that $i$ and some node $j$ are no longer neighbours. 
Doing this is equivalent to requiring all interactions
$\tilde{\beta}^{\Lambda}$, involving both $i$ and $j$ to be zero. As before
we denote the subset of all interactions involving $i$ and $j$ as
$\check{S}_{\{i,j\}} \subseteq \check{S}$ and construct our
approximation set as in Section \ref{sec:pairwise}, defining $\tilde{S} = \check{S}\setminus
\check{S}_{\{i,j\}}$. Our approximation is defined by the
equations corresponding to \eqref{eq:equations} and using the results
from Section \ref{sec:pairwise}, the solution is easily available. 
We can then imagine a scheme where we reduce
$\eta_i$ one at a time until we reach our desired size $\nu$. This
leaves the question of how to choose $j$. One could calculate the
$\mbox{SSE}$ for all possibilities of $j$ and choose the value of $j$
that has the minimum $\mbox{SSE}$. However, this may be
computationally very expensive and would in many cases dominate the
total computation time of our algorithm. Instead we propose to compute 
an approximate upper bound for $|f(x)-\widetilde{f}(x)|$ for all 
values of $j$ and to select the value of $j$ that minimises this 
approximate bound. We define the approximate bound by first defining
a modified version $f^\star(x)$ of $f(x)$, where we set all
first, second and third order interactions for $f^\star(x)$ 
equal to corresponding quantities for $f(x)$, and set all 
fourth and higher order interactions for $f^\star(x)$ equal to zero.
We then define the approximate upper bound as the exact
upper bound for $|f^\star(x)-\widetilde{f}^\star(x)|$. As $f^\star(x)$
has no non-zero fourth or higher order interactions upper and lower
bounds for $f^\star(x)-\widetilde{f}^\star(x)$ can readily be computed
as discussed in Section \ref{sec:maxmin}.

One should note that Theorem \ref{th:2} means that after reducing $\eta_i$ by $\eta_i-\nu$
our approximation is still optimal for the given selection of
$j$'s. However, there is no guarantee that our selection of $j$'s is
optimal, and it is possible that we could have obtained a better set of 
$j$'s by looking at the error from
reducing $\eta_i$ by more than one at the time.

Using this approximate variable elimination algorithm 
we can define a corresponding approximate model through a product of the approximate
conditional distributions,
\begin{equation}\label{eq:pommapprox}
\tilde{p}(x) = \tilde{p}(x_1|x_{2:n}) \hdots  \tilde{p}(x_{n-1}|x_{n})\tilde{p}(x_n),
\end{equation}
which is a POMM \citep{art119}. One of the aspects we wish
to investigate in the results section is to what extent this distribution can
mimic some of the attributes of the original MRF. Clearly, to sample from 
$\tilde{p}(x)$ is easy via a backward pass, first simulating $x_n$ from 
$\tilde{p}(x_n)$, thereafter simulating $x_{n-1}$ from $\tilde{p}(x_{n-1}|x_n)$
and so on. One should note that two versions (\ref{eq:pommapprox}) can be defined.
The first version is obtained by taking $\widetilde{p}(x_i|x_{i+1:n})$ to be the 
resulting conditional distribution after we have (approximately) summed out $x_{1:i-1}$.
In $\widetilde{p}(x_i|x_{i+1:n})$ we then have no guarantee for how many of the 
elements in $x_{i+1:n}$ the variable $x_i$ really depends on, and it may be 
computationally expensive to compute the normalising constant of this 
conditional distribution for all values of $x_{i+1:n}$.
The alternative is to let $\widetilde{p}(x_i|x_{i+1:n})$ be the 
resulting conditional distribution after one has both (approximately) summed
out $x_{1:i-1}$ and done the necessary approximations so that $x_i$ is linked with 
at most $\nu$ of the elements in $x_{i+1:n}$. Then we know that $x_i$ is linked to at most 
$\nu$ of the variables in $x_{i+1:n}$ in the conditional distribution also, and we have 
an upper limit for the computational complexity of computing the 
normalising constant in the conditional distributions. Which of the two 
versions of $\widetilde{p}(x)$ one should use depends on what one intends to use
$\widetilde{p}(x)$ for. One would expect the first version to be the 
best approximation of $p(x)$, but for some applications it may be computationally
infeasible. We discuss this issue further in the examples in Section 
\ref{sec:examples}.

\section{\label{sec:marg}Bounds and alternative marginalisation operations}
In this section we consider some variations of the approximate algorithm defined above.
We fist discuss how the results in Section \ref{sec:maxmin} can be used 
to modify the procedure to get upper and lower bounds for the normalising constant. 
Thereafter we 
consider how approximations (or bounds) for some other quantities can be found by
replacing the summation operation in the above algorithm 
with alternative marginalisation operations.

\subsection{\label{sec:bound}Bounds for the normalising constant}
The approximate value for the normalising constant, $c$, found by the algorithm 
in Section \ref{sec:mrfApprox} comes without any measure of precision. 
Using the results of Section \ref{sec:maxmin} we can modify the algorithm described
above and instead find upper or lower bounds for $c$. 

Our point of origin for finding a bound $c_L$ (or $c_U$) such that $c_L\leq c$
(or $c_U \geq c$) is the approximate variable elimination algorithm described 
in the previous section. An iteration of this algorithm consists of two steps. 
First the energy function is replaced by an approximate
energy function and, second, we sum over the chosen
variable. To construct an upper or lower bound we simply change the first step. 
Instead of replacing the energy function by an approximation we
replace it with a lower (or upper) bound. To define such a bound we
adopt the strategy discussed in Section \ref{sec:maxmin}. 
Letting $x_i$ denote the next variable to sum over, we first have to 
decide which second order interaction to remove, i.e. the value of $j$
in Section \ref{sec:maxmin}. For this we follow
the same strategy as for the approximate variable elimination algorithm
discussed above. Then we use the bound in (\ref{eq:upper1}) 
whenever $d_{ij}\leq \nu$, where $d_{ij}$ is as defined in 
Section \ref{sec:pairwise} and $\nu$ is the same input parameter to the
algorithm as in Section \ref{sec:mrfApprox}. If $d_{ij} > \nu$ we use a coarser 
bound as discussed in Section \ref{sec:mrfApprox}. In the definition of these
coarser bounds it remains to specify how to choose the 
value of $r$ in (\ref{boundalt}), and if necessary also the value of 
$s$ and so forth. For definiteness we here describe how we 
select the value of $r$, but follow the same strategy for all such choices. 
For the choice of $r$ we adopt a similar strategy as for $j$ in Section 
\ref{sec:mrfApprox}, but now include all interactions up to order 
four in the definition of $f^\star(x)$. When selecting a value
for $s$ we include in $f^\star(x)$ interactions up to order five and so fourth.

\subsection{Alternative marginalisation operations}
The exact variable elimination algorithm finds 
the normalising constant of an MRF by 
summing over each variable in turn. As also discussed in \citet{book34} for the 
junction tree algorithm, other quantities of interest can be
found by replacing the summation operation by alternative marginalisation 
operations. Two quantities of particular interest in our setting is to find 
the state $x$ which maximises $U(x)$, and to compute moments of $x$. In the 
following we first consider the maximisation problem and thereafter
the computation of moments

\subsubsection{\label{sec:maximisation}Maximisation}

By replacing the summation over $x_i$ in the exact variable elimination 
algorithm with a maximisation over $x_i$, the algorithm returns
the maximal value of $\exp\{ U(x)\}$ over $x\in\Omega$. By a following 
backward scan it is also possible to find the value of $x$ which
maximises $\exp\{ U(x)\}$. The forward and 
backward passes are together known as the Viterbi 
(1967)\nocite{art140} algorithm. Just as summation over $x_i$ becomes
computationally infeasible if the number of neighbours to 
node $i$ is too large, the maximisation over $x_i$ is also 
computationally infeasible in this situation. 

To see how to construct an approximate Viterbi algorithm, 
recall that our approximate variable elimination algorithm consists 
of two steps in each iteration. First to approximate the energy function and 
then to sum over a variable. To get an approximate Viterbi
algorithm we simply replace the second step, so instead of summing over a 
variable we take the maximum over that variable. Note that a lower or
upper bound for $\exp\{ U(x)\}$ can be found by replacing the 
approximation with a lower or upper bound as discussed 
in Section \ref{sec:bound}.

\subsubsection{Moments}

Consider the problem of computing a moment $\mbox{E}\{\psi(x)\}$,
where $x$ is distributed 
according to an MRF $p(x)$, and $\psi(x)$ is a given function of $x\in\Omega$.
Inserting the expression for $p(x)$ in (\ref{eq:p}) we get
\begin{equation}\label{eq:moment}
\mbox{E}\{\psi(x)\} = \frac{1}{c}\sum_{x\in\Omega} \exp\{U(x)+\ln\psi(x)\}.
\end{equation}
Thereby $\mbox{E}\{\psi(x)\}$ can be found by running the exact variable 
elimination algorithm twice, first as described in Section 
\ref{sec:exactVEA} to find $c$ and thereafter with the energy function 
redefined as $U(x)+\ln\psi(x)$ to find the sum in (\ref{eq:moment}).
In general the computational complexity of the second run of the 
variable elimination algorithm is much higher than for the 
first run, but if the pseudo-Boolean function $\psi(x)$ can be 
represented on the same set $S$ as the energy function $U(x)$ both 
runs are of the same complexity.
We obtain an approximation to $\mbox{E}\{\psi(x)\}$ simply by adopting the 
approximate variable elimination algorithm instead of the exact one.
To obtain a lower (upper) bound for $\mbox{E}\{\psi(x)\}$ we can divide 
a lower (upper) bound for the sum in (\ref{eq:moment}) with an upper
(lower) bound for $c$.

\section{\label{sec:implementation}Some implementational issues}
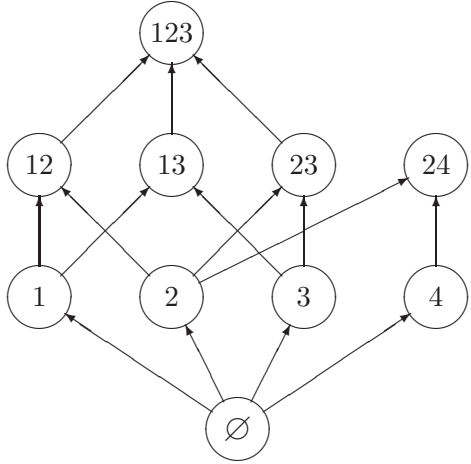
\begin{figure}
\begin{center}
\vspace*{-0.9cm}
\begin{picture}(400,200)(-200,0)
\put(0,0){\circle{23}}
\put(0,0){\makebox(0,0){$\emptyset$}}
\put(-9.57,6.38){\vector(-3,2){55.86}}
\put(9.57,6.38){\vector(3,2){55.86}}
\put(-5.14,10.29){\vector(-1,2){14.71}}
\put(5.14,10.29){\vector(1,2){14.71}}

\put(-75,50){\circle{23}}
\put(-75,50){\makebox(0,0){$1$}}
\put(-75,61.5){\vector(0,1){27}}
\put(-66.87,58.13){\vector(1,1){33.74}}

\put(-25,50){\circle{23}}
\put(-25,50){\makebox(0,0){$2$}}
\put(-33.13,58.13){\vector(-1,1){33.74}}
\put(-16.87,58.13){\vector(1,1){33.74}}
\put(-14.71,55.14){\vector(2,1){79.43}}

\put(25,50){\circle{23}}
\put(25,50){\makebox(0,0){$3$}}
\put(16.87,58.13){\vector(-1,1){33.74}}
\put(25,61.5){\vector(0,1){27}}

\put(75,50){\circle{23}}
\put(75,50){\makebox(0,0){$4$}}
\put(75,61.5){\vector(0,1){27}}

\put(-75,100){\circle{23}}
\put(-75,100){\makebox(0,0){$12$}}
\put(-66.86,108.13){\vector(1,1){33.74}}

\put(-25,100){\circle{23}}
\put(-25,100){\makebox(0,0){$13$}}
\put(-25,111.5){\vector(0,1){27}}

\put(25,100){\circle{23}}
\put(25,100){\makebox(0,0){$23$}}
\put(16.87,108.13){\vector(-1,1){33.74}}

\put(75,100){\circle{23}}
\put(75,100){\makebox(0,0){$24$}}

\put(-25,150){\circle{23}}
\put(-25,150){\makebox(0,0){$123$}}

\end{picture}
\end{center}
\caption{\label{fig:dag}The directed acyclic graph used to represent a
pseudo-Boolean function represented on $S=\{ \emptyset,\{ 1\},\{ 2\},\{ 3\},
\{ 4\},\{ 1,2\},\{ 1,3\},\{ 2,3\},\{ 1,2,3\}\}$.}
\end{figure}

When implementing the exact and approximate algorithms discussed above we
need to use one (or more) data structure(s) for storing our representation 
of pseudo-Boolean functions. The operations we need to perform on 
pseudo-Boolean functions fall into two categories. The first
is to compute and store all interaction parameters of a given 
pseudo-Boolean function without any (known) Markov structure. The 
computation of the interaction parameters $\check{\beta}^\Lambda$ defined
by (\ref{eq:newpbf}) is of this form, and so is the corresponding operation
for the $\max$ terms in (\ref{eq:upper1}) and (\ref{boundalt}). When doing this type
of operations we are simply numbering all the interaction parameters
in some order and storing their values in a vector. The values are then 
fast to assess and the necessary computations can be done efficiently.
The second type of operation we need to do on a pseudo-Boolean function
consists of operations on functions defined as in (\ref{eq:canonical}), 
functions for which a lot of the interaction parameters are zero. The 
approximation operation defined in Theorem \ref{th:6} is of this type.
We then need to adopt a data structure which stores an 
interaction parameter $\beta^\Lambda$ only if $\Lambda\in S$. We use a 
directed acyclic graph (DAG) for this, where we have one node for 
each $\Lambda\in S$ and a node $\lambda\in S$ is a child of 
another node $\Lambda\in S$ if and only if 
$\lambda = \Lambda \cup \{ i\}$ for some $i\in N\subset\Lambda$.
An illustration for $N=\{ 1,2,3,4\}$ and $S=\{ \emptyset,\{ 1\},\{ 2\},\{ 3\},
\{ 4\},\{ 1,2\},\{ 1,3\},\{ 2,3\},\{ 1,2,3\}\}$ is shown in Figure
\ref{fig:dag}. 
The value of $\beta^\Lambda$ is stored in the
node $\Lambda$, and the arrows in the figure are represented as pointers.
To assess the value of an interaction parameter $\beta^\Lambda$ in 
this data structure we need to follow the pointers from the root
to node $\Lambda$. This is clearly less efficient than in the 
vector representation discussed above, but this is the cost one has to
pay to reduce the memory requirements. One should also 
note that such a DAG representation is very convenient when
computing the approximation defined by Theorem \ref{th:6}. What 
one needs to do is first to clip out the subgraph which 
corresponds to $S_{\{ i,j\}}$. Thereafter one should traverse that 
subgraph and for each node in the subgraph add the required quantity
to three interaction parameters in the remaining DAG, as specified by
(\ref{eq:soir}).

\section{\label{sec:examples}Simulation and data examples}
In this section we first present the results of a number of simulation 
exercises to evaluate the quality of the approximations and bounds.
Thereafter we present some simulation examples to demonstrate possible
applications of the approximations and bounds we have introduced.
Finally, we use our approximation in the evaluation of a data set of 
cancer mortality from the United States.
In all the examples we adopt the approximation and the bounds
defined in Sections \ref{sec:mrfApprox} and \ref{sec:marg}.

\subsection{\label{sec:models}Models}
In the simulation examples we consider two classes of MRFs.
The first class we consider is the Ising model \citep{art17}. 
The energy function can then be expressed as
\begin{equation}\label{eq:Ising}
U(x) = \theta\sum_{i\sim j} I(x_i=x_j),
\end{equation}
where the sum is over all first order neighbourhood pairs, $\theta$
is a model parameter, and $I(x_i=x_j)$ is the indicator function
and takes value $1$ if $x_i=x_j$ and $0$ otherwise. We present results 
for $\theta=0.4$, $0.6$, $0.8$ and $-\ln (\sqrt{2}-1)$, where the last 
value is the critical value in the infinite lattice case.
Representing
$U(x)$ as a binary polynomial is done by recursively calculating
the first and second order interactions, for details see
\citet{art130}.

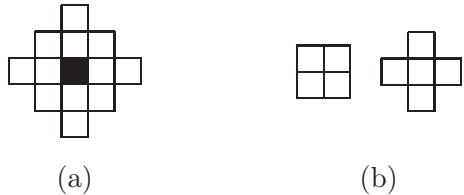
\begin{figure}
\begin{center}

\begin{tabular}{ccc}
\begin{picture}(50,50)(0,-10)
\put(0,-5){\line(0,1){10}}
\put(10,-15){\line(0,1){30}}
\put(20,-25){\line(0,1){50}}
\put(30,-25){\line(0,1){50}}
\put(40,-15){\line(0,1){30}}
\put(50,-5){\line(0,1){10}}

\put(20,-25){\line(1,0){10}}
\put(10,-15){\line(1,0){30}}
\put(0,-5){\line(1,0){50}}
\put(0,5){\line(1,0){50}}
\put(10,15){\line(1,0){30}}
\put(20,25){\line(1,0){10}}

\put(20,-4.5){\line(1,0){10}}
\put(20,-4){\line(1,0){10}}
\put(20,-3.5){\line(1,0){10}}
\put(20,-3){\line(1,0){10}}
\put(20,-2.5){\line(1,0){10}}
\put(20,-2){\line(1,0){10}}
\put(20,-1.5){\line(1,0){10}}
\put(20,-1){\line(1,0){10}}
\put(20,-0.5){\line(1,0){10}}
\put(20,0){\line(1,0){10}}
\put(20,0.5){\line(1,0){10}}
\put(20,1){\line(1,0){10}}
\put(20,1.5){\line(1,0){10}}
\put(20,2){\line(1,0){10}}
\put(20,2.5){\line(1,0){10}}
\put(20,3){\line(1,0){10}}
\put(20,3.5){\line(1,0){10}}
\put(20,4){\line(1,0){10}}
\put(20,4.5){\line(1,0){10}}
\end{picture} & 
~~~~~~~~ &
\begin{tabular}{cc}
\begin{picture}(20,50)(0,-20)
\put(0,0){\line(1,0){20}}
\put(0,10){\line(1,0){20}}
\put(0,20){\line(1,0){20}}
\put(0,0){\line(0,1){20}}
\put(10,0){\line(0,1){20}}
\put(20,0){\line(0,1){20}}
\end{picture} &
\begin{picture}(30,50)(0,-20)
\put(10,-5){\line(0,1){30}}
\put(20,-5){\line(0,1){30}}
\put(0,5){\line(1,0){30}}
\put(0,15){\line(1,0){30}}
\put(0,5){\line(0,1){10}}
\put(30,5){\line(0,1){10}}
\put(10,-5){\line(1,0){10}}
\put(10,25){\line(1,0){10}}
\end{picture}
\end{tabular} \\
(a) & & (b)
\end{tabular}
\end{center}
\caption{\label{fig:maxCliques}(a) The third-order
neighbourhood structure used in the higher-order interaction MRF. The white nodes
are neighbour to the black node. (b) The corresponding two types types of maximal cliques.}
\end{figure}

In the second class of MRFs we assume a third order neighbourhood.
Then each node sufficiently far away from the borders has 12 neighbours
and there exists two types of maximal cliques, see Figure 
\ref{fig:maxCliques}.
In contrast to the Ising model this model includes interactions of higher order
than two, and we denote it the higher-order interaction MRF.
For the two types of maximal cliques we adopt potential functions that are
invariant under rotation, reflection and when interchanging $0$ and $1$. 
With these restrictions the potential functions for the $2\times 2$ and five node
cliques can take four and six values, respectively. We define two models of this 
type and the potentials for the various clique configurations are given in 
Figure \ref{fig:pot}. 
\begin{figure}
\begin{center}
\begin{tabular}{c|cccccccccc} 
Configuration & \begin{picture}(10,15)(0,0)
\put(0,0){\line(1,0){10}}
\put(0,5){\line(1,0){10}}
\put(0,10){\line(1,0){10}}
\put(0,0){\line(0,1){10}}
\put(5,0){\line(0,1){10}}
\put(10,0){\line(0,1){10}}
\end{picture} &
\begin{picture}(10,15)(0,0)
\put(0,0){\line(1,0){10}}
\put(0,5){\line(1,0){10}}
\put(0,10){\line(1,0){10}}
\put(0,0){\line(0,1){10}}
\put(5,0){\line(0,1){10}}
\put(10,0){\line(0,1){10}}

\put(0,5.5){\line(1,0){5}}
\put(0,6.0){\line(1,0){5}}
\put(0,6.5){\line(1,0){5}}
\put(0,7.0){\line(1,0){5}}
\put(0,7.5){\line(1,0){5}}
\put(0,8.0){\line(1,0){5}}
\put(0,8.5){\line(1,0){5}}
\put(0,9.0){\line(1,0){5}}
\put(0,9.5){\line(1,0){5}}
\end{picture} &
\begin{picture}(10,15)(0,0)
\put(0,0){\line(1,0){10}}
\put(0,5){\line(1,0){10}}
\put(0,10){\line(1,0){10}}
\put(0,0){\line(0,1){10}}
\put(5,0){\line(0,1){10}}
\put(10,0){\line(0,1){10}}

\put(0,5.5){\line(1,0){10}}
\put(0,6.0){\line(1,0){10}}
\put(0,6.5){\line(1,0){10}}
\put(0,7.0){\line(1,0){10}}
\put(0,7.5){\line(1,0){10}}
\put(0,8.0){\line(1,0){10}}
\put(0,8.5){\line(1,0){10}}
\put(0,9.0){\line(1,0){10}}
\put(0,9.5){\line(1,0){10}}
\end{picture} &
\begin{picture}(10,15)(0,0)
\put(0,0){\line(1,0){10}}
\put(0,5){\line(1,0){10}}
\put(0,10){\line(1,0){10}}
\put(0,0){\line(0,1){10}}
\put(5,0){\line(0,1){10}}
\put(10,0){\line(0,1){10}}

\put(0,5.5){\line(1,0){5}}
\put(0,6.0){\line(1,0){5}}
\put(0,6.5){\line(1,0){5}}
\put(0,7.0){\line(1,0){5}}
\put(0,7.5){\line(1,0){5}}
\put(0,8.0){\line(1,0){5}}
\put(0,8.5){\line(1,0){5}}
\put(0,9.0){\line(1,0){5}}
\put(0,9.5){\line(1,0){5}}

\put(5,0.5){\line(1,0){5}}
\put(5,1.0){\line(1,0){5}}
\put(5,1.5){\line(1,0){5}}
\put(5,2.0){\line(1,0){5}}
\put(5,2.5){\line(1,0){5}}
\put(5,3.0){\line(1,0){5}}
\put(5,3.5){\line(1,0){5}}
\put(5,4.0){\line(1,0){5}}
\put(5,4.5){\line(1,0){5}}

\end{picture} & 
\begin{picture}(15,15)(0,-3)
\put(5,-2.5){\line(0,1){15}}
\put(10,-2.5){\line(0,1){15}}
\put(0,2.5){\line(1,0){15}}
\put(0,7.5){\line(1,0){15}}
\put(0,2.5){\line(0,1){5}}
\put(15,2.5){\line(0,1){5}}
\put(5,-2.5){\line(1,0){5}}
\put(5,12.5){\line(1,0){5}}
\end{picture} &
\begin{picture}(15,15)(0,-3)
\put(5,-2.5){\line(0,1){15}}
\put(10,-2.5){\line(0,1){15}}
\put(0,2.5){\line(1,0){15}}
\put(0,7.5){\line(1,0){15}}
\put(0,2.5){\line(0,1){5}}
\put(15,2.5){\line(0,1){5}}
\put(5,-2.5){\line(1,0){5}}
\put(5,12.5){\line(1,0){5}}

\put(5,3){\line(1,0){5}}
\put(5,3.5){\line(1,0){5}}
\put(5,4){\line(1,0){5}}
\put(5,4.5){\line(1,0){5}}
\put(5,5){\line(1,0){5}}
\put(5,5.5){\line(1,0){5}}
\put(5,6){\line(1,0){5}}
\put(5,6.5){\line(1,0){5}}
\put(5,7){\line(1,0){5}}
\end{picture} &
\begin{picture}(15,15)(0,-3)
\put(5,-2.5){\line(0,1){15}}
\put(10,-2.5){\line(0,1){15}}
\put(0,2.5){\line(1,0){15}}
\put(0,7.5){\line(1,0){15}}
\put(0,2.5){\line(0,1){5}}
\put(15,2.5){\line(0,1){5}}
\put(5,-2.5){\line(1,0){5}}
\put(5,12.5){\line(1,0){5}}

\put(5,8){\line(1,0){5}}
\put(5,8.5){\line(1,0){5}}
\put(5,9){\line(1,0){5}}
\put(5,9.5){\line(1,0){5}}
\put(5,10){\line(1,0){5}}
\put(5,10.5){\line(1,0){5}}
\put(5,11){\line(1,0){5}}
\put(5,11.5){\line(1,0){5}}
\put(5,12){\line(1,0){5}}
\end{picture} &
\begin{picture}(15,15)(0,-3)
\put(5,-2.5){\line(0,1){15}}
\put(10,-2.5){\line(0,1){15}}
\put(0,2.5){\line(1,0){15}}
\put(0,7.5){\line(1,0){15}}
\put(0,2.5){\line(0,1){5}}
\put(15,2.5){\line(0,1){5}}
\put(5,-2.5){\line(1,0){5}}
\put(5,12.5){\line(1,0){5}}

\put(5,8){\line(1,0){5}}
\put(5,8.5){\line(1,0){5}}
\put(5,9){\line(1,0){5}}
\put(5,9.5){\line(1,0){5}}
\put(5,10){\line(1,0){5}}
\put(5,10.5){\line(1,0){5}}
\put(5,11){\line(1,0){5}}
\put(5,11.5){\line(1,0){5}}
\put(5,12){\line(1,0){5}}

\put(10,3){\line(1,0){5}}
\put(10,3.5){\line(1,0){5}}
\put(10,4){\line(1,0){5}}
\put(10,4.5){\line(1,0){5}}
\put(10,5){\line(1,0){5}}
\put(10,5.5){\line(1,0){5}}
\put(10,6){\line(1,0){5}}
\put(10,6.5){\line(1,0){5}}
\put(10,7){\line(1,0){5}}
\end{picture} &
\begin{picture}(15,15)(0,-3)
\put(5,-2.5){\line(0,1){15}}
\put(10,-2.5){\line(0,1){15}}
\put(0,2.5){\line(1,0){15}}
\put(0,7.5){\line(1,0){15}}
\put(0,2.5){\line(0,1){5}}
\put(15,2.5){\line(0,1){5}}
\put(5,-2.5){\line(1,0){5}}
\put(5,12.5){\line(1,0){5}}

\put(5,3){\line(1,0){5}}
\put(5,3.5){\line(1,0){5}}
\put(5,4){\line(1,0){5}}
\put(5,4.5){\line(1,0){5}}
\put(5,5){\line(1,0){5}}
\put(5,5.5){\line(1,0){5}}
\put(5,6){\line(1,0){5}}
\put(5,6.5){\line(1,0){5}}
\put(5,7){\line(1,0){5}}

\put(5,8){\line(1,0){5}}
\put(5,8.5){\line(1,0){5}}
\put(5,9){\line(1,0){5}}
\put(5,9.5){\line(1,0){5}}
\put(5,10){\line(1,0){5}}
\put(5,10.5){\line(1,0){5}}
\put(5,11){\line(1,0){5}}
\put(5,11.5){\line(1,0){5}}
\put(5,12){\line(1,0){5}}
\end{picture} &
\begin{picture}(15,15)(0,-3)
\put(5,-2.5){\line(0,1){15}}
\put(10,-2.5){\line(0,1){15}}
\put(0,2.5){\line(1,0){15}}
\put(0,7.5){\line(1,0){15}}
\put(0,2.5){\line(0,1){5}}
\put(15,2.5){\line(0,1){5}}
\put(5,-2.5){\line(1,0){5}}
\put(5,12.5){\line(1,0){5}}

\put(5,8){\line(1,0){5}}
\put(5,8.5){\line(1,0){5}}
\put(5,9){\line(1,0){5}}
\put(5,9.5){\line(1,0){5}}
\put(5,10){\line(1,0){5}}
\put(5,10.5){\line(1,0){5}}
\put(5,11){\line(1,0){5}}
\put(5,11.5){\line(1,0){5}}
\put(5,12){\line(1,0){5}}

\put(5,-2){\line(1,0){5}}
\put(5,-1.5){\line(1,0){5}}
\put(5,-1){\line(1,0){5}}
\put(5,-0.5){\line(1,0){5}}
\put(5,0){\line(1,0){5}}
\put(5,0.5){\line(1,0){5}}
\put(5,1){\line(1,0){5}}
\put(5,1.5){\line(1,0){5}}
\put(5,2){\line(1,0){5}}
\end{picture} \\ \hline
Model 1 & $0.5$ & $0.0$ & $0.0$ & $-1.0$ & $0.0$ & $-1.5$ & $0.0$ & $0.0$ & $-0.5$ & $-0.5$ \\
Model 2 & $0.75$ & $0.0$ & $0.0$ & $-1.5$ & $0.0$ & $-2.0$ & $0.0$ & $0.0$ & $-1.0$ & $-1.0$ 
\end{tabular}
\end{center}
\caption{\label{fig:pot}Potential values for the various clique configurations in the higher-order
MRF models. The potentials are invariant under rotation, reflection and inversion of the colours.}
\end{figure}
A realization from each of the models, generated by 
Gibbs sampling, is shown in Figure \ref{fig:homrf-real}.
\begin{figure}
\begin{center}
\begin{tabular}{cc}
\includegraphics[height=3.0cm,width=3.0cm,angle=-90]{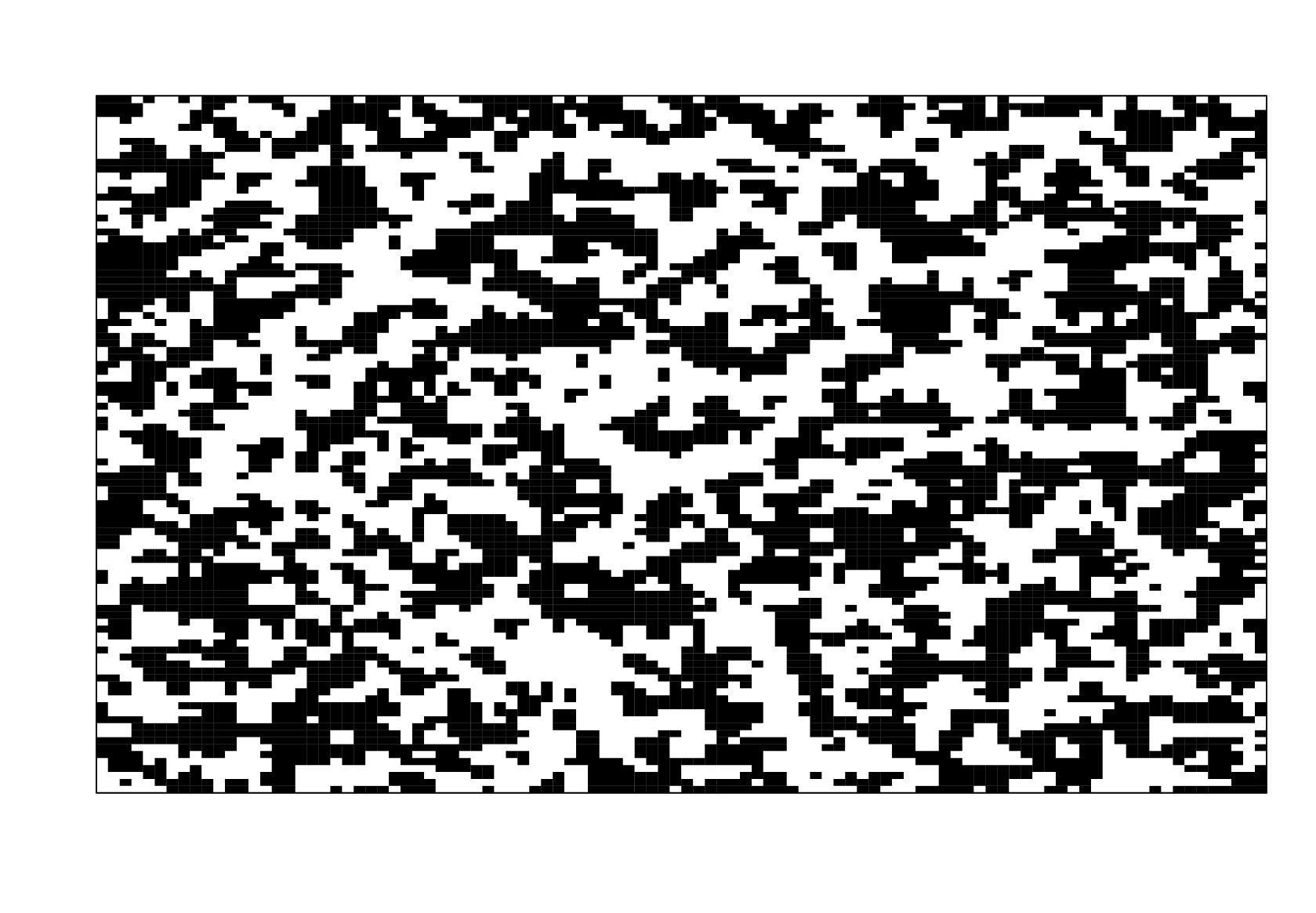} &
\includegraphics[height=3.0cm,width=3.0cm,angle=-90]{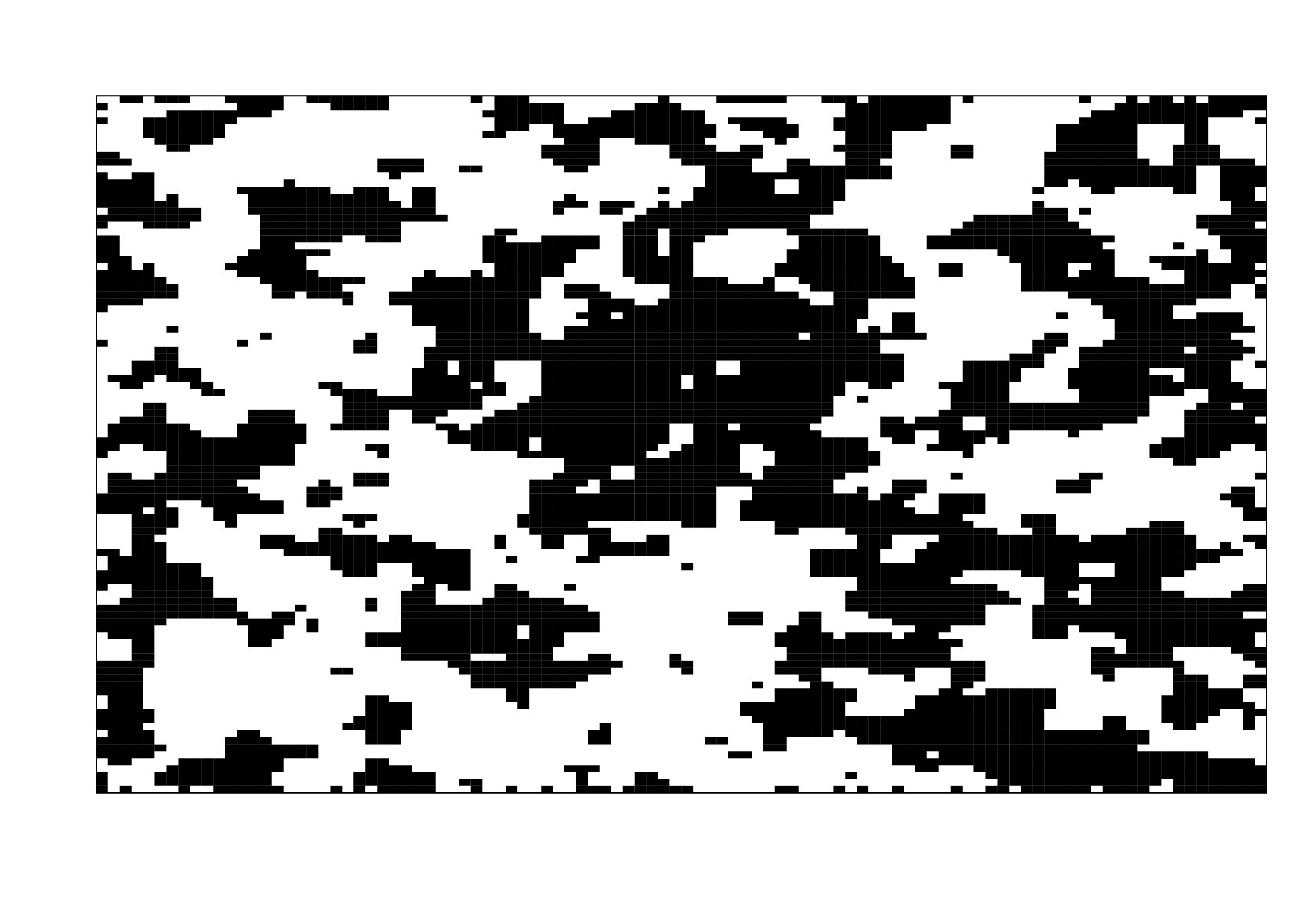} \\
Model 1 & Model 2 \\[-0.5cm]
\end{tabular}
\end{center}
\caption{\label{fig:homrf-real}Realizations from the two higher-order MRF models on a $100\times 100$ lattice, 
generated 
by Gibbs sampling. Potential functions for the models are defined in Figure 
\ref{fig:pot}.}
\end{figure}

\subsection{\label{sec:evaluation}Empirical evaluation}
In this section we first consider the quality of the approximation of 
the normalising constant $c$ and the corresponding bounds. 
Thereafter we evaluate to what extent 
our $\widetilde{p}(x)$ defined in Section \ref{sec:mrfApprox} can be
used as an approximation to the corresponding $p(x)$.

\begin{figure}
\begin{center}
\begin{tabular}{cccc}
\includegraphics[height=3.0cm,width=2.5cm,angle=-90]{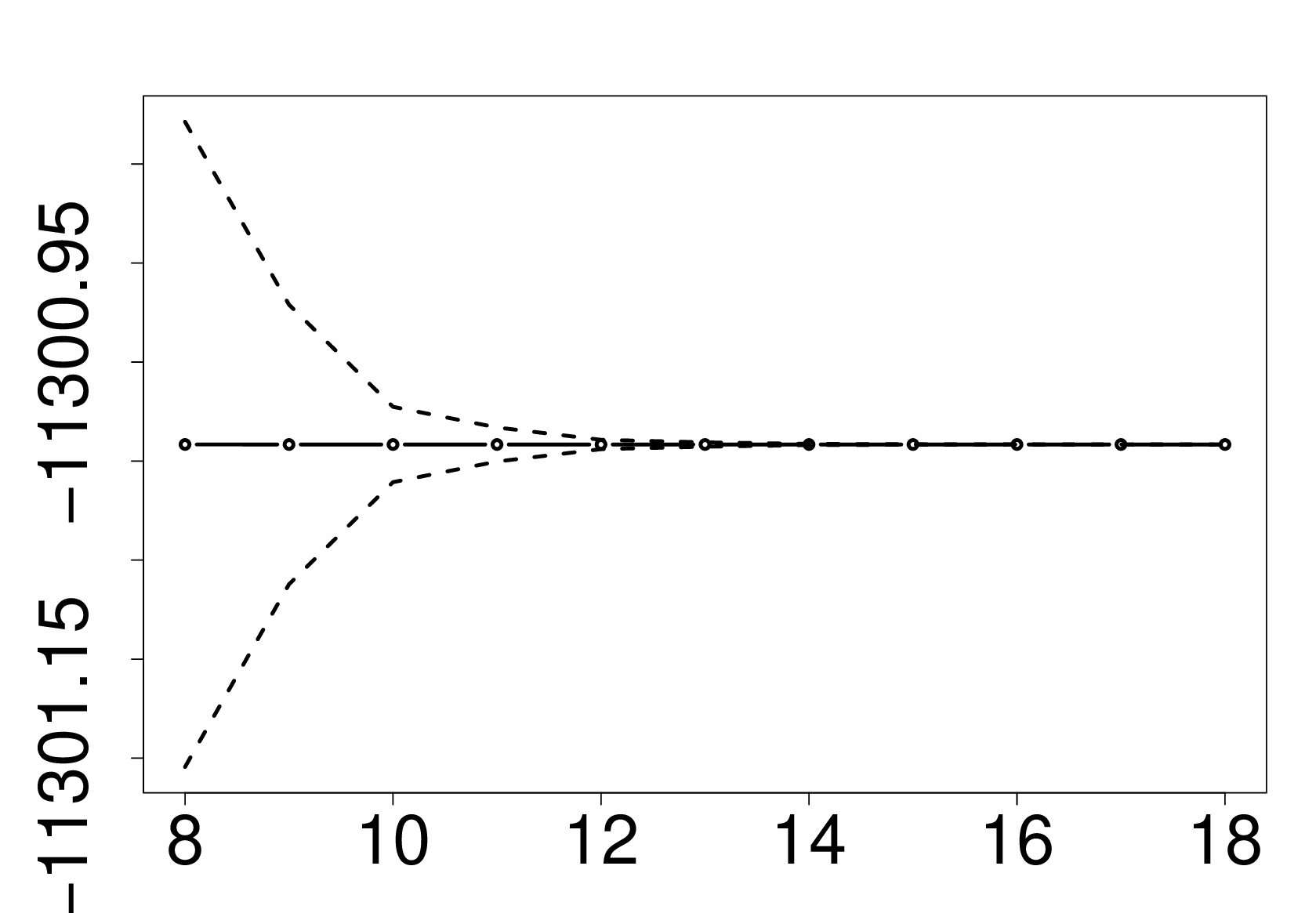} &
\includegraphics[height=3.0cm,width=2.5cm,angle=-90]{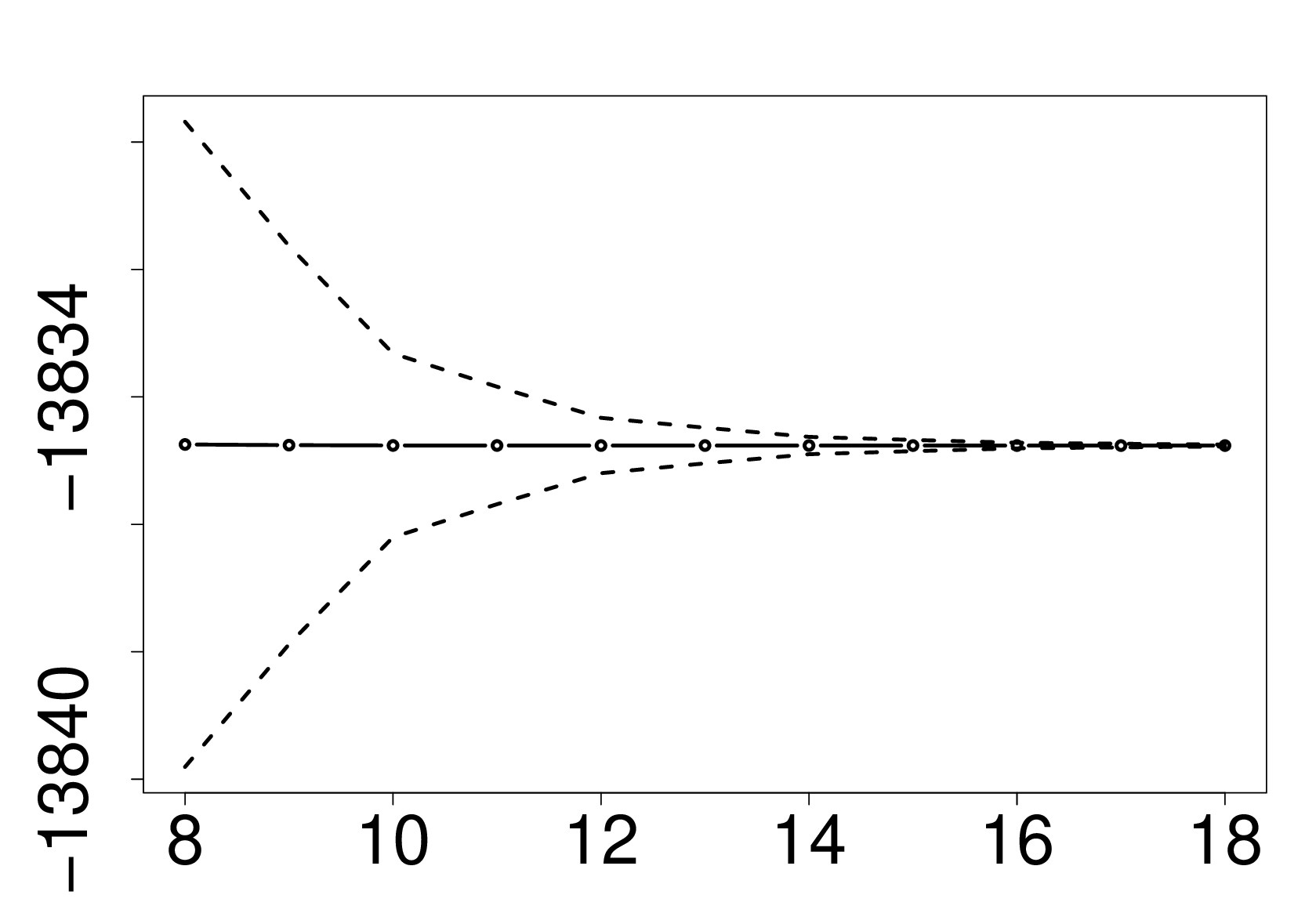} &
\includegraphics[height=3.0cm,width=2.5cm,angle=-90]{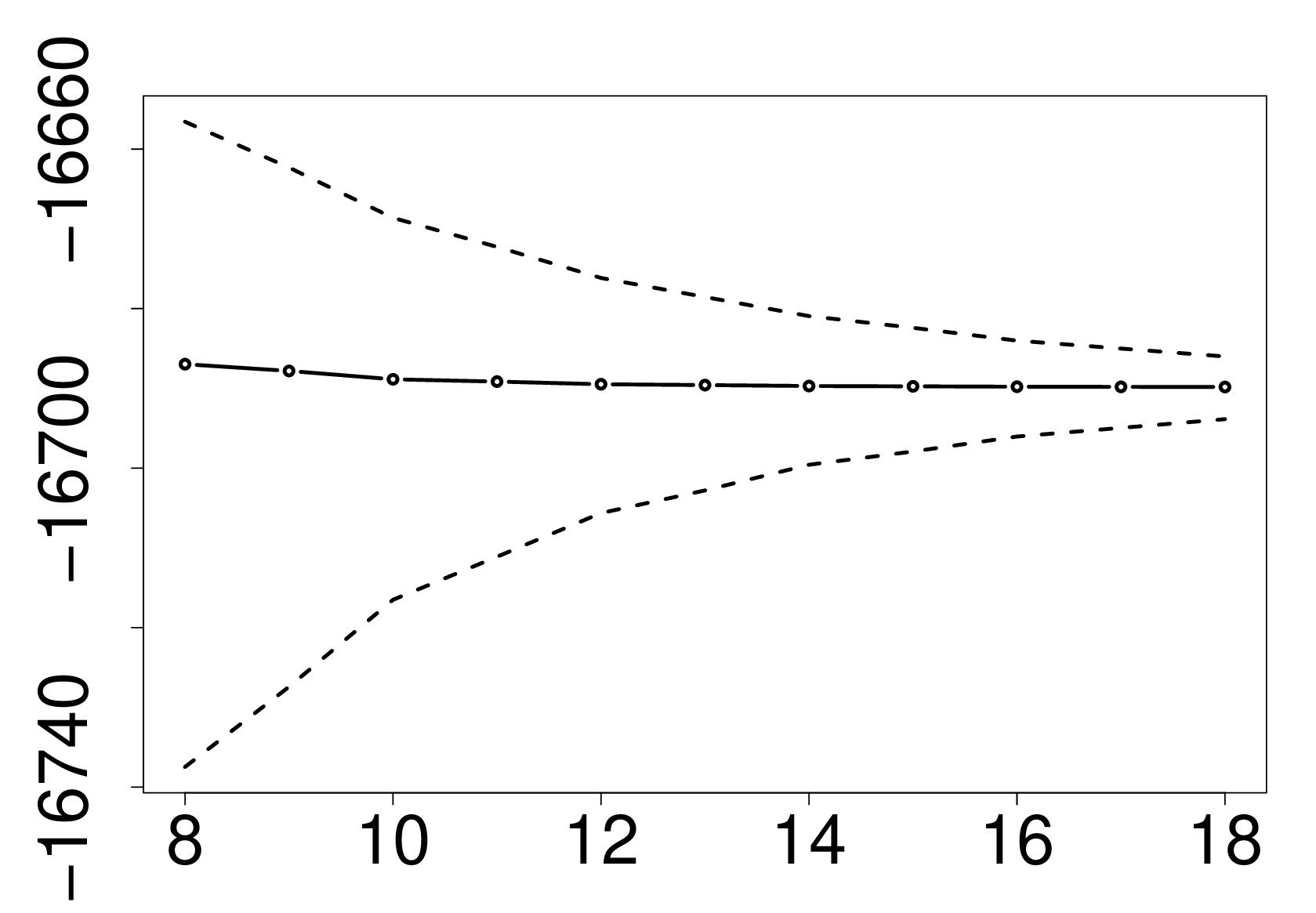} &
\includegraphics[height=3.0cm,width=2.5cm,angle=-90]{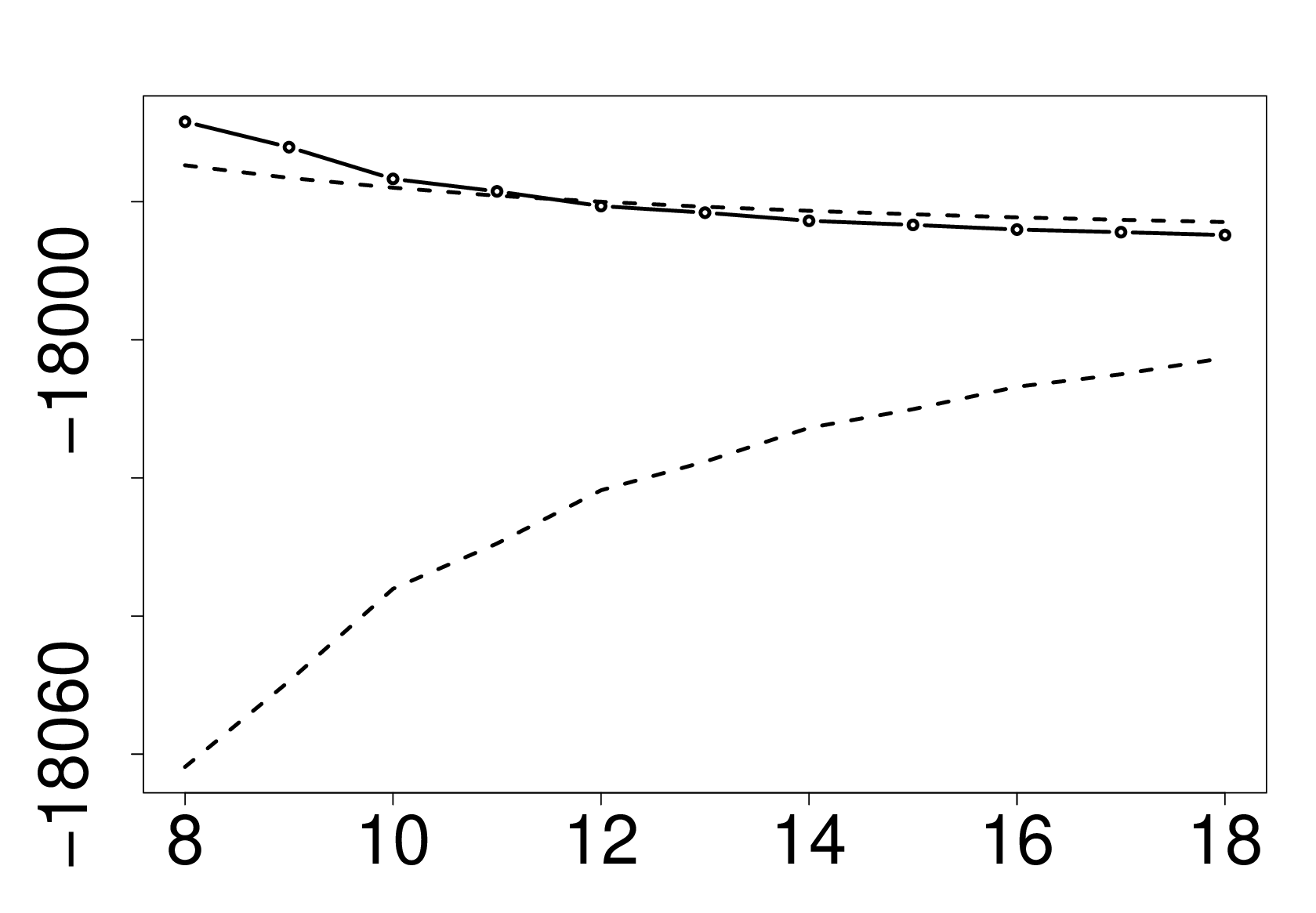} \\
\includegraphics[height=3.0cm,width=2.5cm,angle=-90]{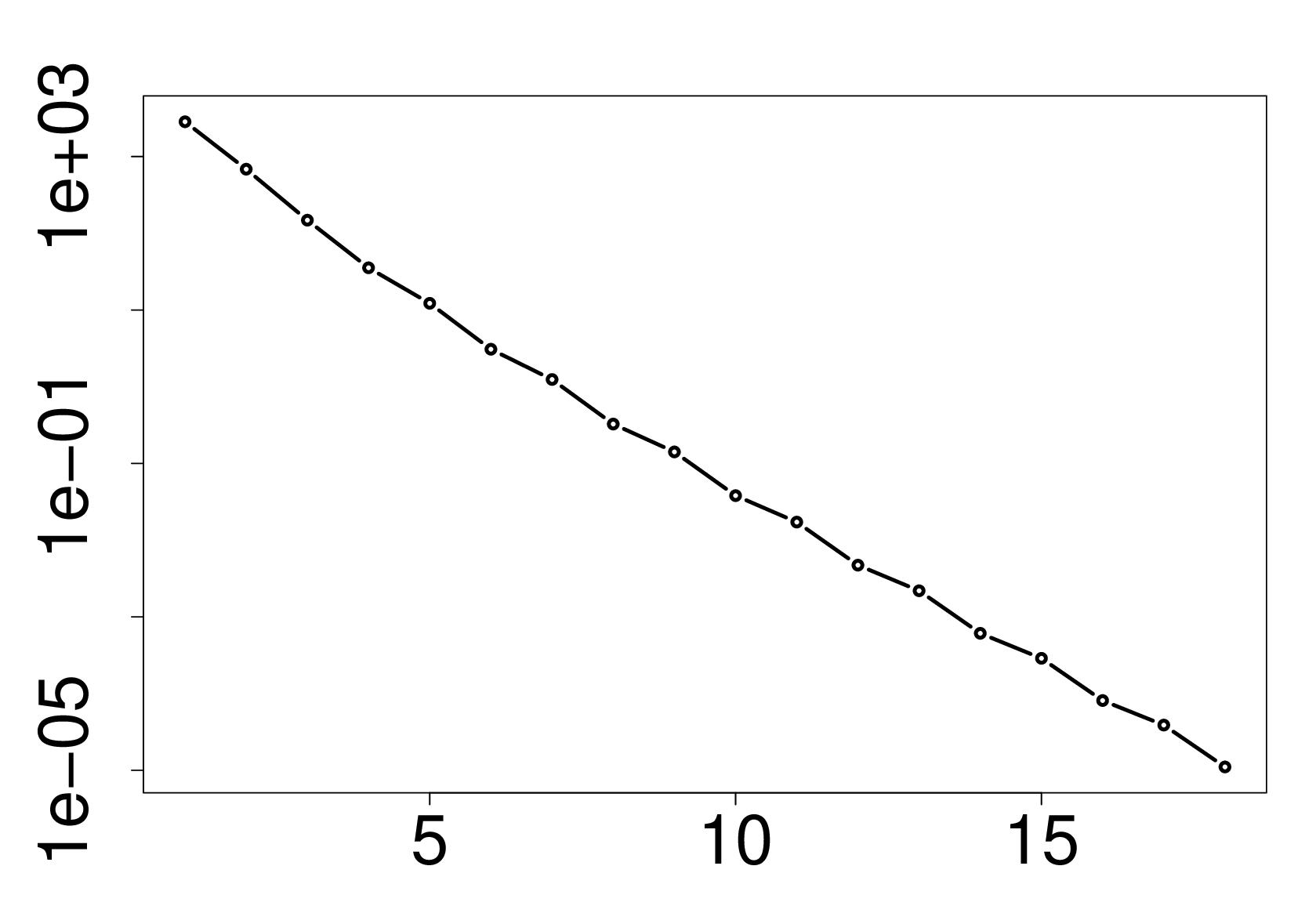} &
\includegraphics[height=3.0cm,width=2.5cm,angle=-90]{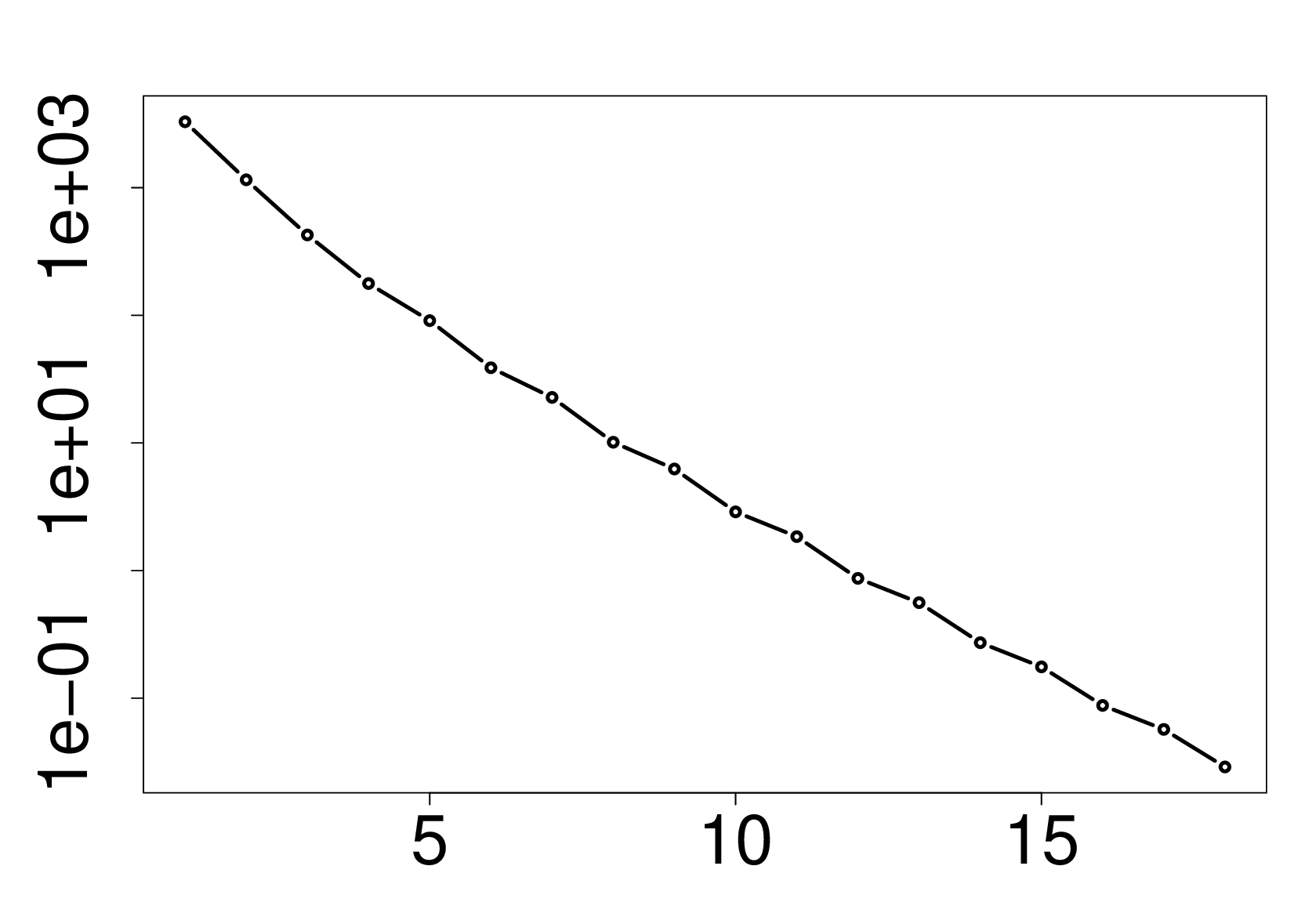} &
\includegraphics[height=3.0cm,width=2.5cm,angle=-90]{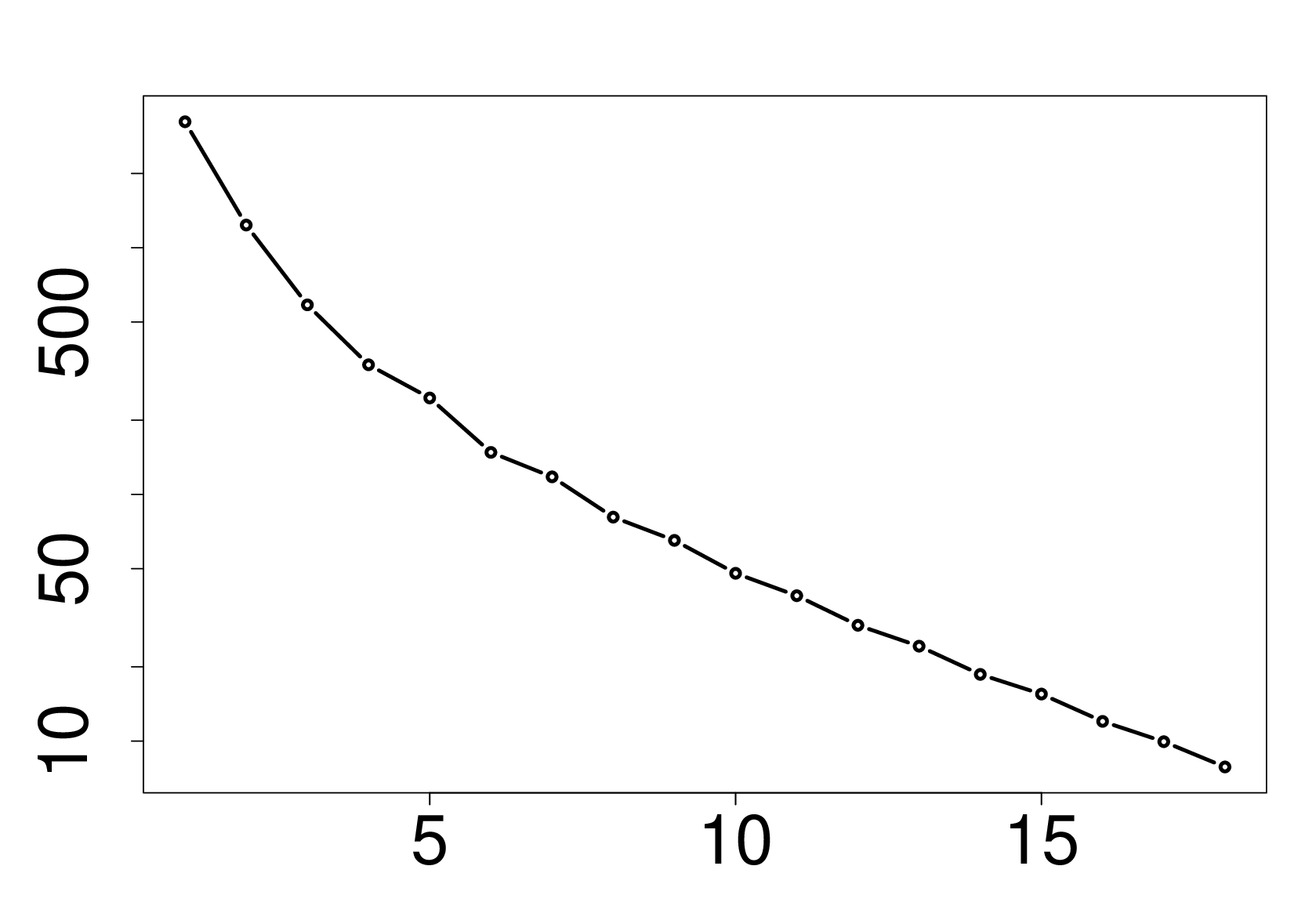} &
\includegraphics[height=3.0cm,width=2.5cm,angle=-90]{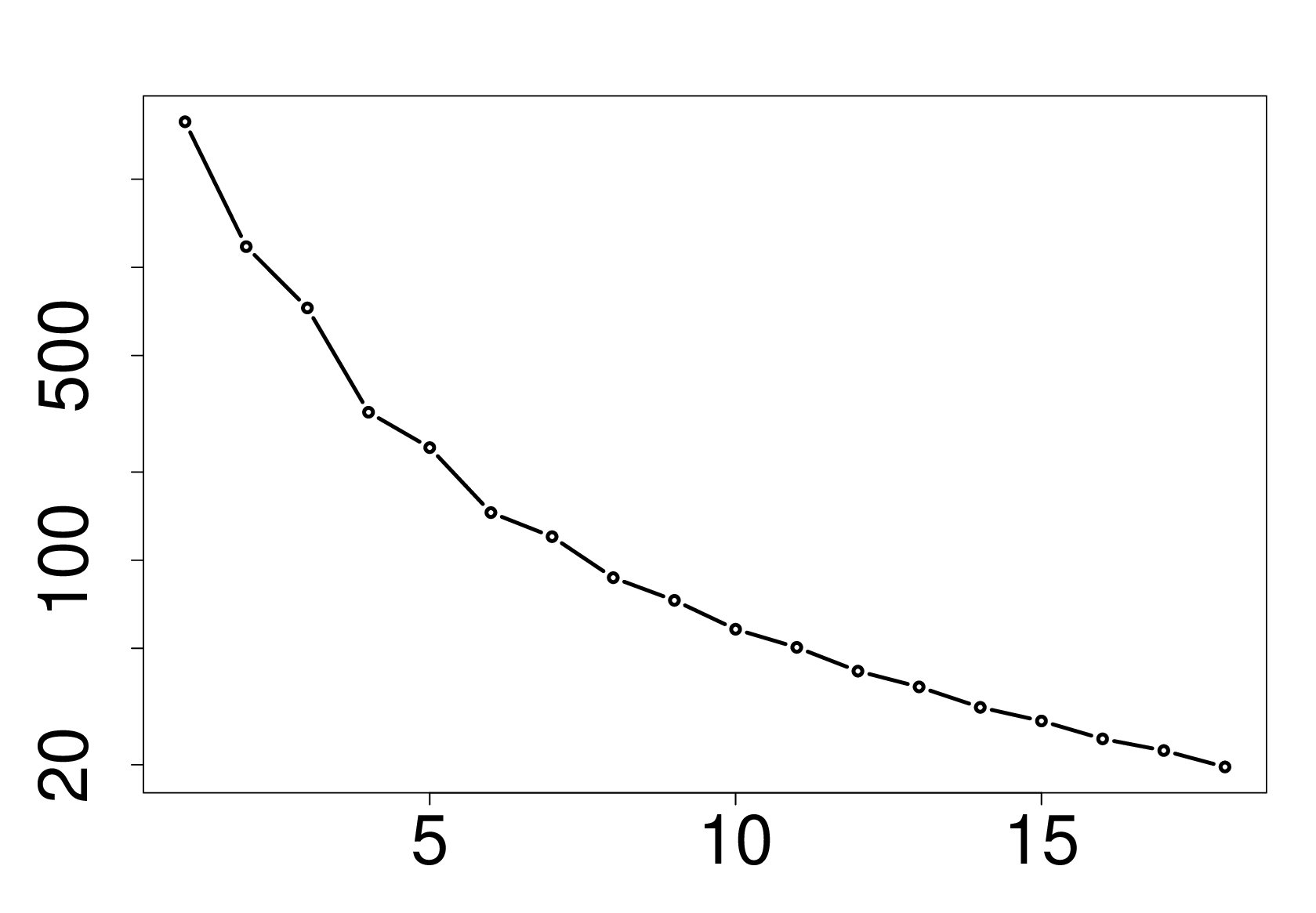} \\
$\theta=0.4$ & $\theta=0.6$ & $\theta=0.8$ & $\theta=-\ln(\sqrt{2}-1)$\\[-0.4cm]
\end{tabular}
\end{center}
\caption{\label{fig:IsingC}Results for the Ising model: The upper row shows
the approximation $\ln (\widetilde{c})$ (solid) and the corresponding
upper and lower bounds $\ln (c_U)$ and $\ln(c_L)$ (dashed) for $\nu=8,\ldots,18$,
and the lower row shows $\ln (c_U)-\ln (c_L)$ for $\nu=1$ to $18$.}
\end{figure}

\begin{figure}[t]
\begin{center}
\begin{tabular}{cc}
\includegraphics[height=4.5cm,width=3.0cm,angle=-90]{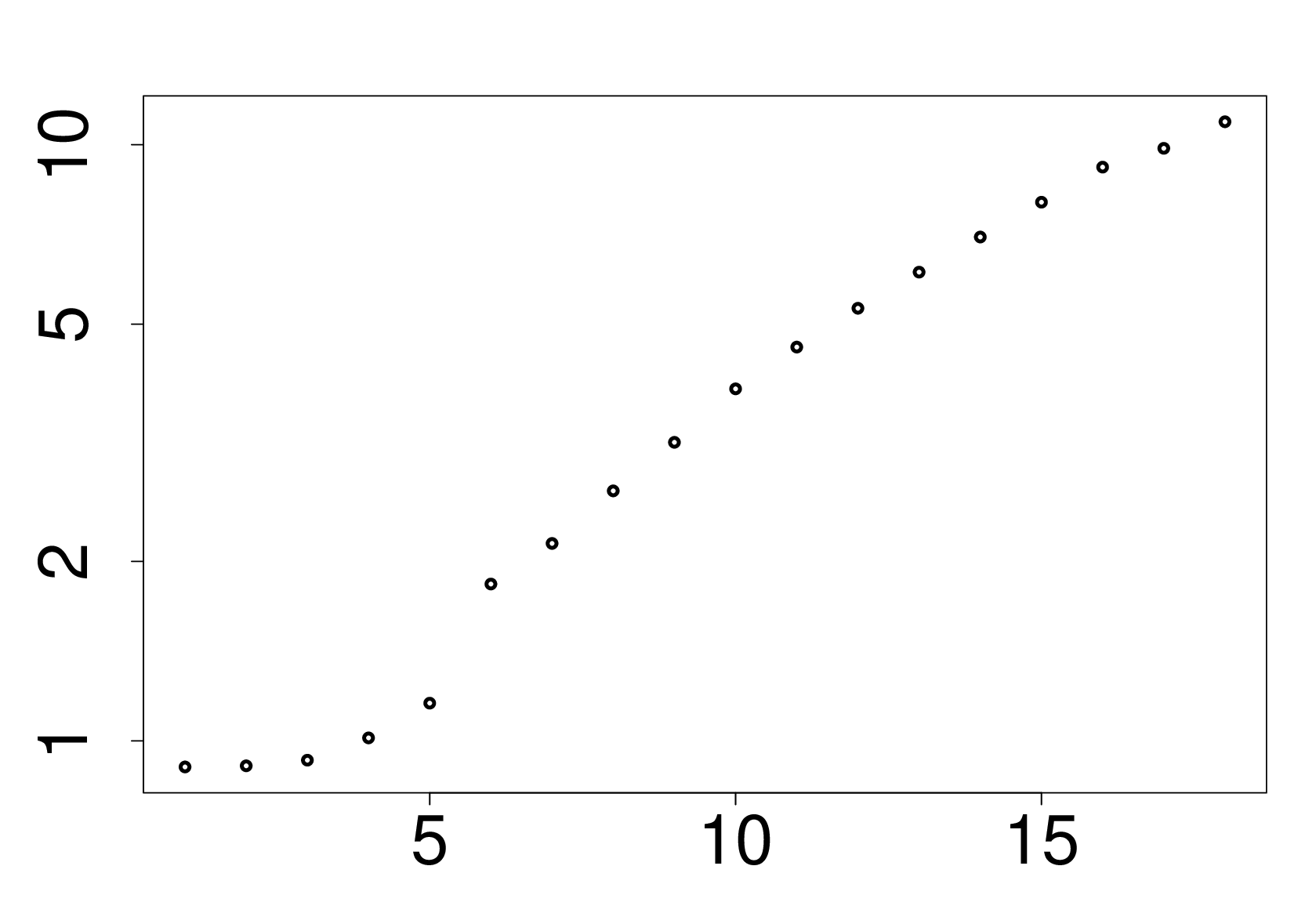} &
\includegraphics[height=4.5cm,width=3.0cm,angle=-90]{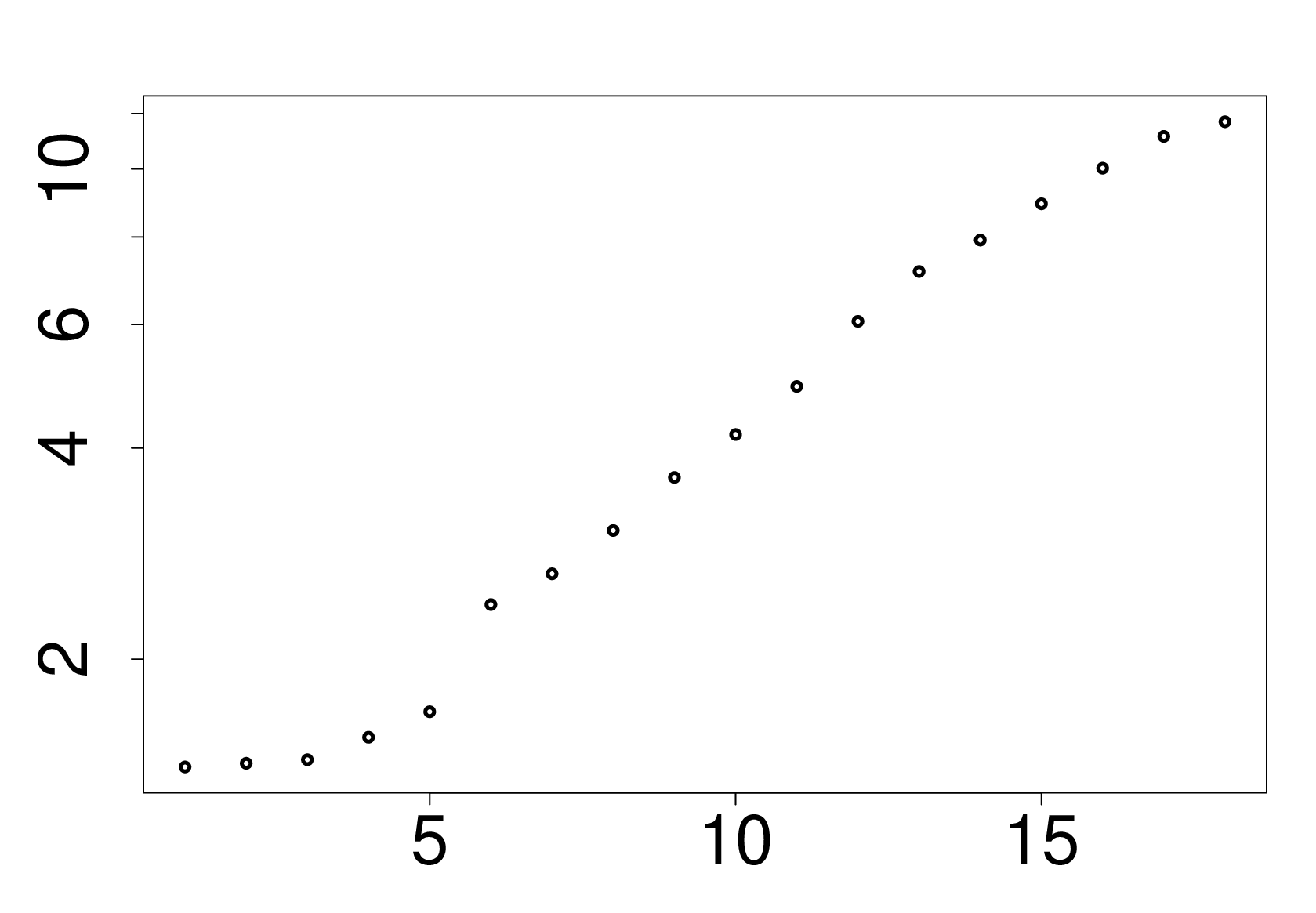} \\
(a) Ising & (b) higher-order MRF\\[-0.4cm]
\end{tabular}
\end{center}
\caption{\label{fig:cpuTime}The natural logarithm of the computation times (in seconds)
used to compute an approximation or a bound in (a) the Ising model, and (b) the 
higher-order interaction MRF.}
\end{figure}

\begin{figure}
\begin{center}
\begin{tabular}{cc}
\includegraphics[height=4.5cm,width=3.0cm,angle=-90]{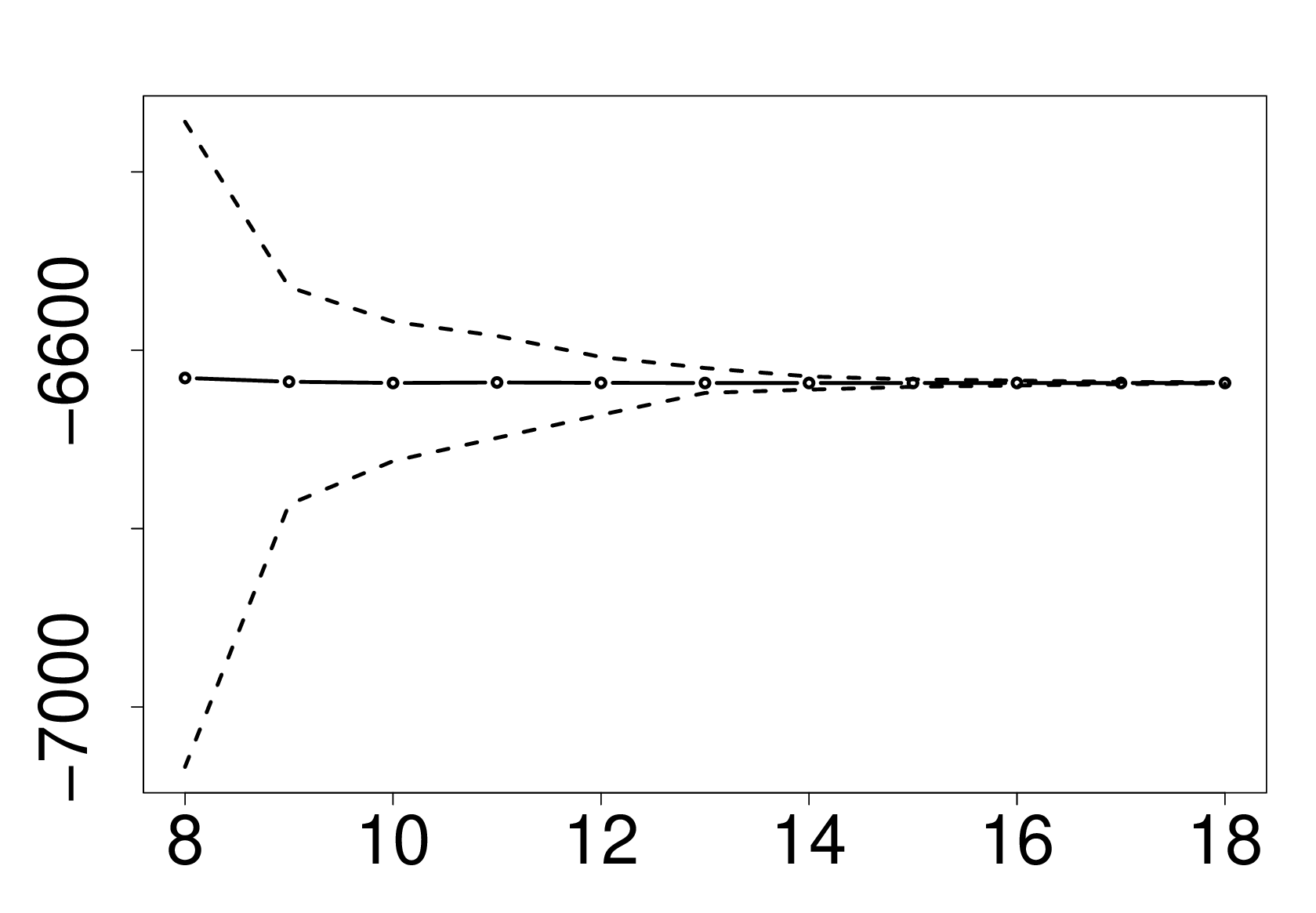} &
\includegraphics[height=4.5cm,width=3.0cm,angle=-90]{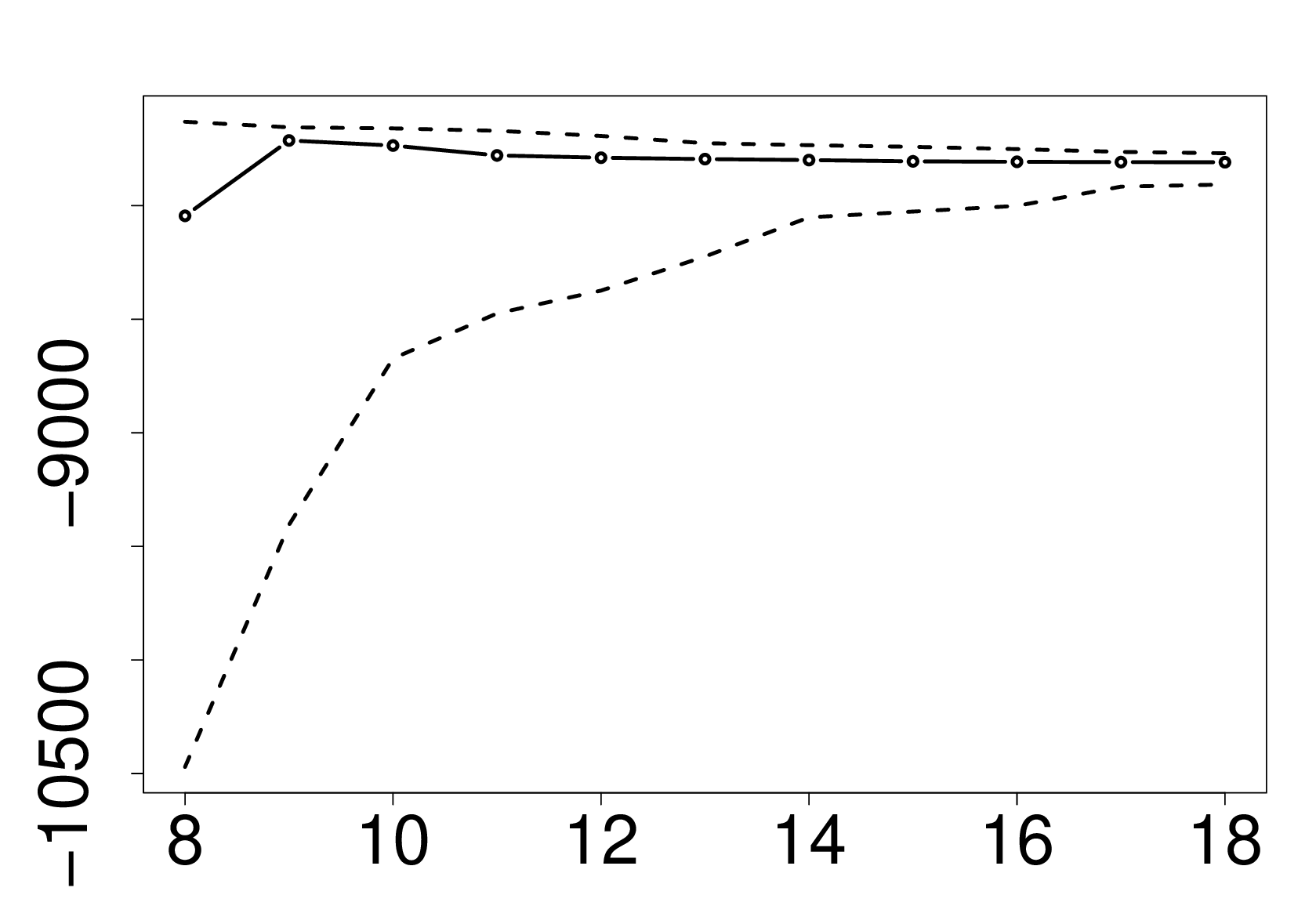} \\
\includegraphics[height=4.5cm,width=3.0cm,angle=-90]{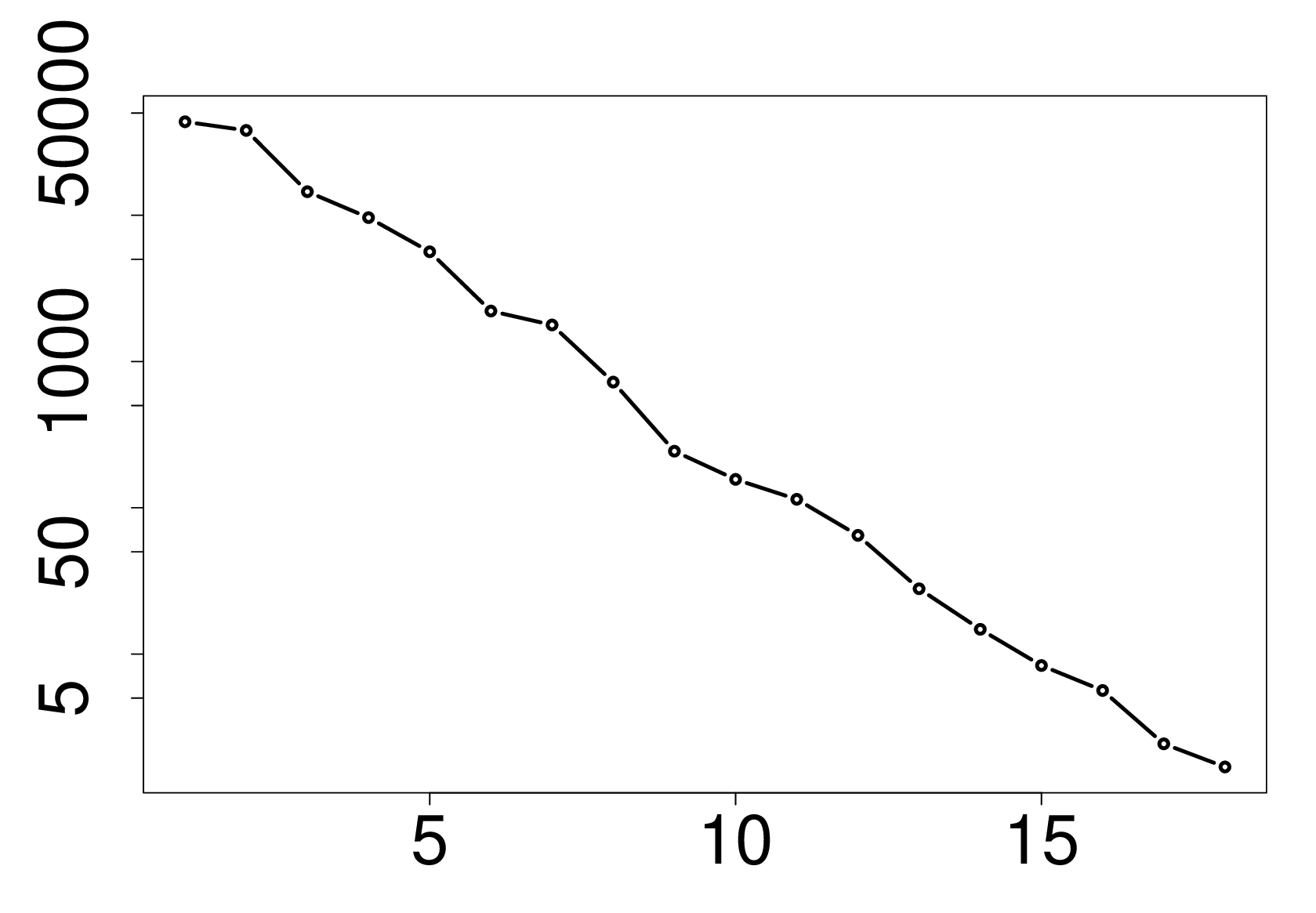} &
\includegraphics[height=4.5cm,width=3.0cm,angle=-90]{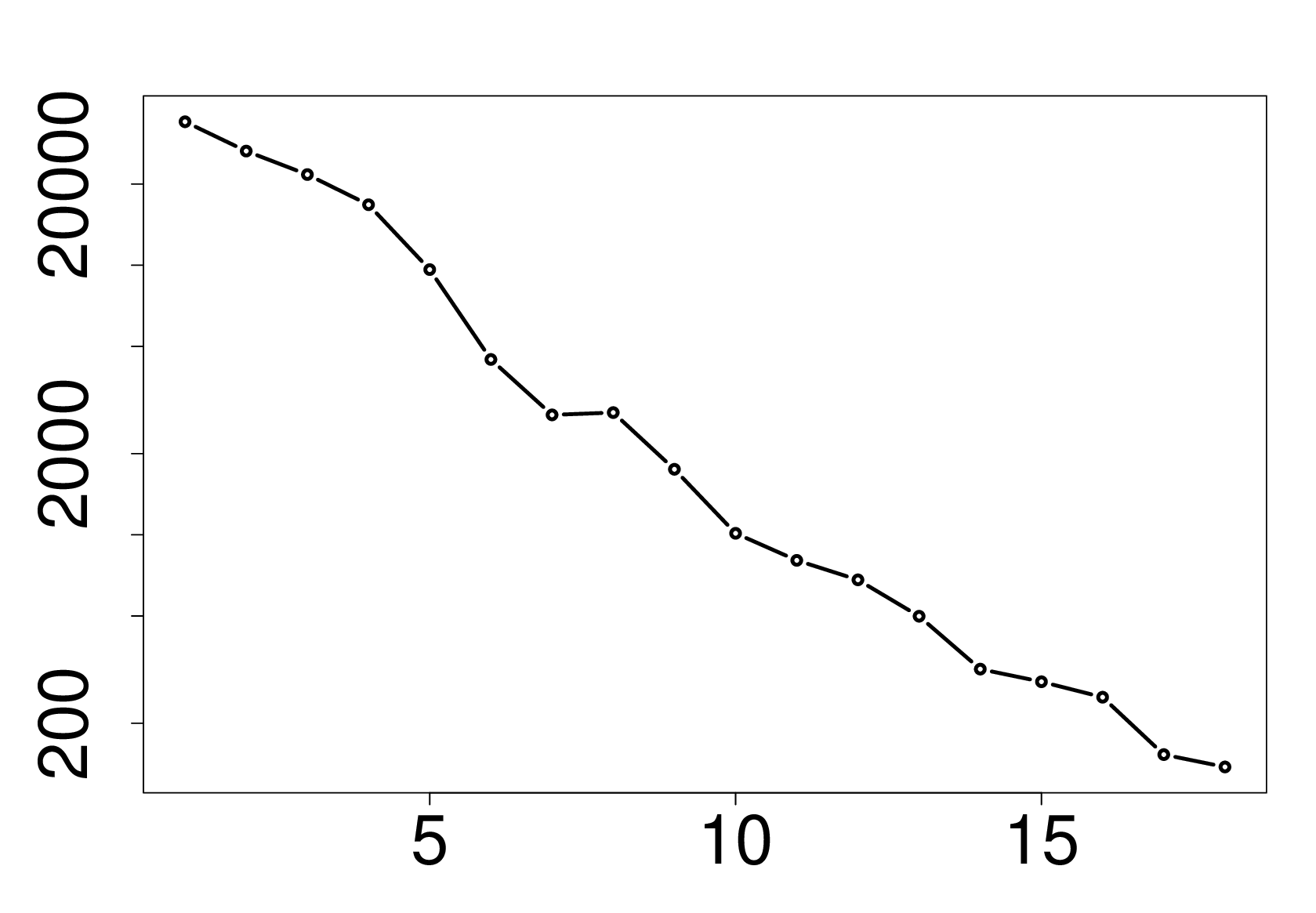} \\
Model 1 & Model 2\\[-0.4cm]
\end{tabular}
\end{center}
\caption{\label{fig:homrfC}Results for the higher-order MRF model: The upper row shows
the approximation $\ln (\widetilde{c})$ (solid) and the corresponding
upper and lower bounds $\ln (c_U)$ and $\ln(c_L)$ (dashed) for $\nu=6,\ldots,18$,
and the lower row shows $\ln (c_U)-\ln (c_L)$ for $\nu=1$ to $18$.}
\end{figure}

We first compute the approximate normalising constant
$\widetilde{c}$ and the corresponding upper and lower bounds
$c_U$ and $c_L$ for $\nu=1,2,\ldots,18$ for each of our four 
$\theta$ values. The results are presented in Figure \ref{fig:IsingC}.
The upper row shows the approximation $\ln (\widetilde{c})$ together
with the bounds $\ln (c_L)$ and $\ln (c_U)$ for $\nu=8$ to $19$, 
whereas the lower row shows $\ln (c_U)-\ln (c_L)$ for $\nu=1$ to $19$.
For a given value of $\nu$, computation of an approximation takes about the 
same time for all values of $\theta$, and computation of a bound takes about the 
same time as evaluating the corresponding approximation. The computation time
for one evaluation as a function of $\nu$ is shown in Figure
\ref{fig:cpuTime}(a).
Figure \ref{fig:homrfC} shows similar results for the higher-order 
MRF models, again for a $100\times 100$ lattice, and 
corresponding computation times are shown in Figure \ref{fig:cpuTime}(b).
Not surprisingly, we see that the quality of the approximation and bounds are best
for models with weak interactions.

Next we evaluate the quality of the POMM approximation $\widetilde{p}(x)$ given in 
(\ref{eq:pommapprox}), still on a $100\times 100$ lattice. 
To do this we consider an independent proposal 
Metropolis--Hastings algorithm
where the MRF $p(x)$ is the target distribution and $\widetilde{p}(x)$ is used as
proposal distribution. We use the acceptance rate in such an algorithm to measure
the quality of the approximation. It should be emphasised that we do not 
propose this Metropolis--Hastings as a way to sample from $p(x)$, we just 
use the acceptance rate of this algorithm to measure the quality of our approximation.
It should be noted that in this evaluation test 
we do not need to compute the normalising constant of the conditional distributions
in (\ref{eq:pommapprox}) for all values of the conditioning variables, so we apply
the first (and best) POMM approximation variant discussed in the paragraph following 
(\ref{eq:pommapprox}).

\begin{figure}
\begin{center}
\begin{tabular}{ccc}
\includegraphics[height=4.5cm,width=3.0cm,angle=-90]{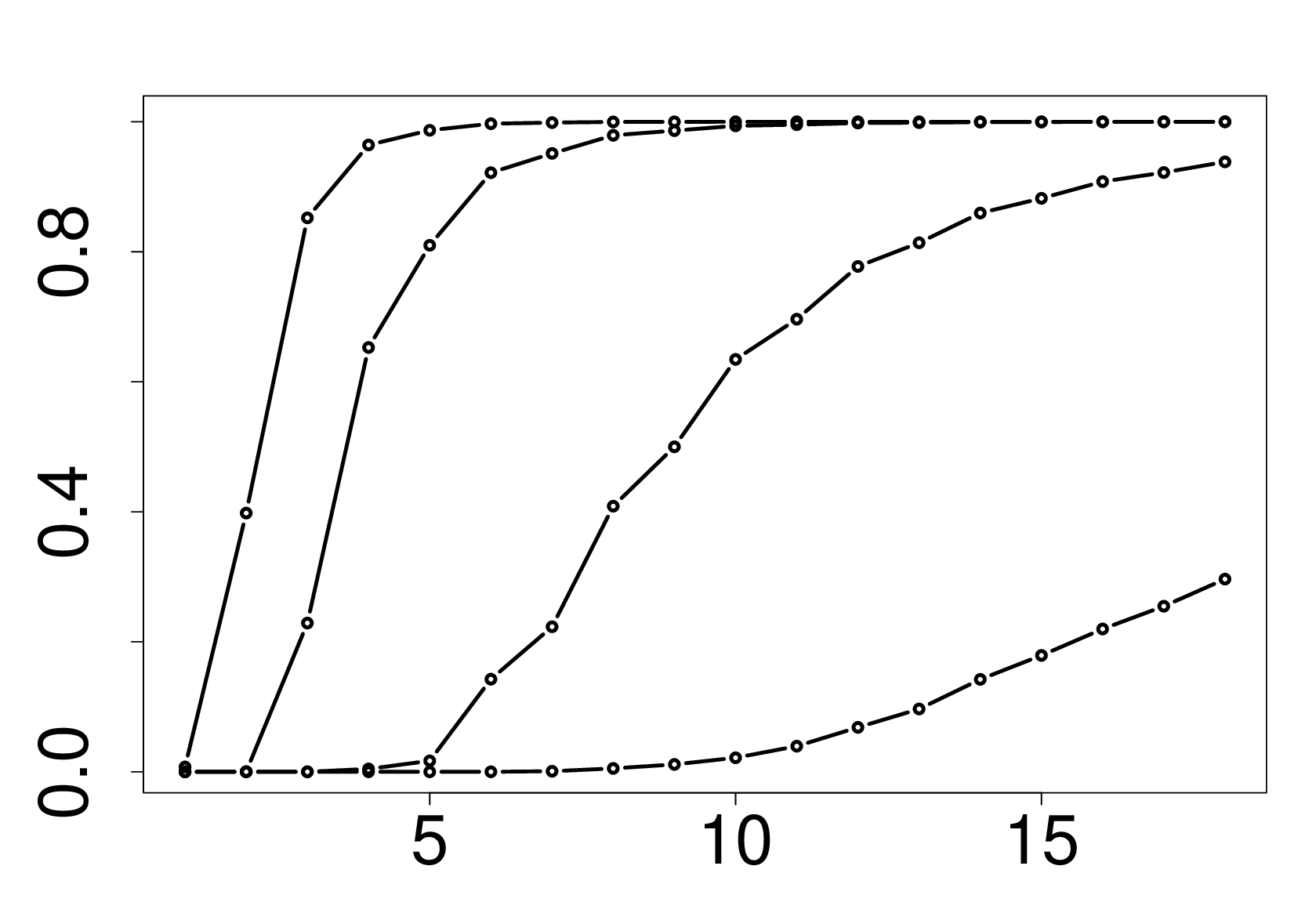} &
~~~~~~&
\includegraphics[height=4.5cm,width=3.0cm,angle=-90]{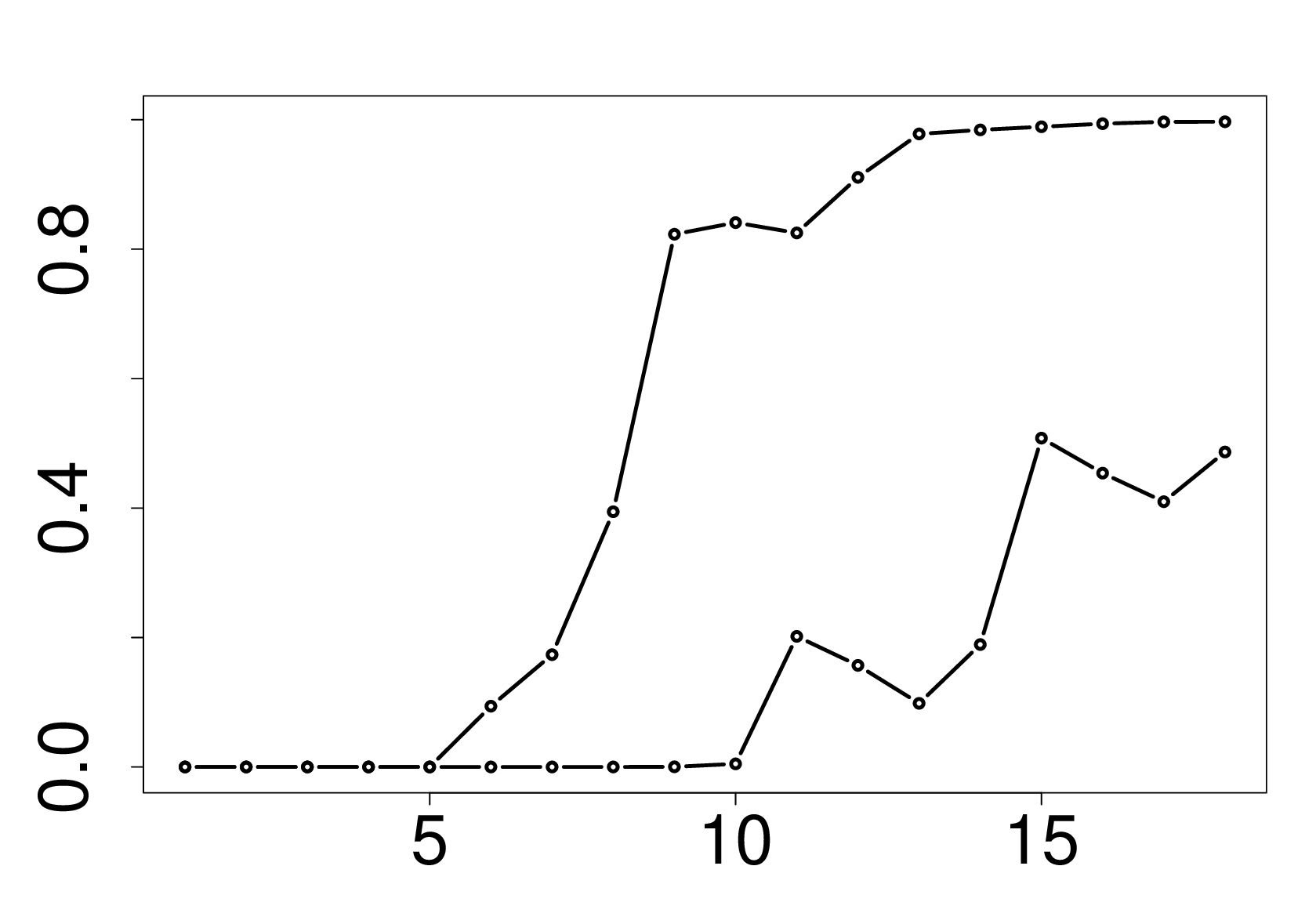} \\
(a) Ising & & (b) higher-order MRF\\[-0.4cm]
\end{tabular}
\end{center}
\caption{\label{fig:MHaccept}Estimated acceptance rates, for $\nu=1,\ldots,18$,
for an independent proposal
Metropolis--Hastings algorithm with target distribution given by the MRF $p(x)$
on a $100\times 100$ lattice
and proposal distribution given by the corresponding POMM approximation 
$\widetilde{p}(x)$ in (\ref{eq:pommapprox}). (a) Results for the 
Ising model, from top to bottom the curves are for $\theta=0.4$, $\theta=0.6$, $\theta=0.8$ and
$\theta=-\ln(\sqrt{2}-1)$. (b) Results for the higher-order MRFs, the upper and 
lower curves are for Model 1 and Model 2, respectively, defined in Figures 
\ref{fig:maxCliques} and \ref{fig:pot}.}
\end{figure}

To estimate the quantity we generate
$1000$ independent samples from $p(x)$ and corresponding $1000$ independent 
samples from $\widetilde{p}(x)$, compute the Metropolis--Hastings acceptance
probability for each pair and use the average of these numbers as our estimate.
For the Ising model we generate perfect samples from $p(x)$ on a 
$100\times 100$ lattice by first 
sampling perfectly from the dual random cluster model by coupling from the 
past \citep{art27}. For the higher-order MRF model we are not able to generate
perfect samples, so we instead run a long Gibbs sampling algorithm, 
and obtain (essentially) independent realizations by sub-sampling this chain. The results for 
the Ising and higher-order MRF models are given in Figure \ref{fig:MHaccept}.
For $\theta=0.4$ and $0.6$ in the Ising model we see that we get very good approximations
even for quite small values of $\nu$, For $\theta=0.8$ a large values for $\nu$ is 
needed to get high acceptance rate. For $\theta=-\ln(\sqrt{2}-1)$ the acceptance
rate ends at $30\%$ for $\nu=18$, and remembering that this is for a block update 
of $100\times 100$ variables we think this is quite impressive. For the higher-order 
MRFs, the acceptance rate for Model 1 becomes very high for the higher values of 
$\nu$, whereas the results for Model 2 resemble the results 
for $\theta = -\ln(\sqrt{2}-1)$ in the Ising model.


\subsection{\label{sec:applications}Some possible applications}
In this section we present some simulation examples that demonstrate some 
possible applications of the proposed approximations and bounds. All 
the examples are for MRFs defined on a rectangular lattice, but similar
applications for MRFs defined on graphs are of course also possible.

\subsubsection{Maximum likelihood estimation}
\begin{figure}[t]
\begin{center}
\begin{tabular}{ccc}
\includegraphics[height=4.0cm,width=2.2cm,angle=-90]{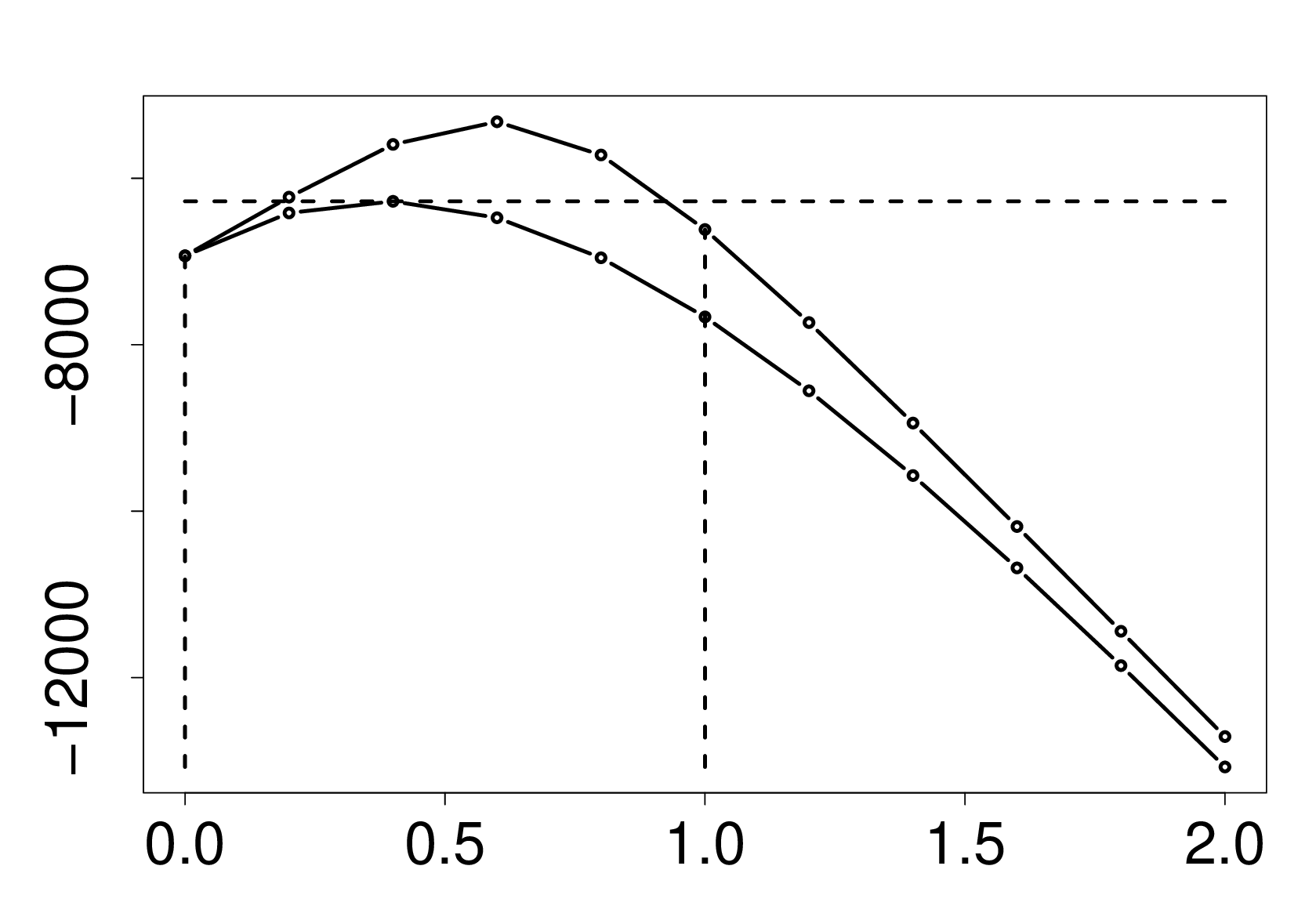} &
\includegraphics[height=4.0cm,width=2.2cm,angle=-90]{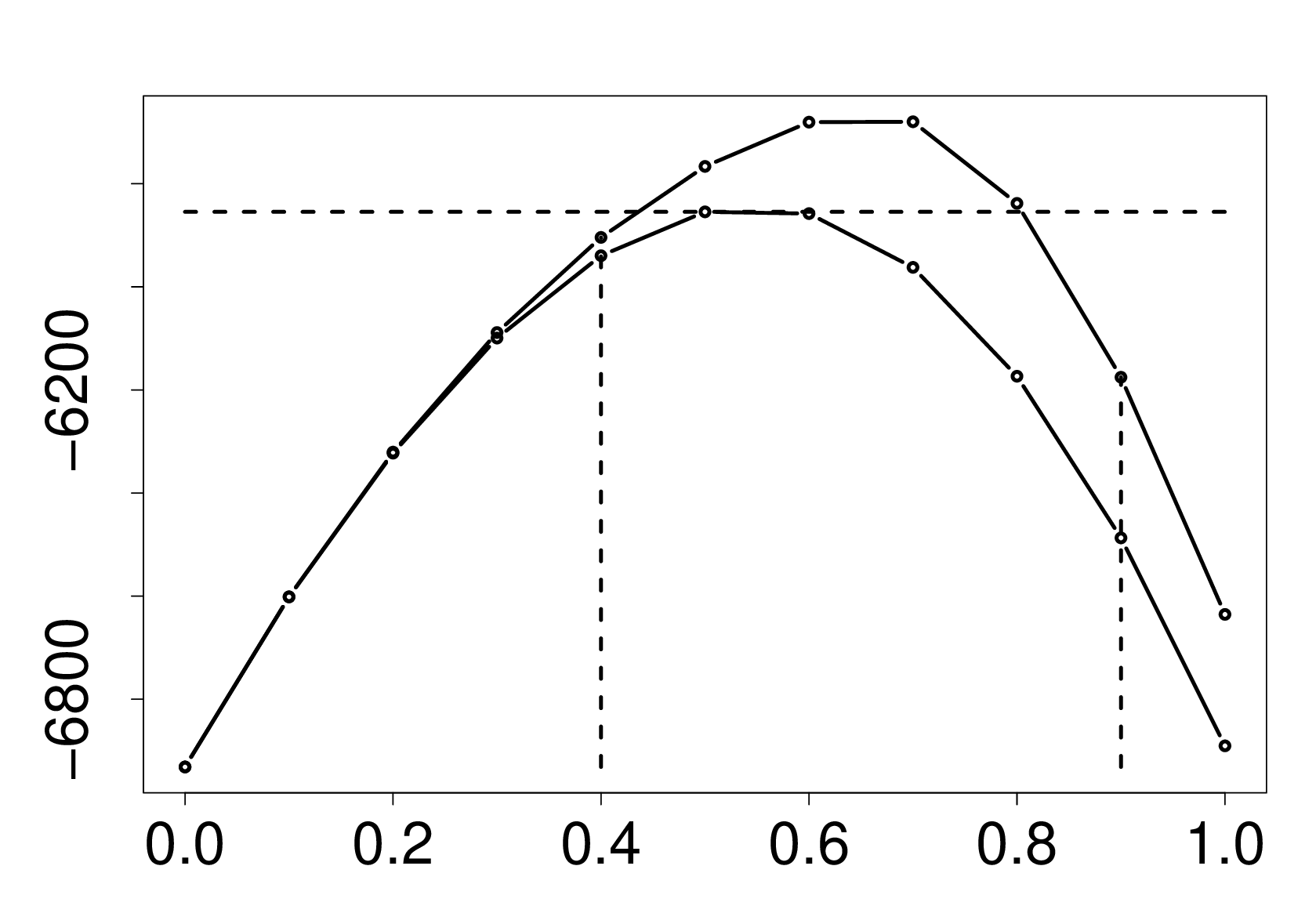} &
\includegraphics[height=4.0cm,width=2.2cm,angle=-90]{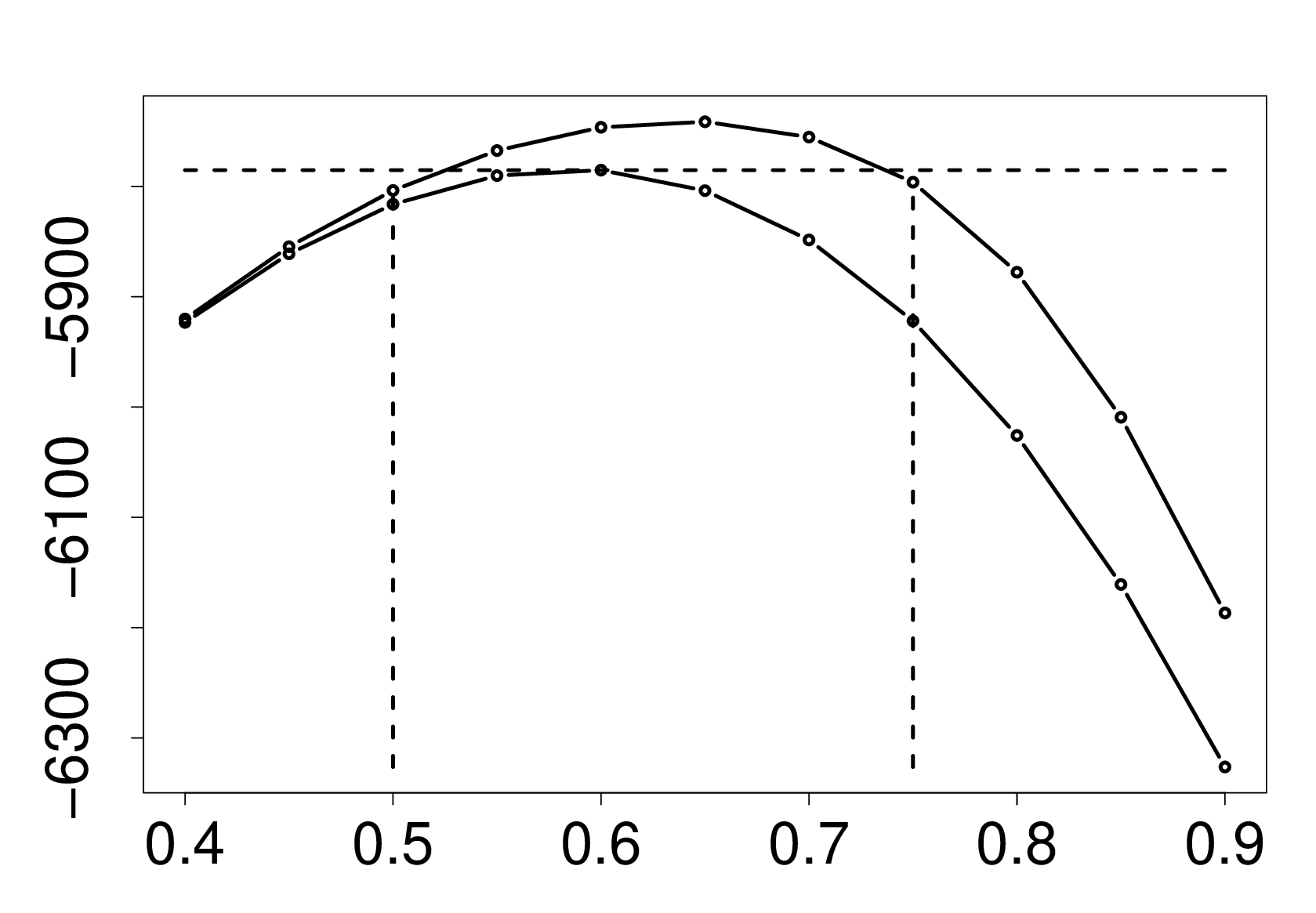} \\[-0.1cm]
$\nu=2$ & $\nu=4$ & $\nu=6$ \\[-0.3cm]
\includegraphics[height=4.0cm,width=2.2cm,angle=-90]{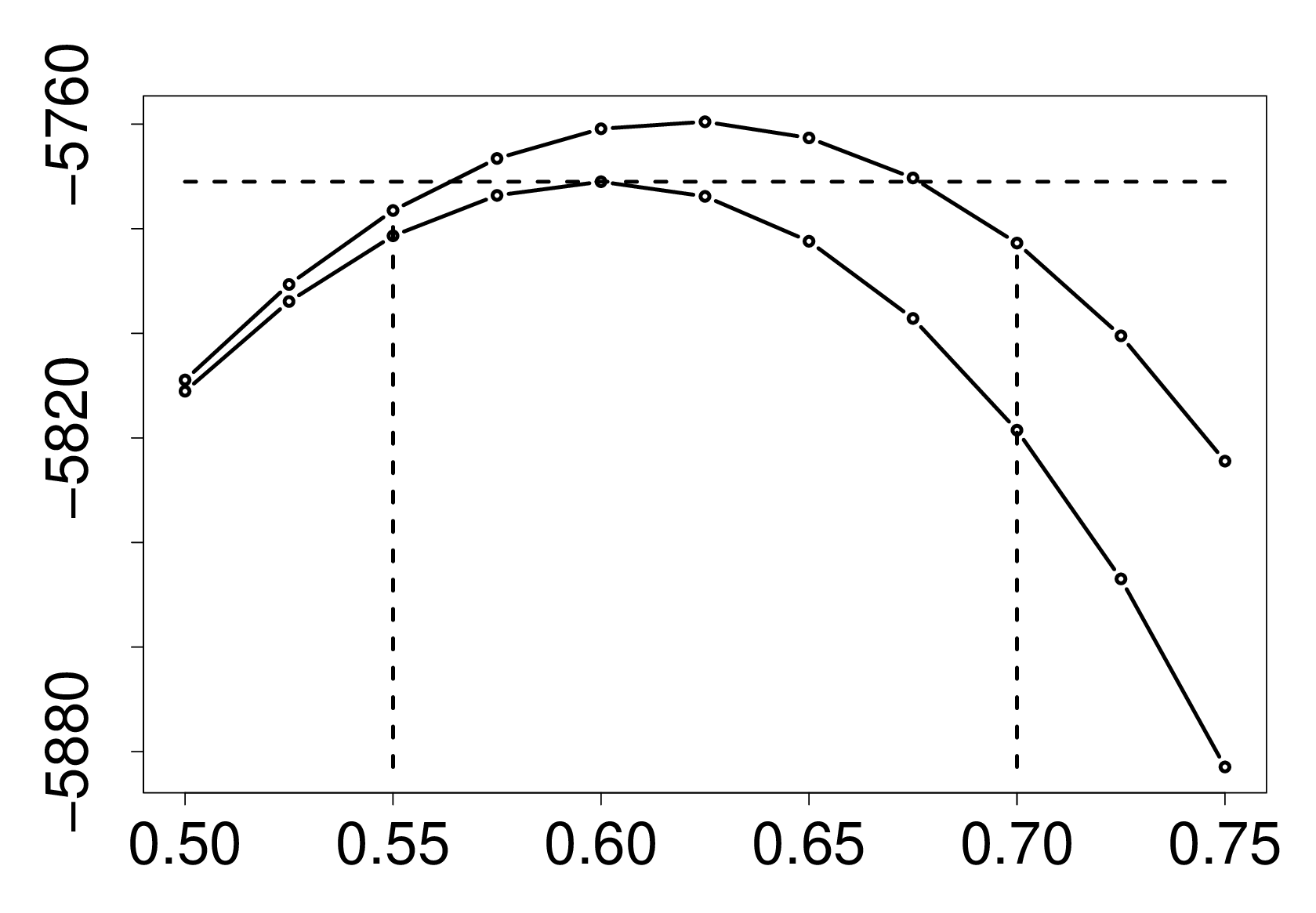} &
\includegraphics[height=4.0cm,width=2.2cm,angle=-90]{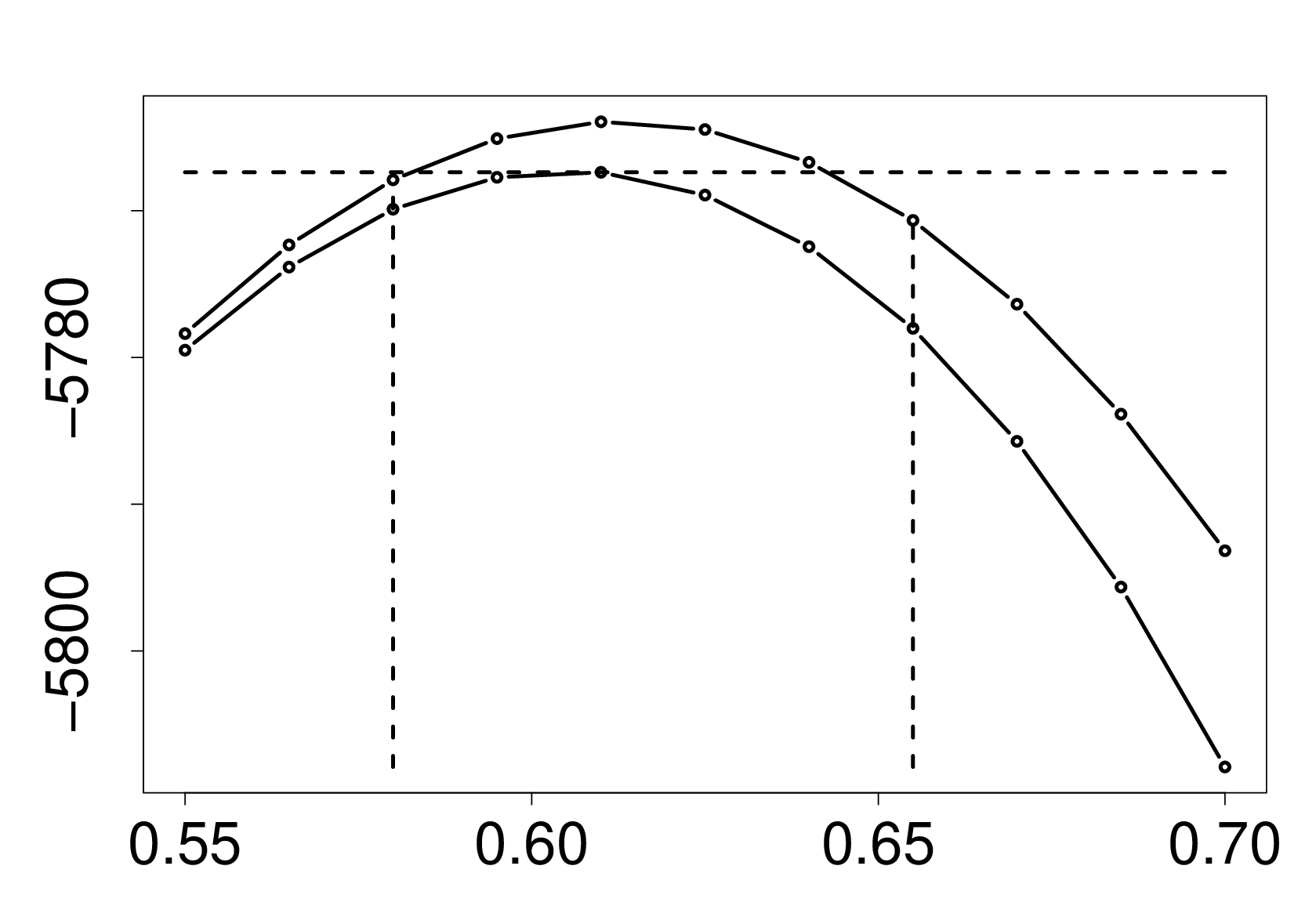} &
\includegraphics[height=4.0cm,width=2.2cm,angle=-90]{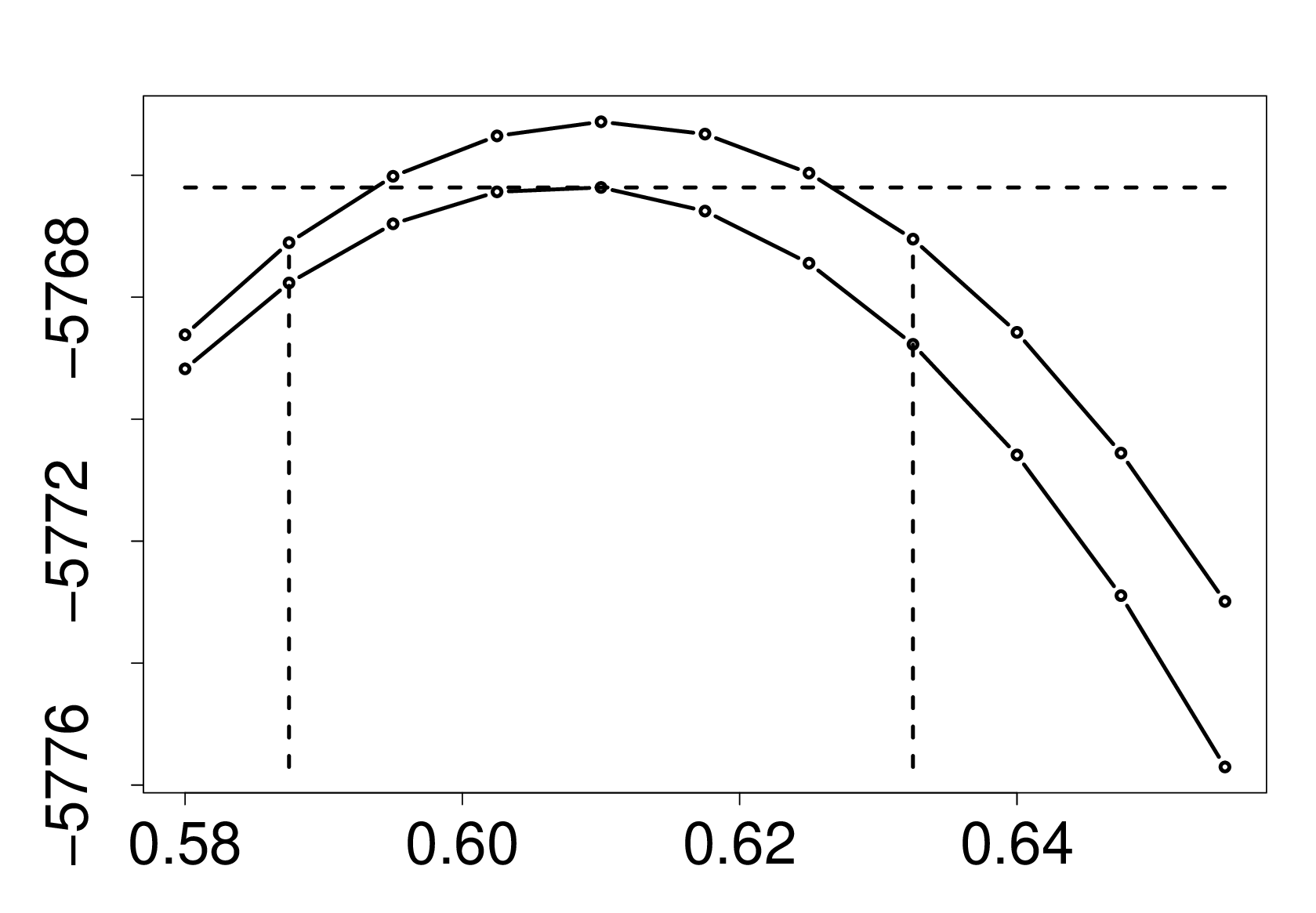} \\[-0.1cm]
$\nu=8$ & $\nu=10$ & $\nu=12$ \\[-0.3cm]
\includegraphics[height=4.0cm,width=2.2cm,angle=-90]{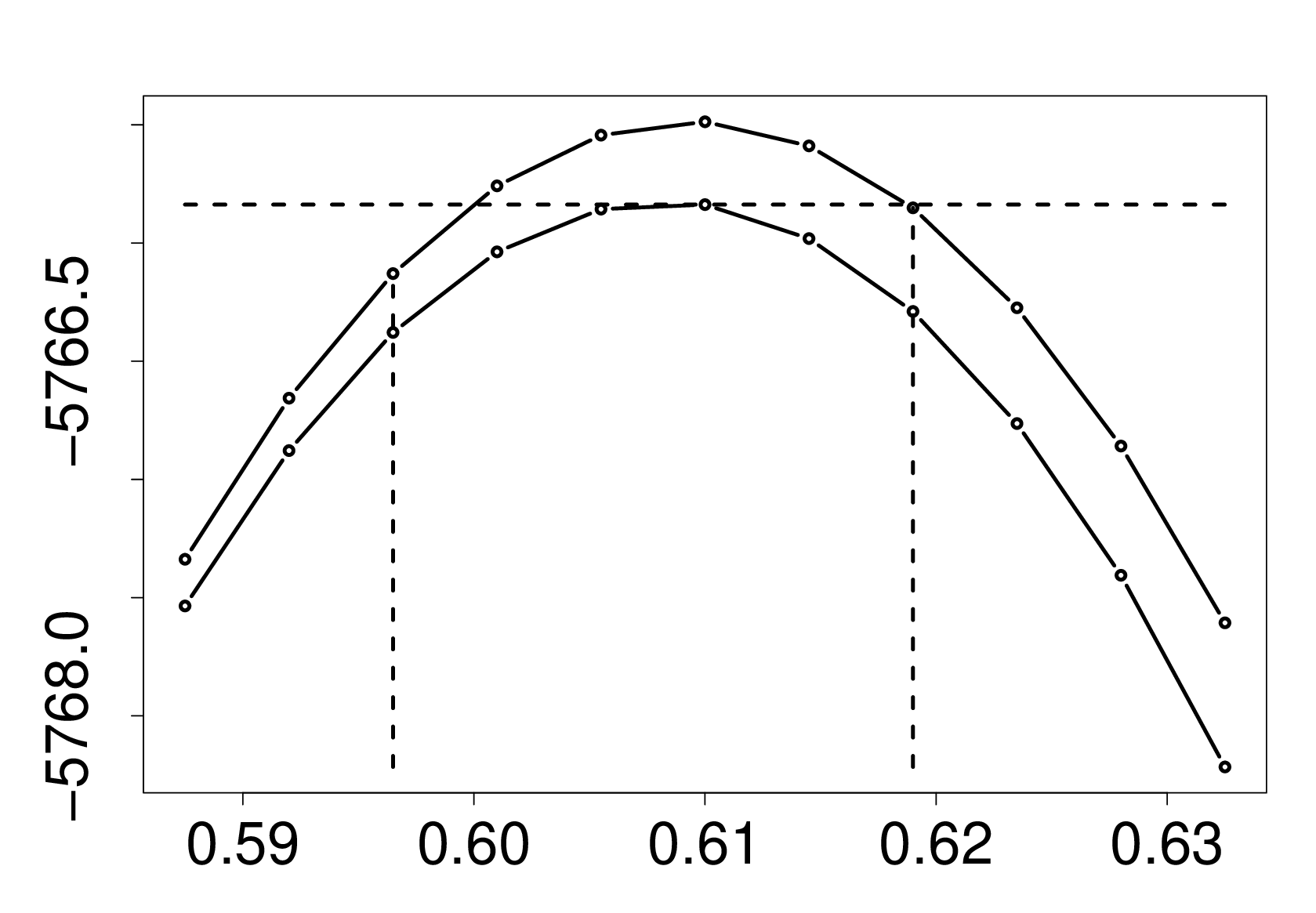} &
\includegraphics[height=4.0cm,width=2.2cm,angle=-90]{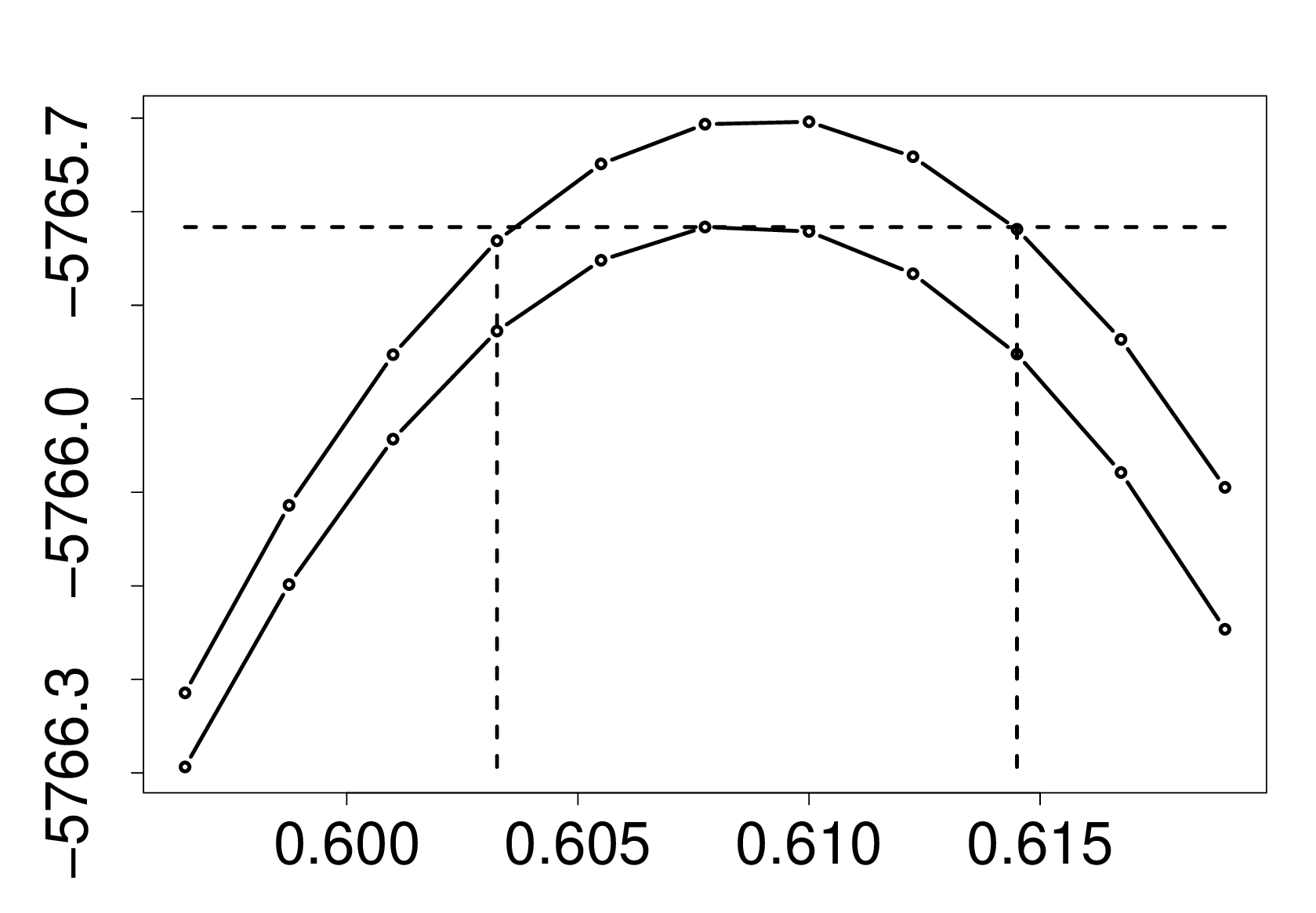} &
\includegraphics[height=4.0cm,width=2.2cm,angle=-90]{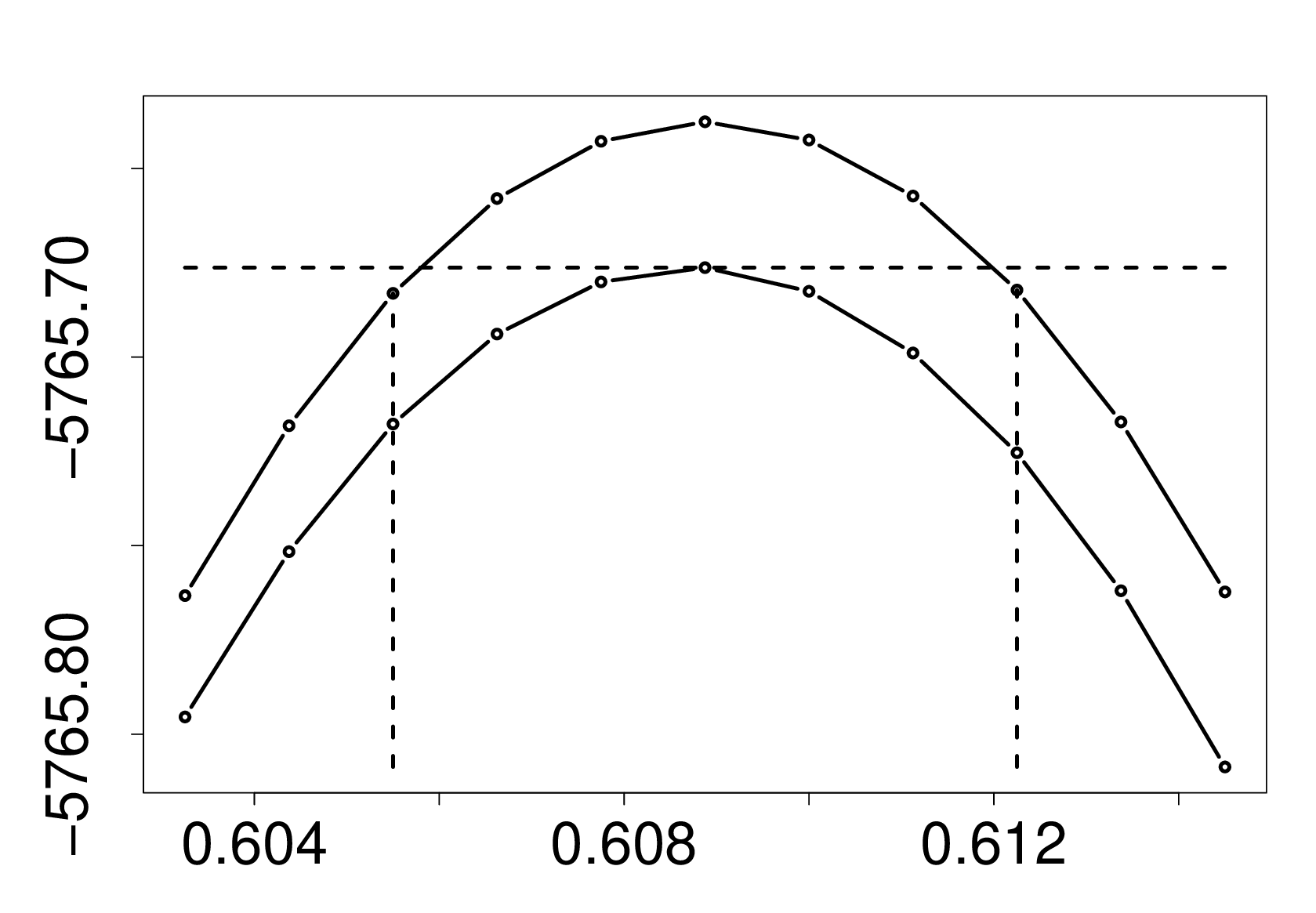} \\[-0.1cm]
$\nu=14$ & $\nu=16$ & $\nu=18$\\[-0.2cm]
\end{tabular}
\end{center}
\caption{\label{mle-0.6}Maximum likelihood example: The identification of an 
interval in which the MLE for $\theta$ in an Ising model must be located. The
estimation is based on a sample from the Ising model with 
$\theta=0.6$. The plots shows computed upper and lower bounds for 
the log-likelihood function for different values of $\nu$. The horizontal dotted
line is the maximum of the computed lower bounds and the two vertical 
dotted lines defines the interval in which the maximum likelihood estimate
must be located.}
\end{figure}

Assume we have observed an image $x$ which we suppose is a realization from 
an MRF $p(x|\theta)$, where $\theta$ is a scalar parameter. The $p(x|\theta)$
may for example be the Ising model. If we want to estimate $\theta$ the maximum
likelihood estimator (MLE) is a natural alternative. As discussed in the 
introduction the computation of the MLE is complicated by the intractable 
normalising constant of the MRF. Figure \ref{mle-0.6} 
illustrates how upper 
and lower bounds for the normalising constant can be used to identify
an interval in which the MLE for $\theta$ must be located. To produce the curves
in the figure we first simulated an $x$ 
from the Ising model with parameter 
$\theta=0.6$, on an $100\times 100$ rectangular lattice. 
Assume we want to find the MLE of $\theta$ based on this $x$. 
We first computed upper and lower bounds for the log-likelihood function by 
replacing the intractable normalising constant with corresponding upper and 
lower bounds with $\nu=2$ for $11$ values of $\theta$ on a mesh from $0$ to $2$. The 
maximum log-likelihood value must clearly be higher than the maximum of the lower
bound values. Assuming
the log-likelihood function to be concave we can then identify an interval in 
which the MLE of $\theta$ must lie. Defining a new mesh of $11$ $\theta$ values over this 
interval we repeated the process for $\nu=4$ and obtain an improved interval, see the 
upper middle plot in Figure \ref{mle-0.6}. 
We then repeated this process for $\nu=6,8,\ldots,18$, and the final interval for the MLE
shown in the lower right plot in Figure \ref{mle-0.6} was 
$(0.6055,0.6123)$. It is important not to confuse this interval with 
confidence or credibility intervals, the interval computed here is just 
an interval which, with certainty, includes the MLE of $\theta$.
It should be noted that computations of the
$2\cdot 11=22$ bounds for the normalising constants can be done in 
parallel, and for larger values of $\nu$ this is essential 
for the strategy to be practical.
In corresponding simulation experiments with true
$\theta$ values equal to $0.4$, $0.8$ and $-\ln(\sqrt{2}-1)$ we obtained
the final intervals $(0.4113,0.4115)$, $(0.7716,0.8490)$ and 
$(0.8234,0.9250)$, respectively. Our bounds are less tight for higher values of $\theta$
and naturally this gives longer intervals for the MLEs when the true $\theta$ value 
is larger.

\subsubsection{\label{sec:mpe}Maximum posterior estimation}
\begin{figure}
\begin{center}
\begin{tabular}{cc}
\includegraphics[height=3.3cm,width=3.3cm,angle=-90]{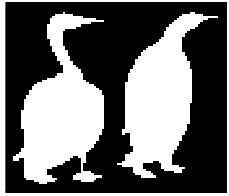}\\[-0.7cm] 
\end{tabular}
\end{center}
\caption{\label{fig:mpe}Maximum posterior estimation example: True scene used in the maximum posterior
estimation and fully Bayesian model examples.}
\end{figure}

\begin{figure}
\begin{center}
\begin{tabular}{@{}c@{}c@{}c@{}c@{}}
\includegraphics[height=3.3cm,width=3.3cm,angle=-90]{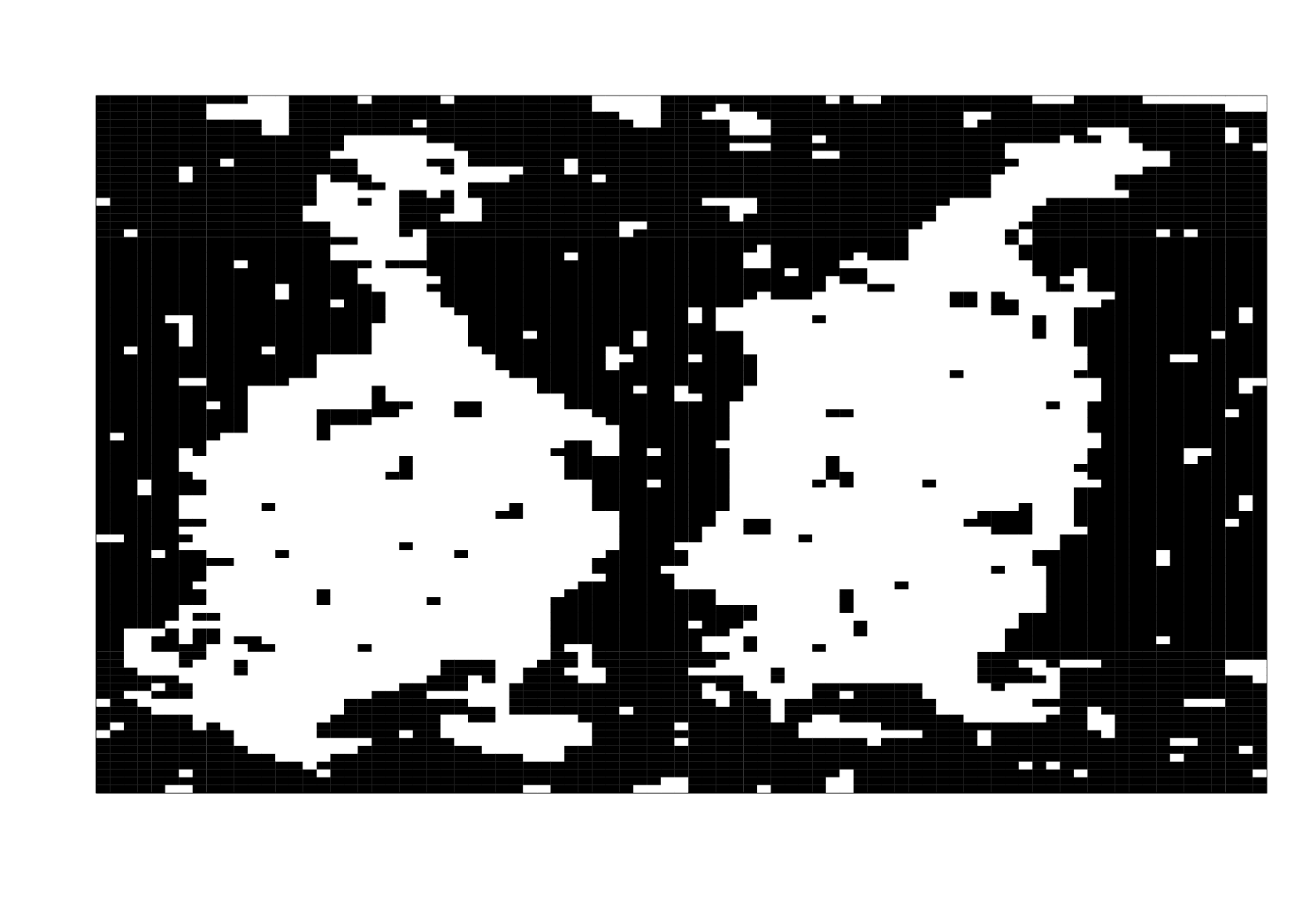} &
\includegraphics[height=3.3cm,width=3.3cm,angle=-90]{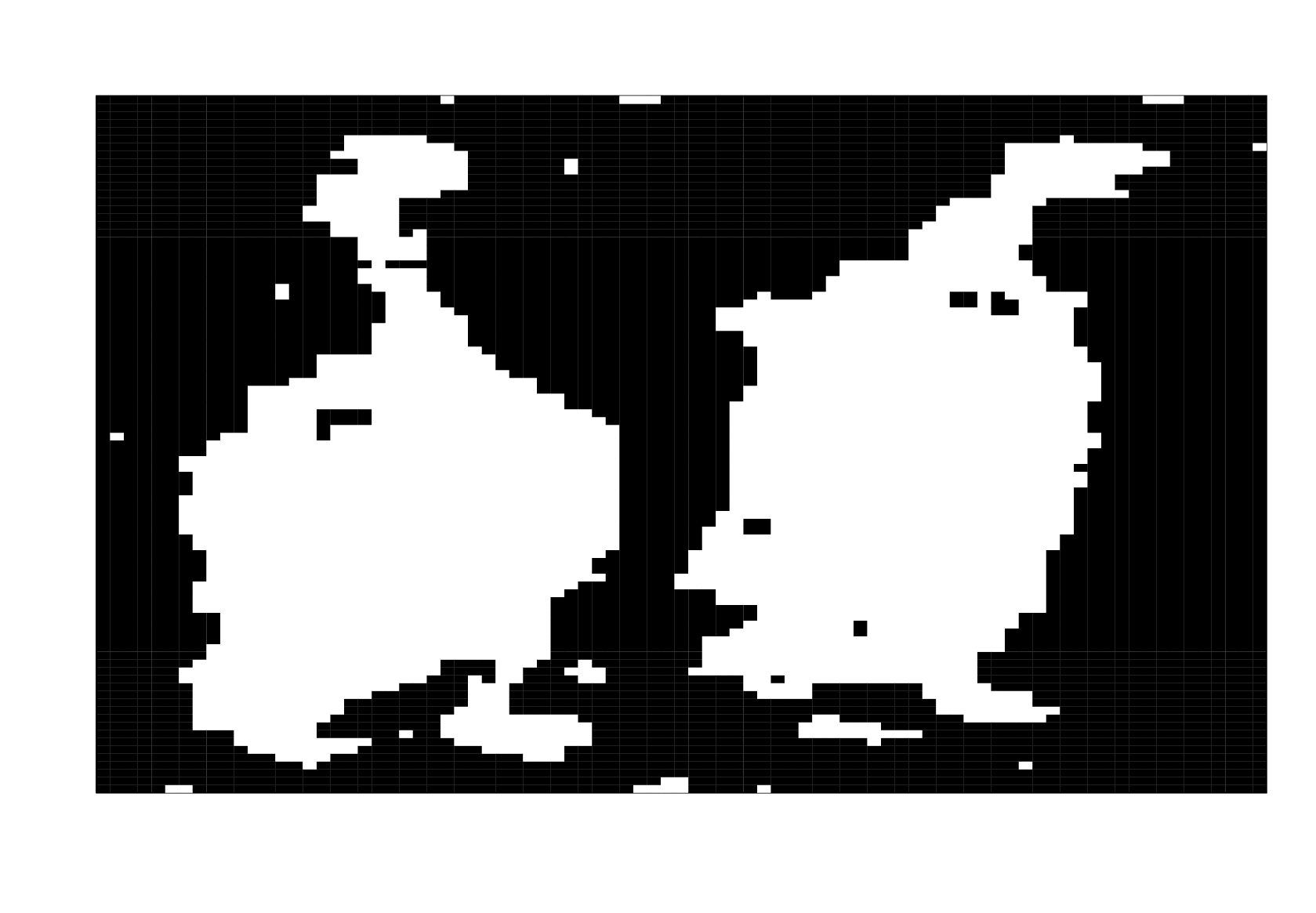} &
\includegraphics[height=3.3cm,width=3.3cm,angle=-90]{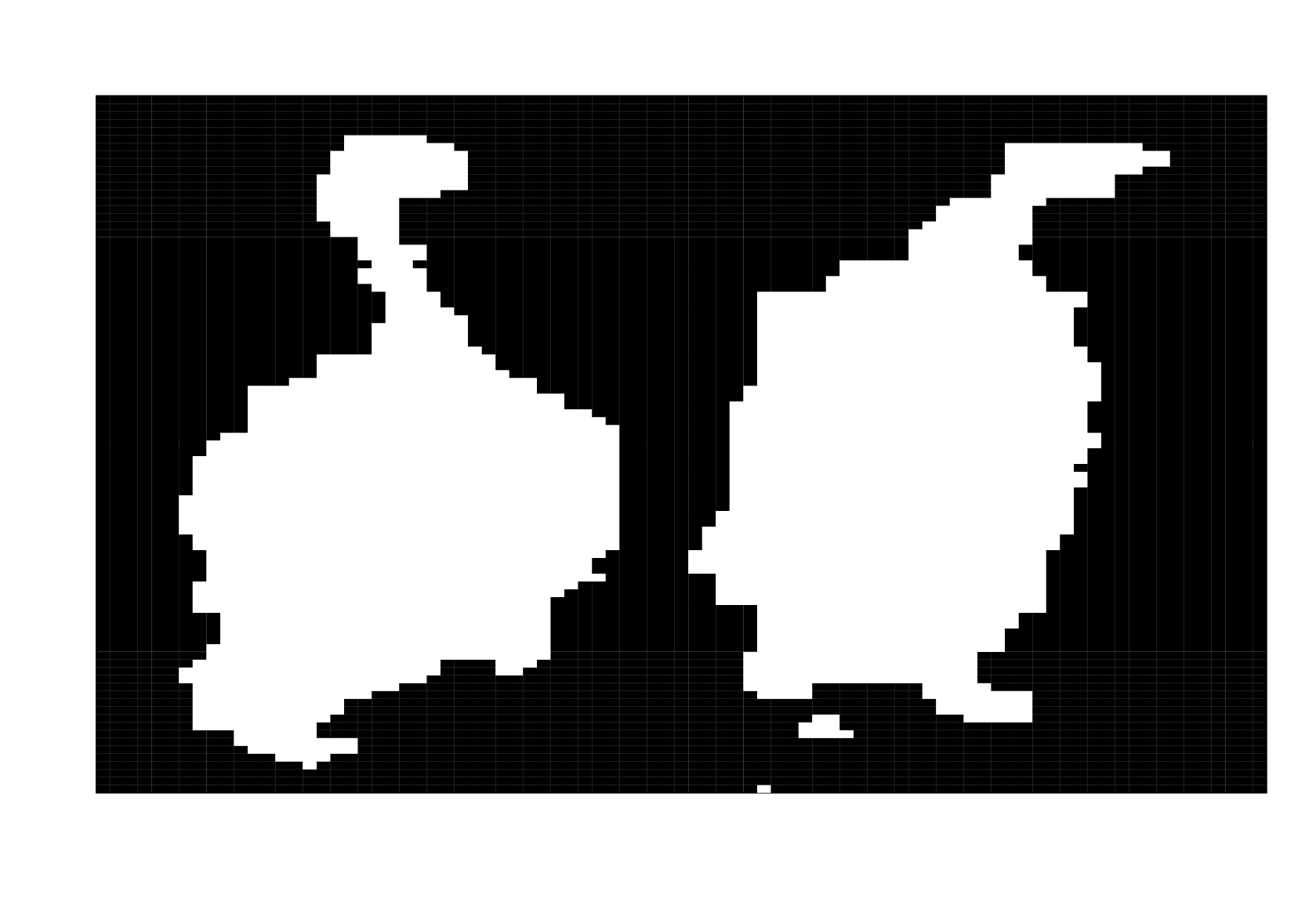} &
\includegraphics[height=3.3cm,width=3.3cm,angle=-90]{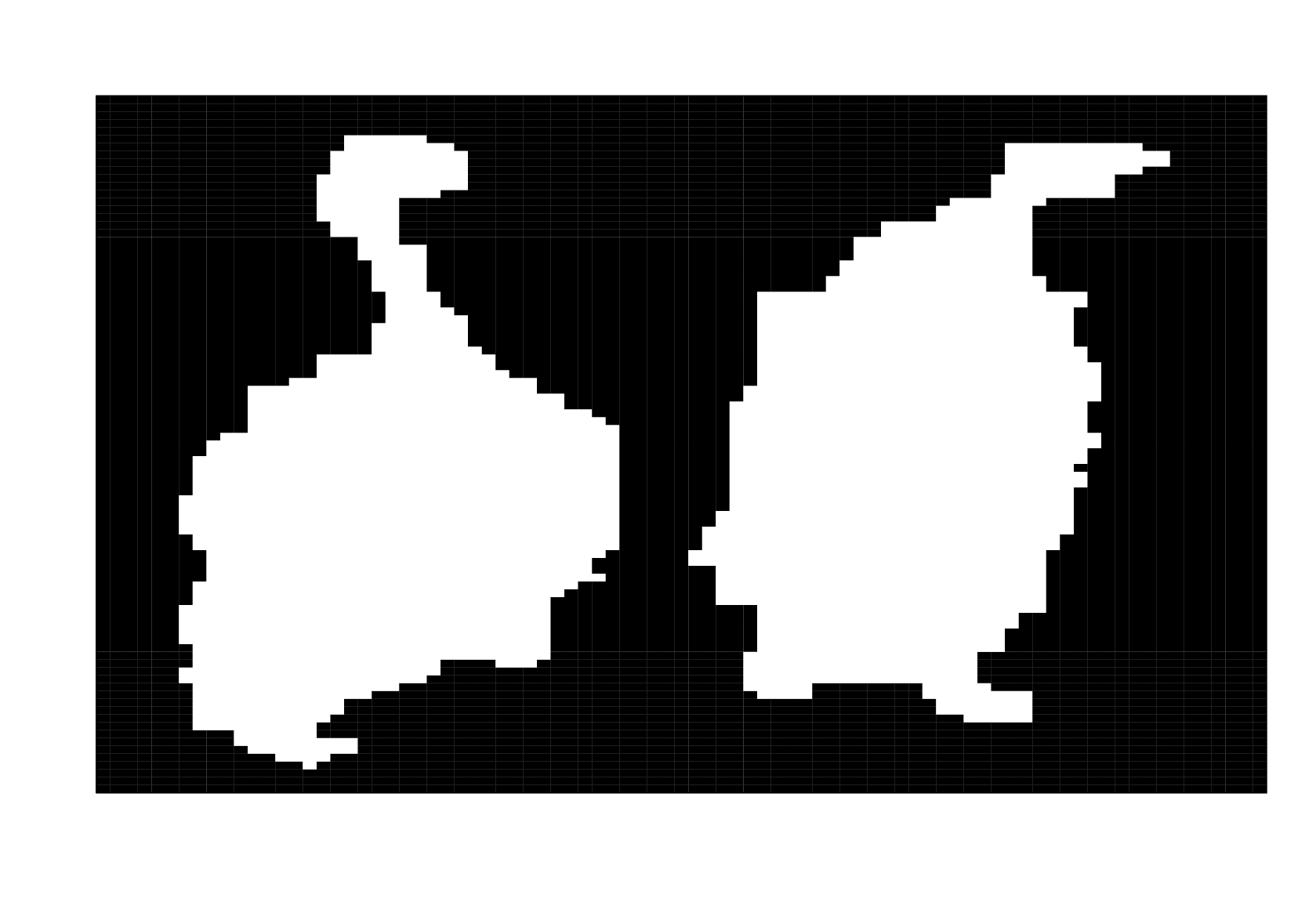}\\[-0.4cm] 
$0.4$ & $0.6$ & $0.8$ & $-\ln(\sqrt{2}-1)$\\[-0.5cm]
& 
\includegraphics[height=3.3cm,width=3.3cm,angle=-90]{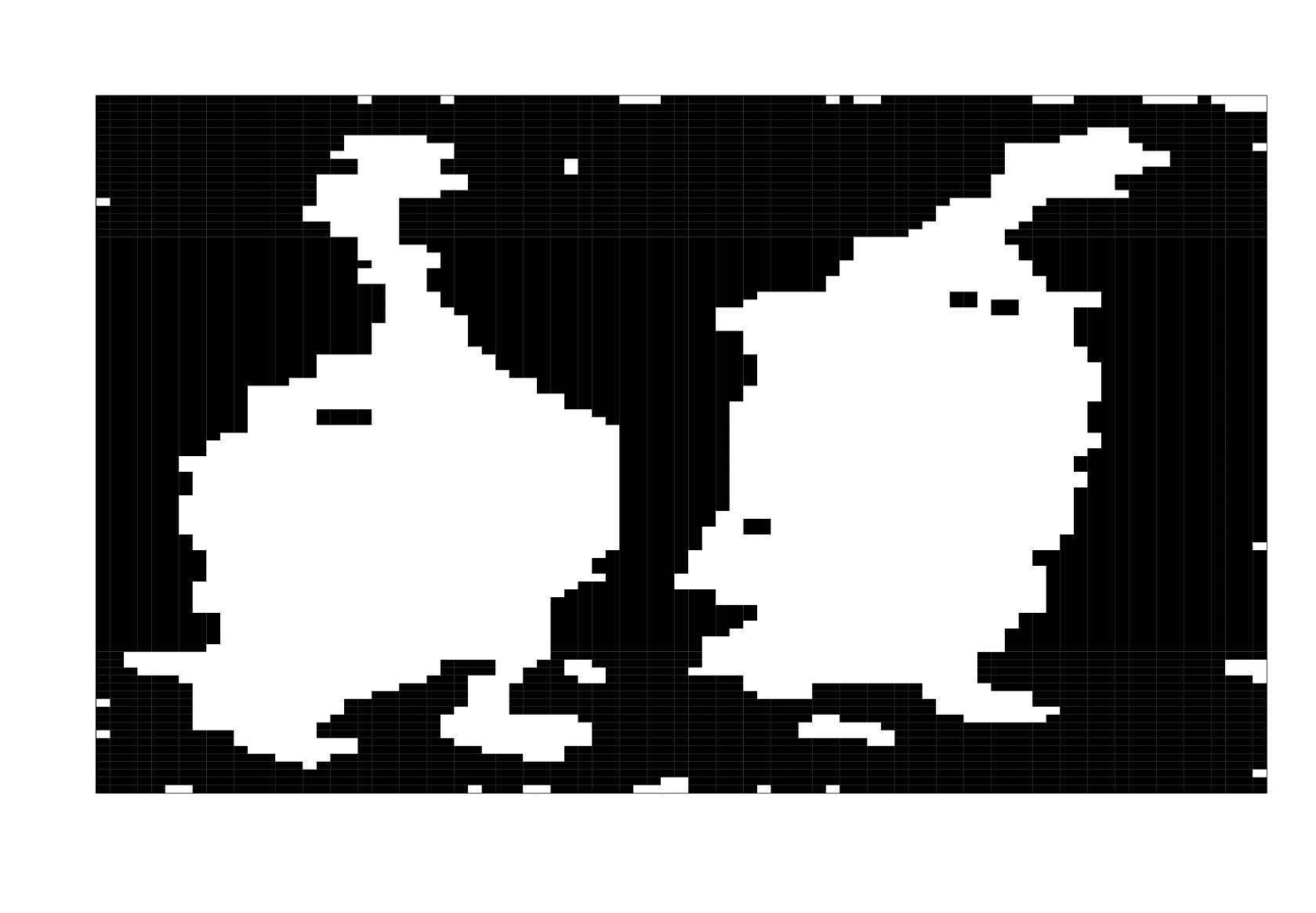} &
\includegraphics[height=3.3cm,width=3.3cm,angle=-90]{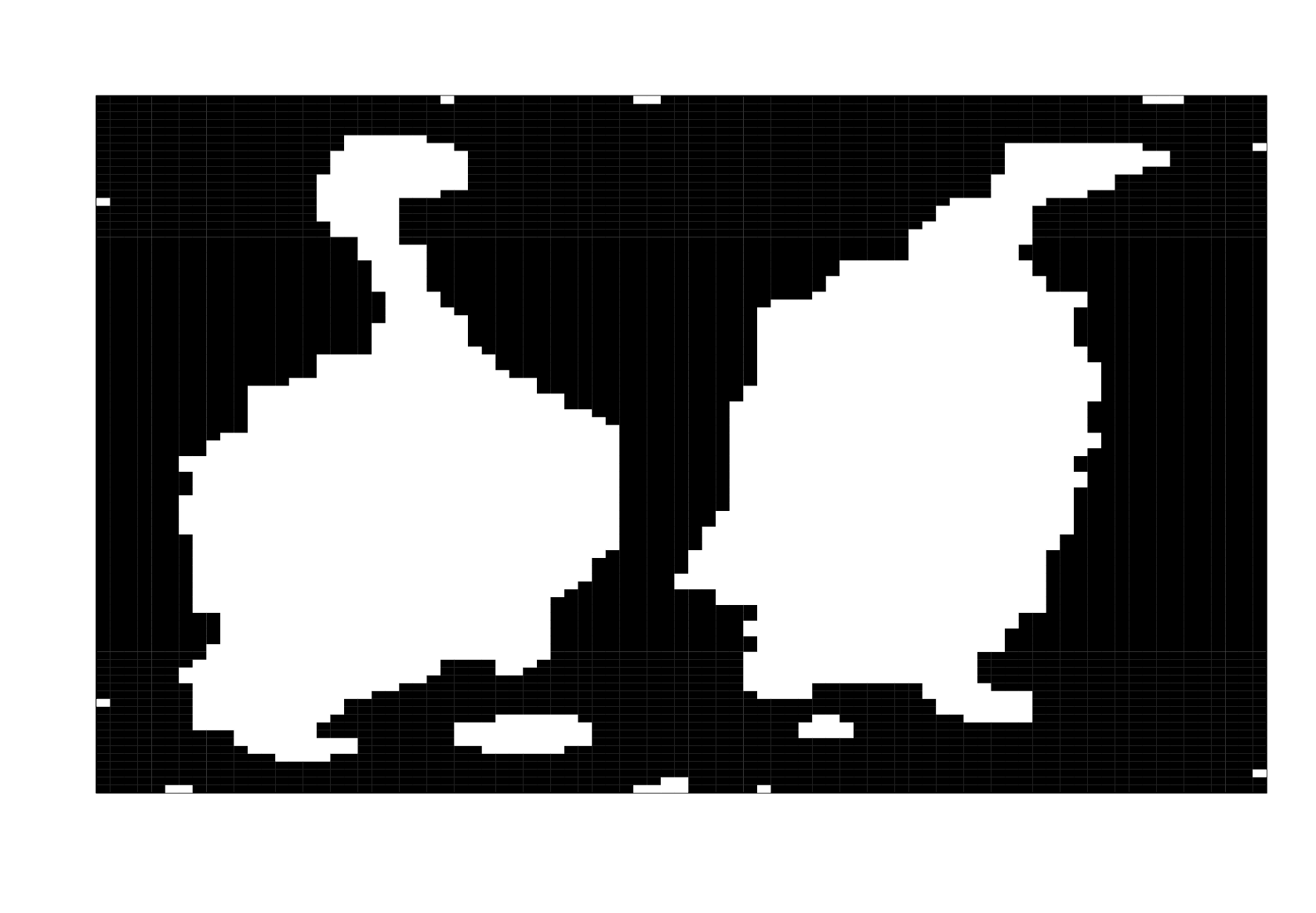} &
\\[-0.4cm]
& Model 1 & Model 2 & \\[-0.5cm]
\end{tabular}
\end{center}
\caption{\label{fig:mpeResults}Maximum posterior estimation example: Approximations
to $\widehat{x}$ when $\nu=18$ for the four Ising priors (upper row) and 
the two higher order MRFs (lower row).}
\end{figure}

The approximate Viterbi algorithm discussed in Section \ref{sec:maximisation}
can be used in image analysis applications. Suppose the constructed scene in Figure 
\ref{fig:mpe}(a), which is in an $89\times 85$ lattice and which we denote by $x$,
is an unobserved true scene, and assume that a corresponding noisy version $y$ of $x$ is 
observed. Here we will use a $y$ 
generated from $x$ by drawing each element of $y$ independently from a normal 
distribution with unit standard deviation and with means $0$ and $1$ for the 
white and black areas of $x$, respectively. Assuming the likelihood parameters
to be known, and assigning an MRF prior $p(x)$ to $x$, we can estimate $x$ from 
$y$ by maximising the posterior distribution with respect to $x$, i.e.
\begin{equation}
\widehat{x} = \mbox{arg}\max_x\left\{ p(x)p(y|x)\right\}.
\end{equation}
To compute $\widehat{x}$ is computationally intractable, 
but by adopting the procedure discussed in Section \ref{sec:maximisation}
we obtain an approximation to $\widehat{x}$. We have done this for the six priors
defined above and for values of $\nu$ from $1$ to $18$. The scenes in Figure
\ref{fig:mpeResults}
show the results for $\nu=18$. It is interesting to observe that the 
results for the Ising prior with $\beta=0.6$ and $0.8$ are very similar
to the results for the higher-order interaction MRFs Models 1 and 2,
respectively.

Comparing the results for different values of 
$\nu$ we find that for the Ising model with $\beta=0.4$ the results are 
identical for all $\nu=5$ to $18$. For the other seven priors the resulting
scenes continued to vary slightly for all the values we used for $\nu$, but 
in all cases the differences became smaller for higher values of $\nu$.
For example, for all these seven priors less than $0.6\%$ of the nodes were 
assigned different values when $\nu=16$ and $\nu=18$ were used. When 
comparing the results with the true scene the Ising model with $\beta=0.8$ 
has the lowest number of misclassifications, with the higher-order MRF Model 2
slightly behind.

\subsubsection{\label{sec:fullyBayesian}Fully Bayesian modelling}
Assume again that we have observed a noisy scene $y$ corresponding to a unobserved
true image $x$. An alternative to the strategy for estimating $x$ from $y$ discussed
above is to adopt a fully Bayesian approach. Here we consider the same simulated $y$
as we did in Section \ref{sec:mpe}. Then we consider the unobserved $x$ to 
be a sample from an MRF prior $p(x|\theta)$, where $\theta$ is a parameter vector. For 
$p(x|\theta)$ we try both the Ising prior (\ref{eq:Ising}) and the higher order MRF defined in 
Section \ref{sec:models}. The higher order MRF has the ten parameters specified in Figure \ref{fig:pot}, 
but to make the model identifiable we fix one parameter corresponding to each of the two 
maximal cliques. We set the 
potentials to zero for the configurations where all nodes in a maximal clique are equal.
Thus, $\theta$ gets eight elements in this prior. To fully specify the Bayesian
model we need to adopt a parametric form for the likelihood $p(y|x,\varphi)$, where
$\varphi$ is a parameter vector, and to specify priors for $\theta$ and $\varphi$. For the 
likelihood we assume the product of normals given in Section \ref{sec:mpe} and
$\varphi$ contains a mean value and a standard deviation for each of the two possible
values of the elements in $x$, i.e. $\varphi = [\mu_0,\mu_1,\sigma_0,\sigma_1]$.
To avoid problems if all elements of $x$ are assigned to the same value we need to 
adopt proper priors for the likelihood parameters. For $\mu_0$ and $\mu_1$ we use independent
normal priors with zero mean and standard deviation ten, but
to make the model identifiable we add the restriction $\mu_0\leq \mu_1$. For 
$\sigma_0,\sigma_1\geq 0$ we a priori assume independent exponential densities
with means equal to ten.

\begin{figure}
\begin{center}
\begin{tabular}{cccc}
\includegraphics[height=3.0cm,width=3.0cm,angle=-90]{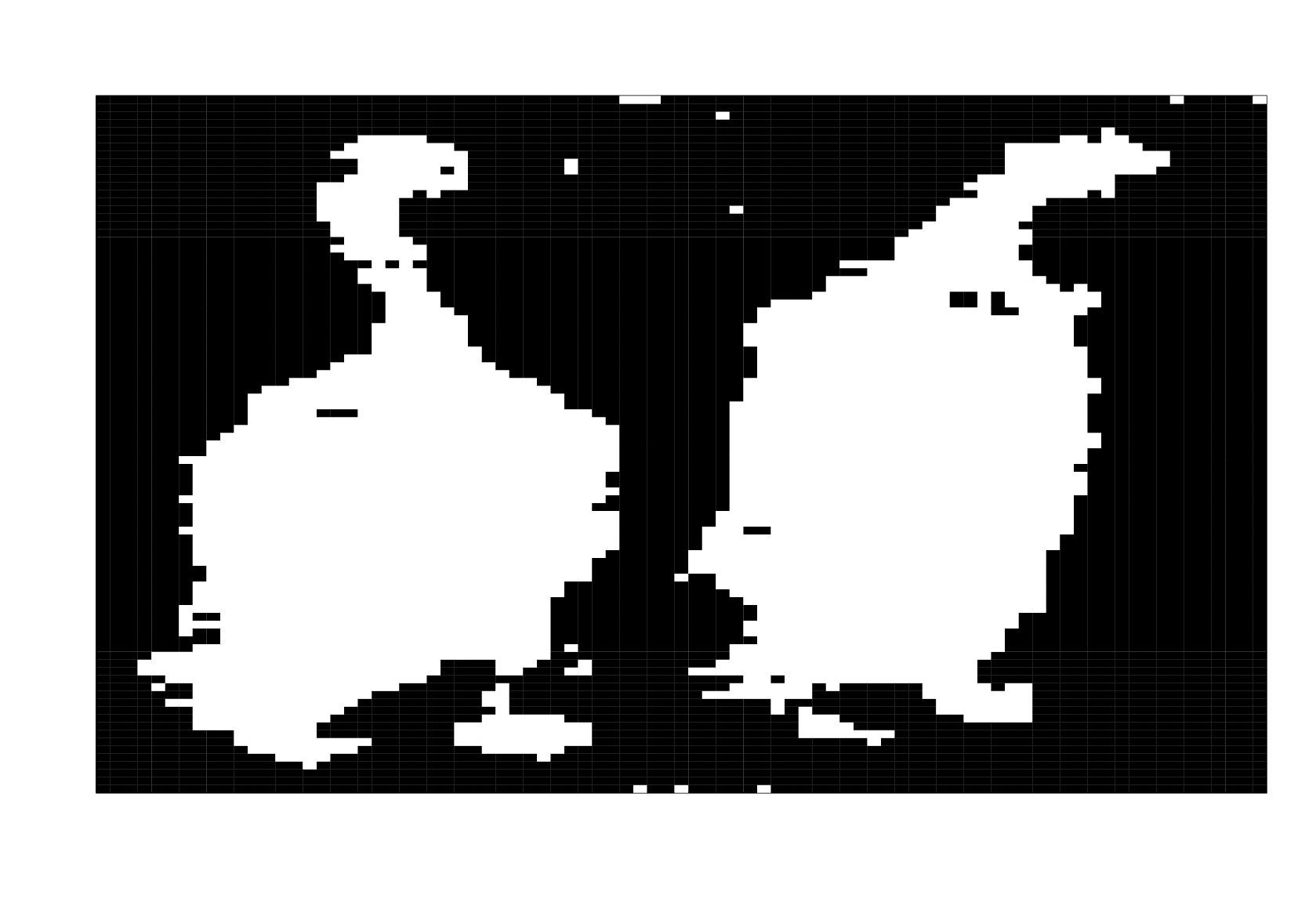} &
\includegraphics[height=3.0cm,width=3.0cm,angle=-90]{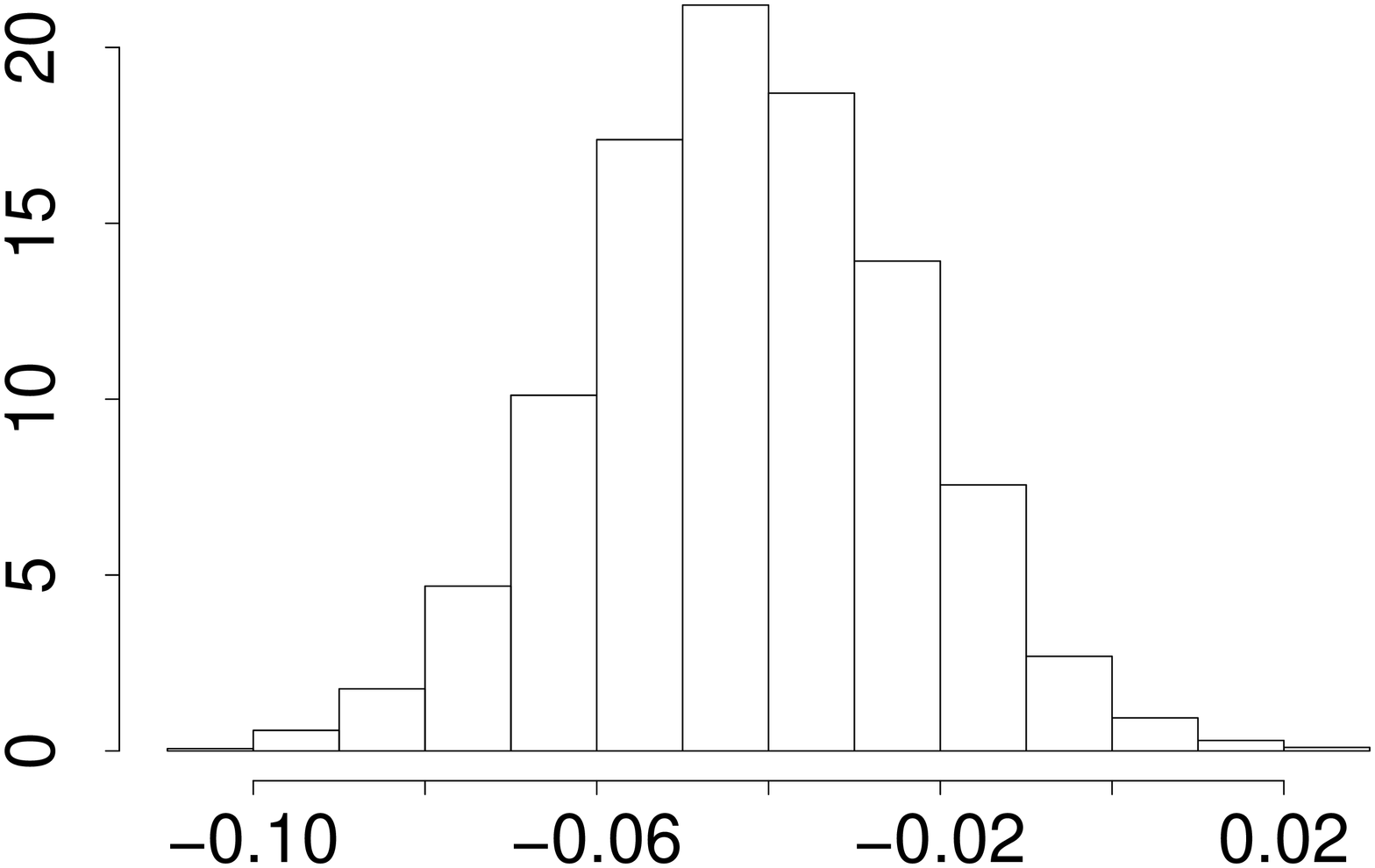} &
\includegraphics[height=3.0cm,width=3.0cm,angle=-90]{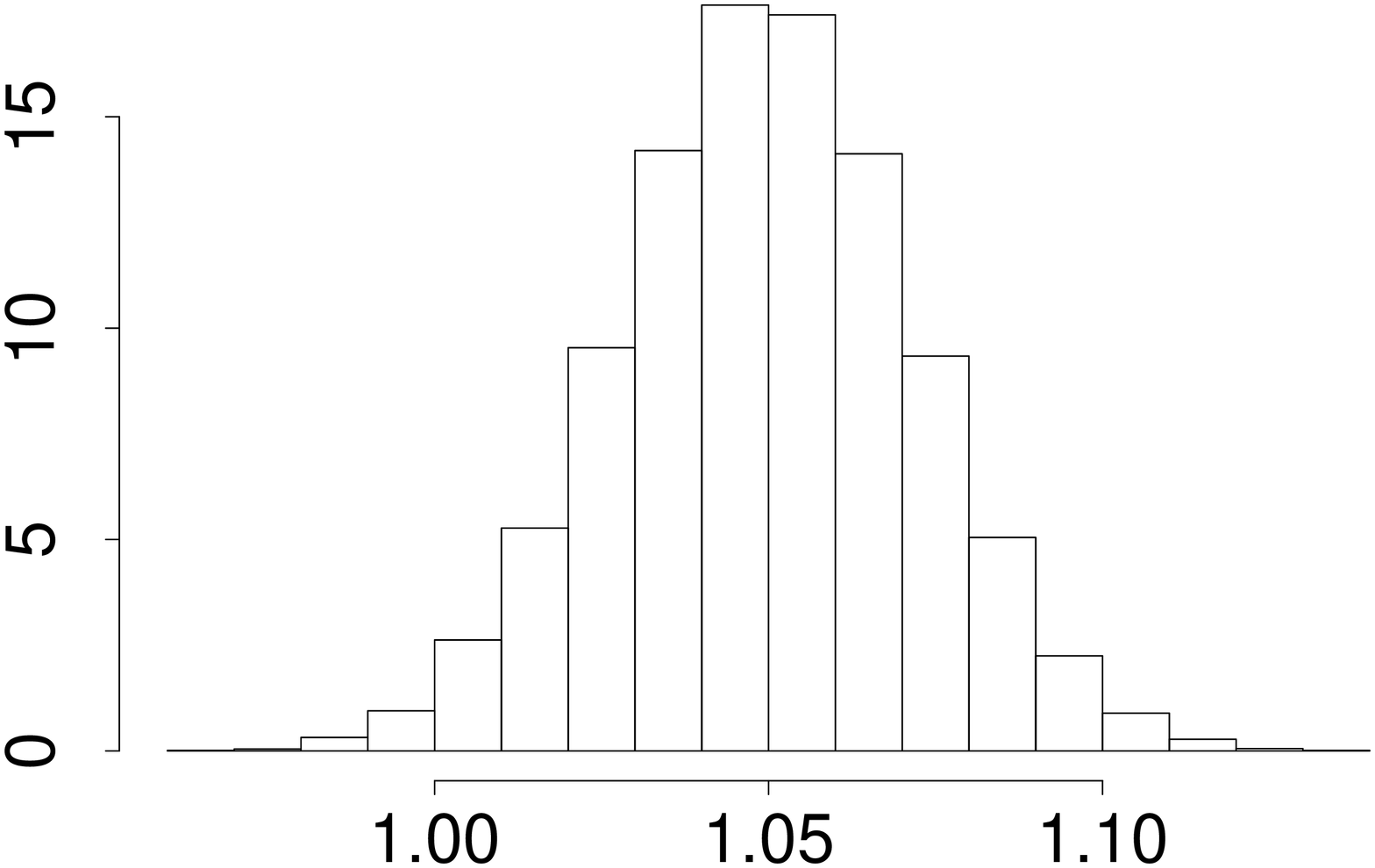} &
\includegraphics[height=3.0cm,width=3.0cm,angle=-90]{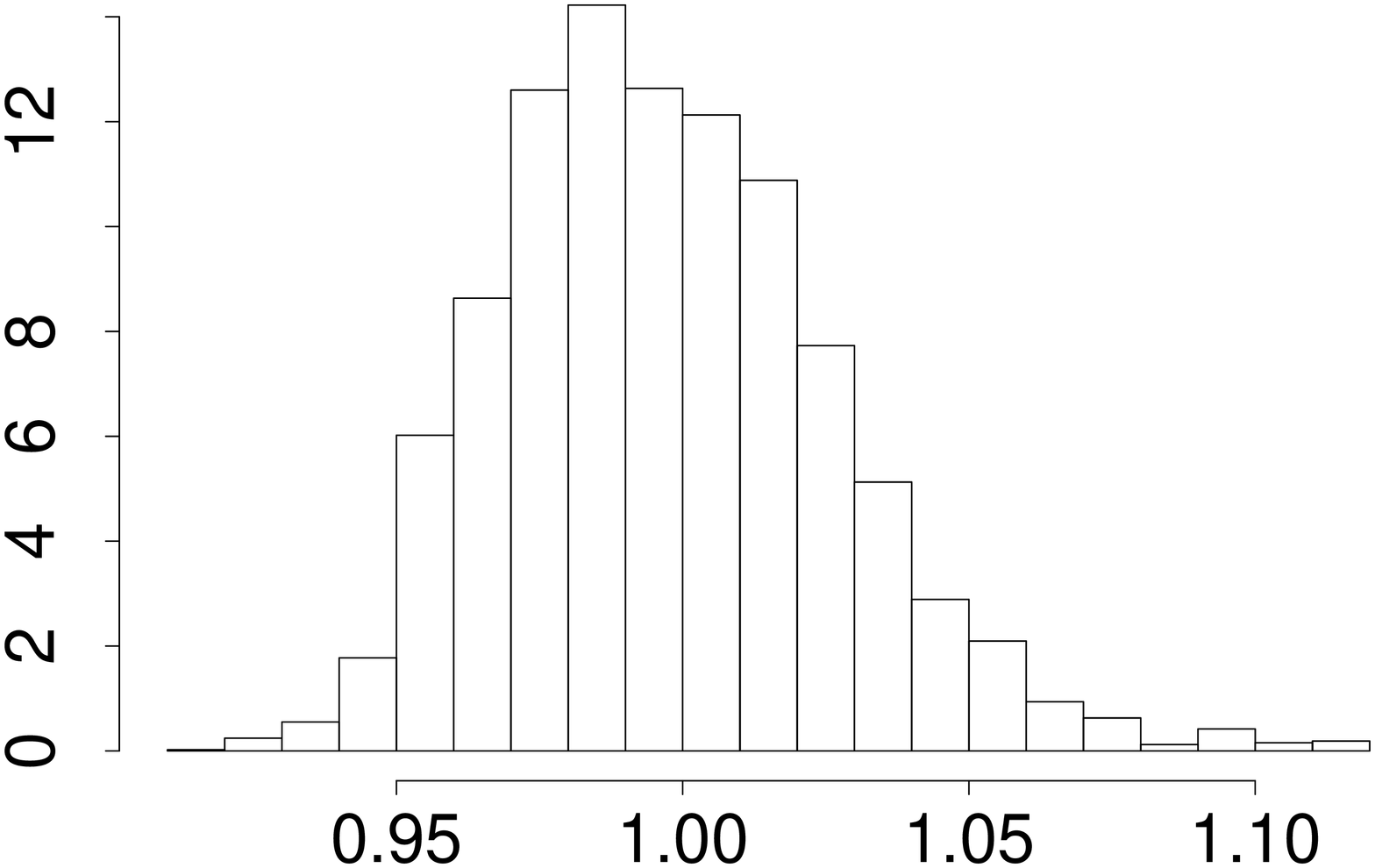} \\
\includegraphics[height=3.0cm,width=3.0cm,angle=-90]{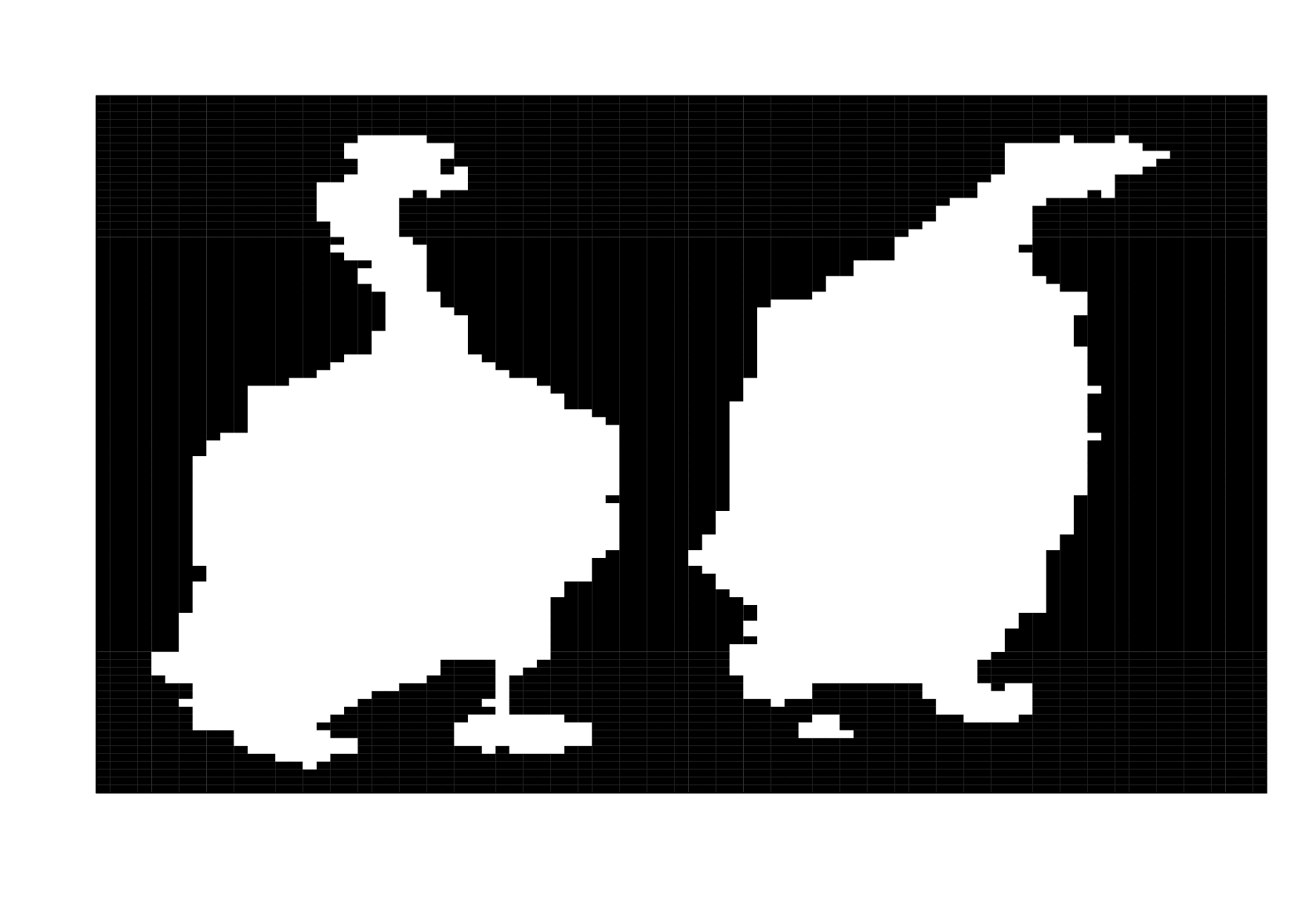} &
\includegraphics[height=3.0cm,width=3.0cm,angle=-90]{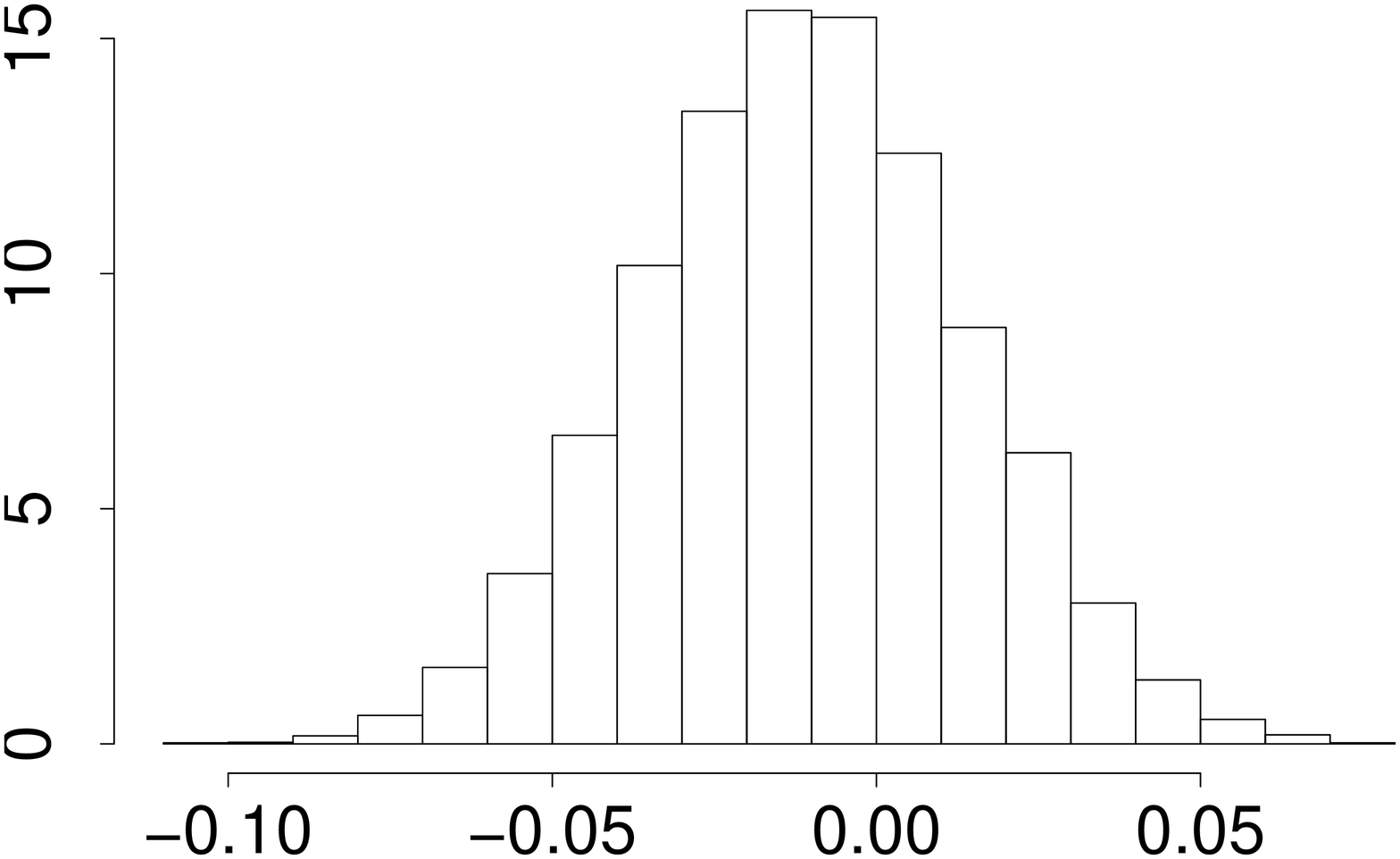} &
\includegraphics[height=3.0cm,width=3.0cm,angle=-90]{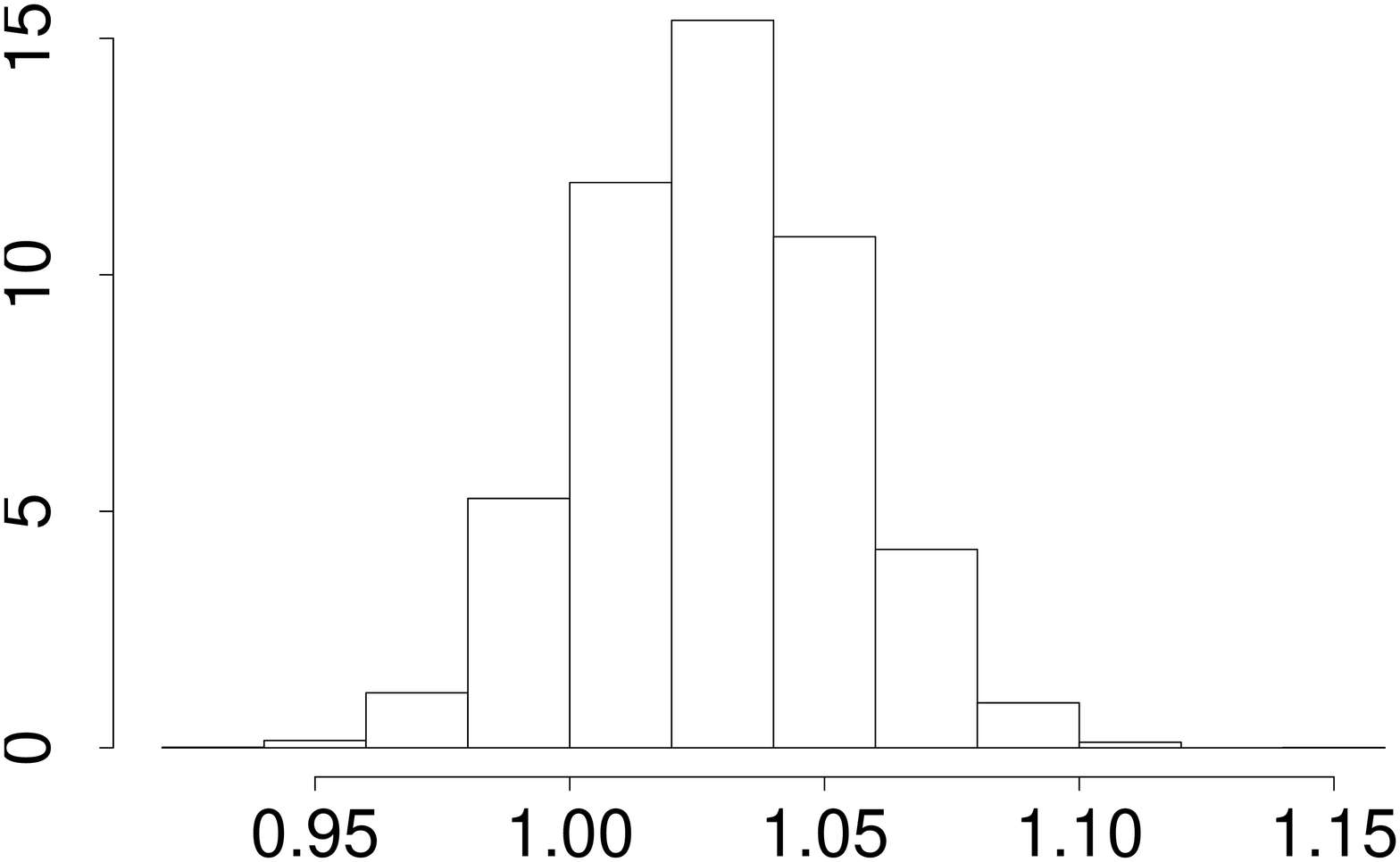} &
\includegraphics[height=3.0cm,width=3.0cm,angle=-90]{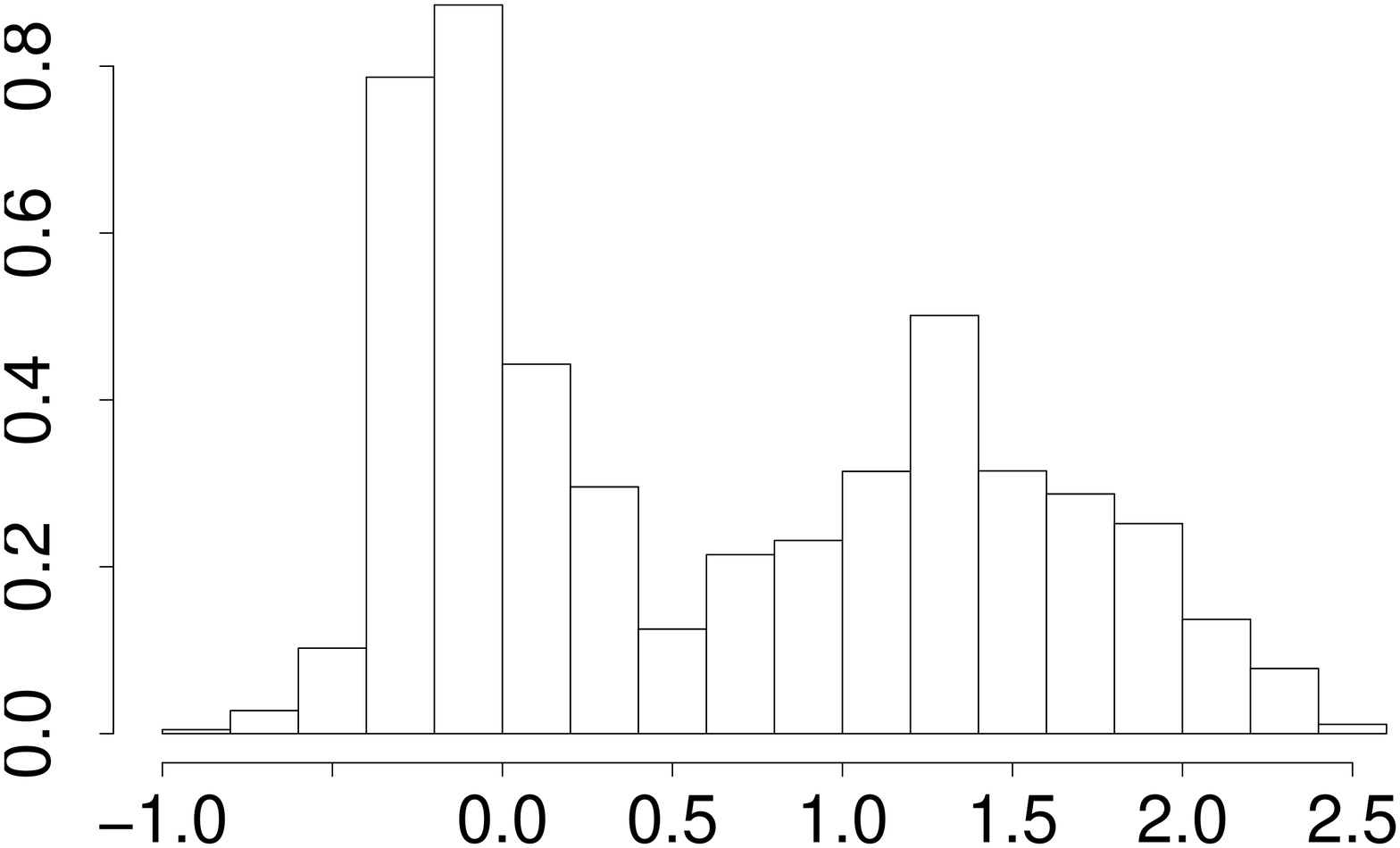}\\[-0.6cm]
\end{tabular}
\end{center}
\caption{\label{fig:fullyhomrf}Fully Bayesian model example: Results for a fully Bayesian model, using
an Ising prior (upper row) and the higher-order interaction MRF prior (lower row). The leftmost column shows
the maximum marginal posterior probability estimate of the underlying scene. The two following columns show
estimated posterior distributions for $\mu_0$ and $\mu_1$, respectively. The rightmost column 
shows estimated posterior distribution for $\theta$ for the Ising prior and for the eighth parameter
in Figure \ref{fig:pot} for the higher-order interaction prior.}
\end{figure}

To simulate from the resulting posterior $p(x,\theta,\varphi|y) \propto 
p(\theta)p(\varphi)p(x|\theta)p(y|x,\varphi)$ is computationally infeasible due 
to the normalising 
constant of $p(x|\theta)$, but we can simulate from the approximation 
$\widetilde{p}(x,\theta,\varphi|y) \propto 
p(\theta)p(\varphi)\widetilde{p}(x|\theta)p(y|x,\varphi)$, where $\widetilde{p}(x|\theta)$
is defined as in (\ref{eq:pommapprox}) and here we adopt the second POMM 
approximation variant discussed in the paragraph following (\ref{eq:pommapprox}).

. To simulate from $\widetilde{p}(x,\theta,\varphi|y)$
we adopt a Metropolis--Hastings algorithm and alternate between single site updates for the
likelihood parameters and the for
the components of $x$ and a joint block update for the MRF parameters and $x$. To do 
Gibbs updates for the likelihood parameters is not feasible, but we use proposals 
distributions which are close to the full conditionals. More precisely, for each of $\mu_0$ and $\mu_1$
we find the full conditionals when ignoring the restriction $\mu_0<\mu_1$ and use 
these as proposals distributions, and for $\sigma_0$ and $\sigma_1$ we propose
potential new values by proposing values for the corresponding variances
from their full conditionals when a very vague inverse gamma prior is assumed.
The proposals are then accepted of rejected according to the standard Metropolis--Hastings
procedure. In practice essentially all proposals are accepted. 
In the single site updates for the components of $x$
we generate the potential new state by changing the value of a randomly 
chosen element in $x$, and in the block update we use a proposal distribution 
\begin{equation}
q(\theta^\prime,x^\prime|\theta,\varphi,x) = 
q(\theta^\prime|\theta) \widetilde{p}(x^\prime|y,\theta^\prime,\varphi^\prime),
\end{equation}
where $q(\theta^\prime|\theta)$ is a
random walk proposal for $\theta$, and 
$\widetilde{p}(x^\prime|y,\theta^\prime,\varphi^\prime)$ is the POMM approximation
of the posterior MRF $p(x^\prime|y,\theta^\prime,\varphi^\prime)\propto \widetilde{p}(x|\theta^\prime)
p(y|x^\prime,\varphi^\prime)$. In the random walk proposals for $\theta$ we sample the 
elements independently from normal distributions centered at the current values and with all
standard deviations equal to $0.1$ both for the Ising and higher-order MRF prior 
cases. We define one MH iteration to consist of one update for each of the 
likelihood parameters, 
$89\cdot 85$ single site updates for elements of $x$ and one block update as defined above.
Using $\nu=8$ to define
the POMM approximations, Figure \ref{fig:fullyhomrf}
summarises some simulation results. The left column shows the estimated true scene using the 
marginal posterior probability estimator for each of the two priors. We observe that the 
higher-order MRF looks less noisy, and the fraction of wrongly classified values are
$4.7\%$ and $4.2\%$ for the Ising and higher-order MRF priors, respectively. The 
corresponding means of the marginal posterior probabilities for the wrong class labels are $0.102$ and
$0.083$, respectively. Thus, the posterior probability for the wrong class is on average
$18.5\%$ lower when using the higher-order MRF prior than with the Ising prior. The two 
middle columns in Figure \ref{fig:fullyhomrf} show the estimated posterior distributions
for $\mu_0$ and $\mu_1$ and we observe that the true values, zero for the second column and one 
for the third, are far out in the tail of the posterior when using the Ising prior and more centrally
located when using the higher-order MRF. All of this demonstrates the potential advantage of using 
a higher-order MRF prior. The last column in Figure \ref{fig:fullyhomrf} shows the 
estimated posterior distribution for $\theta$ when using the Ising prior, and the for the eighth 
parameter in Figure \ref{fig:pot} for the higher-order interaction prior. We observe that the 
variance in the posterior for $\theta$ is quite small, but its value is also
well above the phase transition limit.
The posterior variability for the parameter in the 
higher-order interaction prior, however, is very large, and the same is true for the other
parameters in this prior. This may indicate that the higher-order interaction prior is 
over-parameterised and that better results could perhaps have been obtained by adopting a prior
with fewer parameters. The best alternative would perhaps be to put a prior also on the 
interaction order of the model, so that the complexity of the model could adapt automatically to the 
problem in focus.

\subsubsection{Perfect sampling by rejection sampling}
The last application of our approximation and bounds we discuss is how it can be 
used to construct a rejection sampling algorithm generating perfect samples
from some MRFs $p(x)$. Let $p(x)=\frac{1}{c}\exp\{ U(x)\}$ 
be a given binary MRF, and let $\widetilde{p}(x)$
denote the corresponding POMM approximation. 
One can then imagine a rejection sampling
algorithm generating candidate samples from $\widetilde{p}(x)$ and accepting a 
candidate $x$ with probability
\begin{equation}\label{eq:rejectionSampling}
\alpha(x) = k\cdot\frac{p(x)}{\widetilde{p}(x)} = \widetilde{k}\cdot \frac{\exp\{ U(x)\}}{\widetilde{p}(x)},
\end{equation}
where $k$ is a constant so that $\alpha(x)\leq 1$ for all $x$, and $\widetilde{k}=k/c$. 
Clearly the optimal choice for 
$\widetilde{k}$ is 
\begin{equation}
\widetilde{k}_{\mbox{\tiny opt}} = \min_x\left\{ \widetilde{p}(x)e^{-U(x)}\right\},
\end{equation}
but to compute $\widetilde{k}_{\mbox{\tiny opt}}$ is computationally intractable. However, noting that 
$\ln\left\{\widetilde{p}(x)e^{-U(x)}\right\}$ is a pseudo-Boolean function we can find a lower bound
$\ln \widetilde{k}_{\mbox{\tiny bound}} \leq \ln \widetilde{k}_{\mbox{\tiny opt}}$ as discussed in 
Section \ref{sec:mpe}, and use $\widetilde{k}=\widetilde{k}_{\mbox{\tiny bound}}$ in 
(\ref{eq:rejectionSampling}). 
It should be noted this procedure implies that we need to compute the normalising constant
of the conditional distributions in (\ref{eq:pommapprox}) for all values of 
the conditioning variables, so when defining the POMM approximation $\widetilde{p}(x)$
we must use the second approximation variant discussed in the paragraph 
following (\ref{eq:pommapprox}).

The acceptance rate of such a rejection sampling procedure will depend both 
on the quality of the approximation $\widetilde{p}(x)$ and the ratio
$\widetilde{k}_{\mbox{\tiny bound}}/\widetilde{k}_{\mbox{\tiny opt}}$. One should note that the goal 
in this setting is not to get a very high rejection sampling acceptance rate, but rather to 
find a good trade off between the acceptance rate and the computation time for generating the proposals,
for example by finding the smallest value of $\nu$ that give an acceptance rate above a given threshold.
We have tried the procedure on the posterior distributions defined in Section \ref{sec:mpe}.
For the posteriors based on Ising priors with $\beta=0.4$, $0.6$, $0.8$ and $-\ln(\sqrt{2}-1)$
we needed $\nu=5$, $7$, $9$ and $11$, respectively, to obtain acceptance rates above $0.1$, 
and to get acceptance rates above $0.5$ the corresponding values for $\nu$ are 
$6$, $8$, $11$ and $13$. Perfect samples from these four posterior distributions can of course
alternatively be generated by the coupling from the past procedure of \citet{art27}. As the
rejection sampling procedure requires an initiation step of establishing $\widetilde{p}(x)$
and computing $\widetilde{k}_{\mbox{\tiny bound}}$, coupling from the past is the 
most efficient alternative whenever only one or a few independent samples are required,
but if a large number of samples are wanted the rejection sampling algorithm 
becomes the best alternative. We have tried the rejection sampling procedure also for 
the posteriors when adopting the higher order models defined in Section \ref{sec:models} as priors,
but for this prior we were not able to get useful acceptance rates within reasonable computation times.

\subsection{\label{sec:dataExample}Real data example}
\begin{figure}
\begin{center}
\includegraphics[height=6.0cm,width=6.0cm,angle=-90]{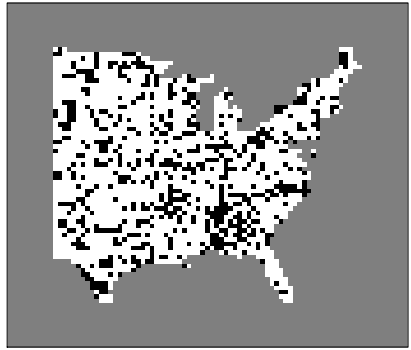}
\end{center}

\vspace*{-0.8cm}

\caption{\label{fig:mortalityMap}Real data example: U.S. liver and gallbladder cancer mortality map
for white males during 1950-1959. Black and white squares denote counties with high
and low cancer mortality rates, respectively. The outer box shows the boundary of the extended lattice.}
\end{figure}

In this section we consider a United States cancer mortality
map compiled by \citet{book36}. Figure \ref{fig:mortalityMap} shows
the mortality map for liver and gallbladder cancers for white males in 1950-1959,
where black and white squares denote counties with high and low cancer mortality rates,
respectively. The data set is previously analysed in 
\citet{art148} and \citet{art149}, see also the discussion in 
\citet{book37}. In these studies a free boundary autologistic model 
is compared with a model where the values in the nodes are assumed to 
be independent, and the conclusions in both studies are that the spatial model
gives a much better fit to the data than the independence model.

We define a fully Bayesian model for the data set and a priori assign 
equal probabilities for three possible models. The two first models are the 
independence and autologistic models adopted in the previous studies,
whereas the third model is an MRF with a $3\times 3$ neighbourhood
and higher-order interactions.
To reduce the effect of the boundary assumption we include the observed
nodes in a larger rectangular lattice as illustrated in Figure 
\ref{fig:mortalityMap}, and adopt a free boundary assumption 
for the extended lattice. We let $z$ denote a vector of the observed values and
$y$ a vector of the unobserved values, and let $x=(y,z)$ be a vector of 
all the values in the extended lattice. The dimensions of the rectangular
lattice is chosen so that every observed node is at least $10$ nodes 
away from the border, and the extended lattice is then $78\times 87$.
Of the nodes in the extended lattice $34.6\%$ is observed. In the following
we first give a more precise specification of the three possible models we allow 
on the extended lattice and thereafter define prior distributions
for the parameters in these models.

The independence model has only one parameter, which we denote by $\theta$,
and
\begin{equation}
p_1(x|\theta) \propto \exp\left( \theta\sum_{i\in S} x_i\right).
\end{equation}
In the autologistic model we assume a first-order neighbourhood and assume the 
horizontal and vertical interactions to be equal. The model then has two 
parameters, which we denote by $\theta_0$ and $\theta_1$ and set 
$\theta = (\theta_0,\theta_1)$. Our MRF is then given by
\begin{equation}
p_2(x|\theta) \propto \exp\left\{ \theta_0\sum_{i\sim j}I(x_i\neq x_j)
+ \theta_1\sum_{i\sim j}I(x_i=1 \cap x_j=1)\right\}.
\end{equation}
In the $3 \times 3$ neighbourhood MRF the maximal cliques are $2\times 2$ blocks of
nodes. Also for this model we assume the potential function $V_\Lambda(x_\Lambda)$ 
to be invariant under rotation of the values in $x_C$. The 
rotational invariance restriction groups the $2^{2\cdot 2}=16$ possible
configurations of $x_C$ into the six groups illustrated in 
Figure \ref{fig:model3}.
\begin{figure}
\begin{center}
\begin{tabular}{c|cccccc} 
$x_C$ & \begin{picture}(10,15)(0,0)
\put(0,0){\line(1,0){10}}
\put(0,5){\line(1,0){10}}
\put(0,10){\line(1,0){10}}
\put(0,0){\line(0,1){10}}
\put(5,0){\line(0,1){10}}
\put(10,0){\line(0,1){10}}
\end{picture} &
\begin{picture}(10,15)(0,0)
\put(0,0){\line(1,0){10}}
\put(0,5){\line(1,0){10}}
\put(0,10){\line(1,0){10}}
\put(0,0){\line(0,1){10}}
\put(5,0){\line(0,1){10}}
\put(10,0){\line(0,1){10}}

\put(0,5.5){\line(1,0){5}}
\put(0,6.0){\line(1,0){5}}
\put(0,6.5){\line(1,0){5}}
\put(0,7.0){\line(1,0){5}}
\put(0,7.5){\line(1,0){5}}
\put(0,8.0){\line(1,0){5}}
\put(0,8.5){\line(1,0){5}}
\put(0,9.0){\line(1,0){5}}
\put(0,9.5){\line(1,0){5}}
\end{picture} &
\begin{picture}(10,15)(0,0)
\put(0,0){\line(1,0){10}}
\put(0,5){\line(1,0){10}}
\put(0,10){\line(1,0){10}}
\put(0,0){\line(0,1){10}}
\put(5,0){\line(0,1){10}}
\put(10,0){\line(0,1){10}}

\put(0,5.5){\line(1,0){10}}
\put(0,6.0){\line(1,0){10}}
\put(0,6.5){\line(1,0){10}}
\put(0,7.0){\line(1,0){10}}
\put(0,7.5){\line(1,0){10}}
\put(0,8.0){\line(1,0){10}}
\put(0,8.5){\line(1,0){10}}
\put(0,9.0){\line(1,0){10}}
\put(0,9.5){\line(1,0){10}}
\end{picture} &
\begin{picture}(10,15)(0,0)
\put(0,0){\line(1,0){10}}
\put(0,5){\line(1,0){10}}
\put(0,10){\line(1,0){10}}
\put(0,0){\line(0,1){10}}
\put(5,0){\line(0,1){10}}
\put(10,0){\line(0,1){10}}

\put(0,5.5){\line(1,0){5}}
\put(0,6.0){\line(1,0){5}}
\put(0,6.5){\line(1,0){5}}
\put(0,7.0){\line(1,0){5}}
\put(0,7.5){\line(1,0){5}}
\put(0,8.0){\line(1,0){5}}
\put(0,8.5){\line(1,0){5}}
\put(0,9.0){\line(1,0){5}}
\put(0,9.5){\line(1,0){5}}

\put(5,0.5){\line(1,0){5}}
\put(5,1.0){\line(1,0){5}}
\put(5,1.5){\line(1,0){5}}
\put(5,2.0){\line(1,0){5}}
\put(5,2.5){\line(1,0){5}}
\put(5,3.0){\line(1,0){5}}
\put(5,3.5){\line(1,0){5}}
\put(5,4.0){\line(1,0){5}}
\put(5,4.5){\line(1,0){5}}
\end{picture} & 
\begin{picture}(10,15)(0,0)
\put(0,0){\line(1,0){10}}
\put(0,5){\line(1,0){10}}
\put(0,10){\line(1,0){10}}
\put(0,0){\line(0,1){10}}
\put(5,0){\line(0,1){10}}
\put(10,0){\line(0,1){10}}

\put(0,5.5){\line(1,0){5}}
\put(0,6.0){\line(1,0){5}}
\put(0,6.5){\line(1,0){5}}
\put(0,7.0){\line(1,0){5}}
\put(0,7.5){\line(1,0){5}}
\put(0,8.0){\line(1,0){5}}
\put(0,8.5){\line(1,0){5}}
\put(0,9.0){\line(1,0){5}}
\put(0,9.5){\line(1,0){5}}

\put(5,5.5){\line(1,0){5}}
\put(5,6.0){\line(1,0){5}}
\put(5,6.5){\line(1,0){5}}
\put(5,7.0){\line(1,0){5}}
\put(5,7.5){\line(1,0){5}}
\put(5,8.0){\line(1,0){5}}
\put(5,8.5){\line(1,0){5}}
\put(5,9.0){\line(1,0){5}}
\put(5,9.5){\line(1,0){5}}

\put(0,0.5){\line(1,0){5}}
\put(0,1.0){\line(1,0){5}}
\put(0,1.5){\line(1,0){5}}
\put(0,2.0){\line(1,0){5}}
\put(0,2.5){\line(1,0){5}}
\put(0,3.0){\line(1,0){5}}
\put(0,3.5){\line(1,0){5}}
\put(0,4.0){\line(1,0){5}}
\put(0,4.5){\line(1,0){5}}
\end{picture} & 
\begin{picture}(10,15)(0,0)
\put(0,0){\line(1,0){10}}
\put(0,5){\line(1,0){10}}
\put(0,10){\line(1,0){10}}
\put(0,0){\line(0,1){10}}
\put(5,0){\line(0,1){10}}
\put(10,0){\line(0,1){10}}

\put(0,5.5){\line(1,0){5}}
\put(0,6.0){\line(1,0){5}}
\put(0,6.5){\line(1,0){5}}
\put(0,7.0){\line(1,0){5}}
\put(0,7.5){\line(1,0){5}}
\put(0,8.0){\line(1,0){5}}
\put(0,8.5){\line(1,0){5}}
\put(0,9.0){\line(1,0){5}}
\put(0,9.5){\line(1,0){5}}

\put(5,5.5){\line(1,0){5}}
\put(5,6.0){\line(1,0){5}}
\put(5,6.5){\line(1,0){5}}
\put(5,7.0){\line(1,0){5}}
\put(5,7.5){\line(1,0){5}}
\put(5,8.0){\line(1,0){5}}
\put(5,8.5){\line(1,0){5}}
\put(5,9.0){\line(1,0){5}}
\put(5,9.5){\line(1,0){5}}

\put(0,0.5){\line(1,0){5}}
\put(0,1.0){\line(1,0){5}}
\put(0,1.5){\line(1,0){5}}
\put(0,2.0){\line(1,0){5}}
\put(0,2.5){\line(1,0){5}}
\put(0,3.0){\line(1,0){5}}
\put(0,3.5){\line(1,0){5}}
\put(0,4.0){\line(1,0){5}}
\put(0,4.5){\line(1,0){5}}

\put(5,0.5){\line(1,0){5}}
\put(5,1.0){\line(1,0){5}}
\put(5,1.5){\line(1,0){5}}
\put(5,2.0){\line(1,0){5}}
\put(5,2.5){\line(1,0){5}}
\put(5,3.0){\line(1,0){5}}
\put(5,3.5){\line(1,0){5}}
\put(5,4.0){\line(1,0){5}}
\put(5,4.5){\line(1,0){5}}
\end{picture} 

\\ \hline
$V_\Lambda(x_\Lambda;\theta)$ & $0$ & $\theta^0$ & $\theta^1$ & $\theta^2$ & $\theta^3$ & $\theta^4$ \\[-0.4cm]
\end{tabular}
\end{center}
\caption{\label{fig:model3}The real data example: The six possible configurations in a $2\times 2$
clique (up to rotation), and the corresponding values of the 
potential function $V_\Lambda(x_\Lambda;\theta)$ in the $3\times 3$ neighbourhood MRF.}
\end{figure}
Without loss of generality we can set the potential value for the all zero 
configuration to zero, and we define one parameter for each of the other five
configuration sets, again as shown in Figure \ref{fig:model3}. The 
parameter vector in the model is thereby 
$\theta=(\theta^0,\ldots,\theta^4)$ and the MRF is given as
\begin{equation}
p_3(x|\theta) \propto \exp\left\{ \sum_{\Lambda\in{\cal C}} U_\Lambda(x_\Lambda;\theta)\right\},
\end{equation}
where $U_\Lambda(x_\Lambda;\theta)$ is as specified in Figure \ref{fig:model3}.

As specified above we assume a prior probability $p(m)=1/3$ for each model $m=1, 2, 3$.
Given any of the three models we let the prior for the associated parameters,
$p_m(\theta)$,  be a Gaussian distribution where the components of $\theta$ are 
independent and Gaussian distributed with zero mean and some variance $\sigma^2$.
A priori we do not expect very strong interactions for this type of data so we
set $\sigma^2=1$. The resulting posterior distribution of interest is
$p(m,\theta|z)$, but the unobserved vector $y$ makes it hard to simulate
from this distribution so in the simulation we focus on
$p(m,\theta,y|z) \propto p(m)p_m(\theta)p_m(y,z|\theta)$.
For $m=2$ and $3$ the MRF $p_m(y,z|\theta)$ of course contains a computationally intractable
normalising constant, so as in Section \ref{sec:fullyBayesian} we replace 
$p_m(y,z|\theta)$ with a corresponding approximation $\widetilde{p}_m(y,z|\theta)$
as defined in (\ref{eq:pommapprox}). As in the fully Bayesian example in Section 
\ref{sec:fullyBayesian} we here use the second POMM approximation variant
discussed in the paragraph following (\ref{eq:pommapprox}) as this alternative
is somewhat faster than the alternative POMM approximation.

The distribution we want to sample is then
\begin{equation}
\widetilde{p}(m,\theta,y|z) \propto p(m)p_m(\theta)\widetilde{p}_m(y,z|\theta),
\end{equation}
and to sample from this distribution we adopt a reversible jump MCMC algorithm
\citep{art22}. To get an acceptable trans-dimensional mixing in 
the MCMC algorithm we found it necessary to include more auxiliary variables.
For each of the three models $k=1,2,3$ we added a parameter $\theta_k$ and 
vector $y_k$, where $\theta_1\in\mathbb{R}$, $\theta_2\in\mathbb{R}^2$ and 
$\theta_3\in\mathbb{R}^5$ and each of the $y_k$'s are vectors of the same 
size as $y$. We assume 
$(\theta_k,y_k)\sim p_k(\theta_k) \widetilde{p}_k(y_k|\theta_k,z)$
independently for each $k$ and independent of $(m,\theta,y)$, where 
$\widetilde{p}_k(y_k|\theta_k,z)$ is the POMM approximation of 
$p_k(y_k|\theta_k,z)\propto p_k(y_k,z|\theta_k)$. 
To simulate from 
\begin{equation}
\widetilde{p}(m,\theta,y,\theta_1,y_1,\theta_2,y_2,\theta_3,y_3|z) = 
\widetilde{p}(m,\theta,y|z)\prod_{k=1}^3 p_k(\theta_k)\widetilde{p}_k(y_k|\theta_k,z)
\end{equation}
we adopt the reversible jump setup with 
three types of proposals. The first proposal is a random walk proposal 
for each of $\theta$, $\theta_1$, $\theta_2$ and $\theta_3$. A separate update is 
performed for each of these four variables, so a proposed new value for one of them
is accepted or rejected before a change for another of the vectors
is proposed. The components in the potential new vector are generated
independently from Gaussian distributions centered at the current value
and with the same variance $\tau^2$ for all components. By trial and error
we tuned the value of the proposal variance and ended up using $\tau=0.025$.
The second proposal type generate potential new values for each of $y$, $y_1$, $y_2$ and 
$y_3$. Again a proposal followed by an acceptance or rejection is done for 
each of the vectors separately. For $y_k,k=1,2,3$ the potential new
value is generated from $\widetilde{p}_k(y_k|\theta_k,z)$, and the 
potential new value for $y$ is generated from $\widetilde{p}_m(y|\theta_m,z)$.
One can note that the updates for $y_1$, $y_2$ and $y_3$ are 
Gibbs updates, whereas the update for $y$ is not. The acceptance rate
for the $y$ updates is, however, very close to one. The third update type
is the only trans-dimensional update. Here new values are proposed
for $(m,\theta,y)$ and for one of $(\theta_1,y_1)$, $(\theta_2,y_2)$
and $(\theta_3,y_3)$. First a new value $m^\prime$ is proposed for the model indicator
$m$. If $m=1$ or $3$  we always set $m^\prime=2$, and if $m=2$ we set
$m^\prime=1$ or $3$ with probabilities a half for each. 
Thereafter potential new values for $\theta$ and $y$ is
deterministically set as $\theta^\prime=\theta_{m^\prime}$ and 
$y^\prime=y_{m^\prime}$. Finally potential new values for $\theta_{m^\prime}$
and $y_{m^\prime}$ are sampled as $\theta^\prime_{m^\prime}\sim\mbox{N}(\theta_{m^\prime},
\tau^2 I)$ and $y^\prime_{m^\prime}\sim \widetilde{p}_{m^\prime}(y_m|\theta^\prime_{m^\prime},z)$,
respectively. Here we use the same variance $\tau^2$ as in the random walk proposals
discussed above, and $I$ is the identity matrix of the suitable dimension.
The proposed new values are then accepted or rejected according to the 
standard reversible jump MCMC acceptance probability. In particular it is
easy to show that the Jacobian determinant occurring in the expression for 
the acceptance probability equals unity for this kind of proposal.

\begin{figure}
\begin{center}
\includegraphics[height=6.0cm,width=3.0cm,angle=-90]{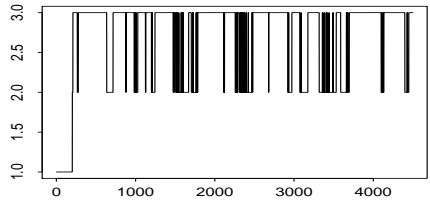}
\end{center}

\vspace*{-0.3cm}

\caption{\label{fig:mchains}Real data example: For the Markov chain simulation starting 
at $m=1$, trace plot of the simulated values for $m$.}
\end{figure}

\begin{figure}
\begin{center}
\begin{tabular}{cc}
\includegraphics[height=4.0cm,width=3.0cm,angle=-90]{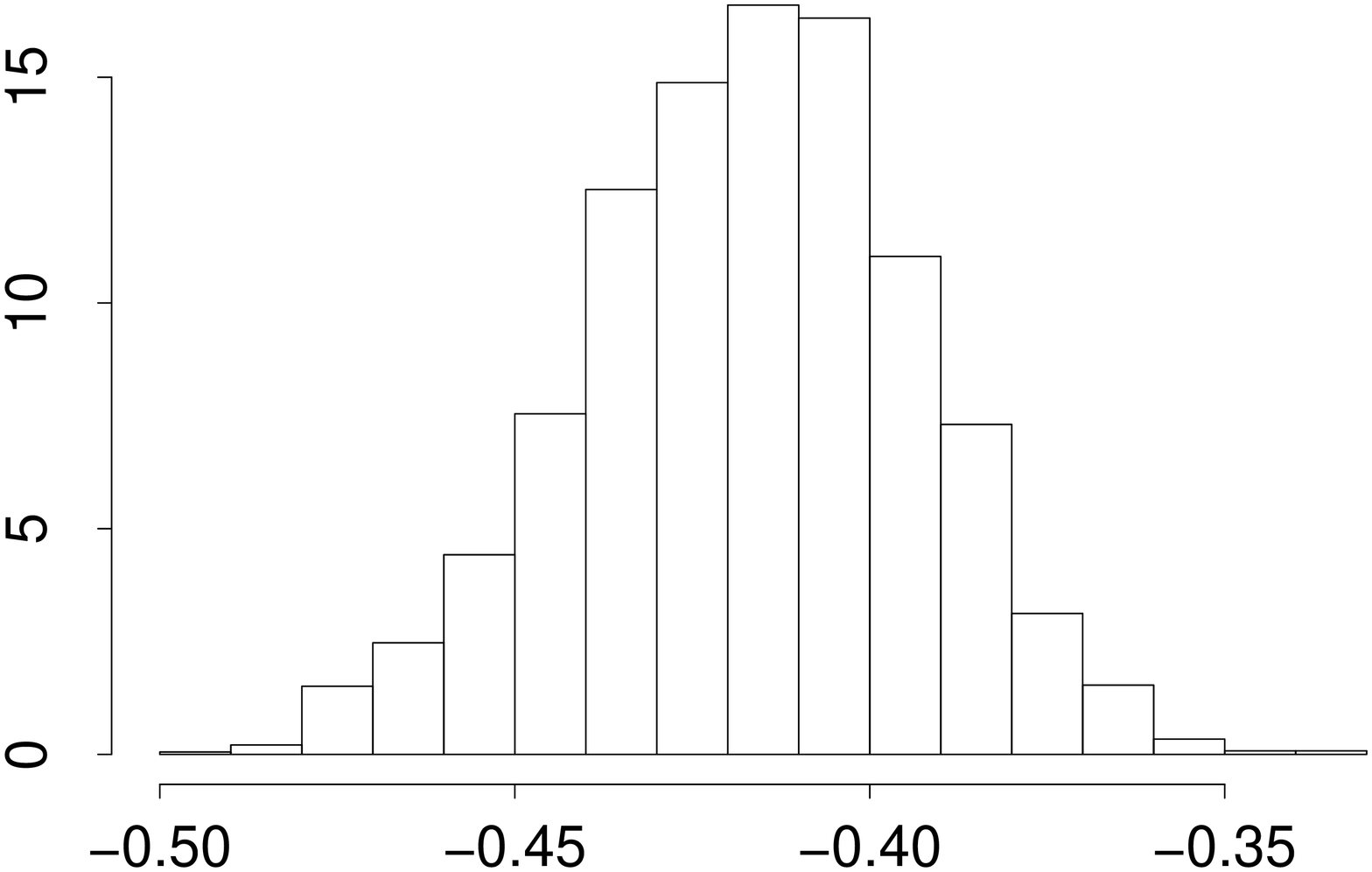} &
\includegraphics[height=4.0cm,width=3.0cm,angle=-90]{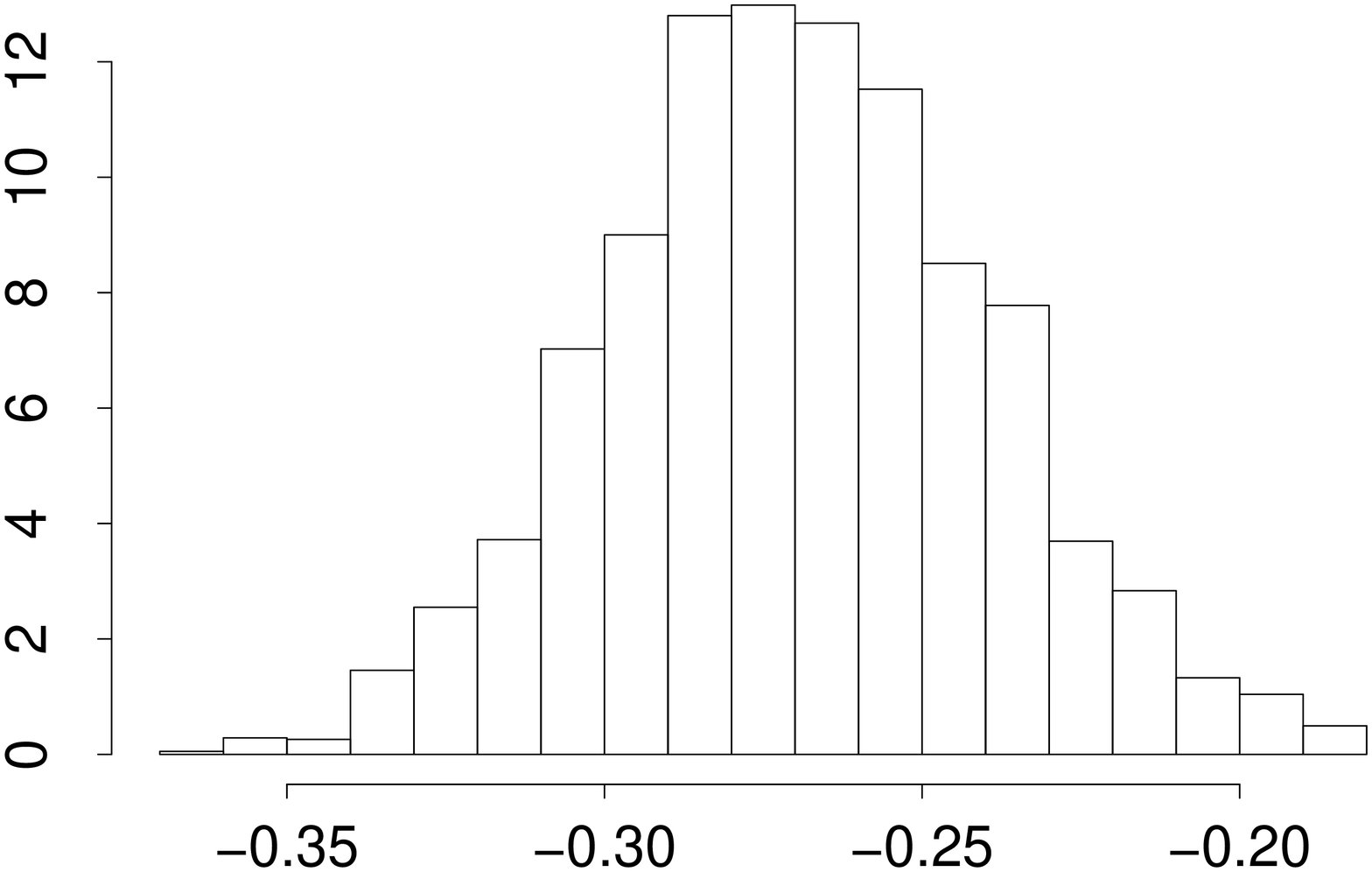} \\
$\theta_0$ & $\theta_1$
\end{tabular}
\end{center}

\vspace*{-0.7cm}

\caption{\label{fig:resultModel2}Real data example: Histograms of the simulated posterior 
values for $\theta_0$ and $\theta_1$ when $m=2$.}
\end{figure}

\begin{figure}
\begin{center}
\begin{tabular}{ccc}
\includegraphics[height=4.0cm,width=3.0cm,angle=-90]{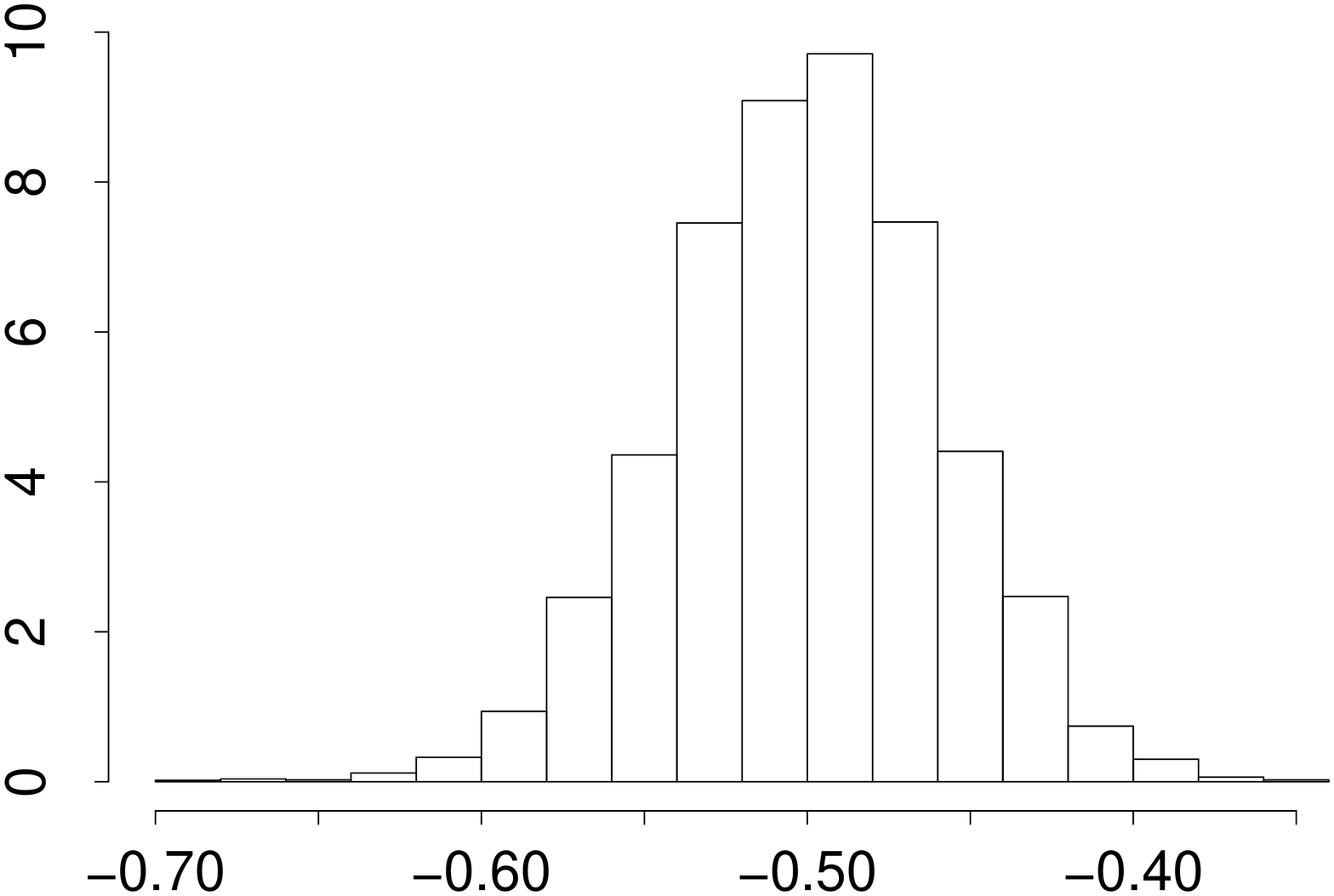} &
\includegraphics[height=4.0cm,width=3.0cm,angle=-90]{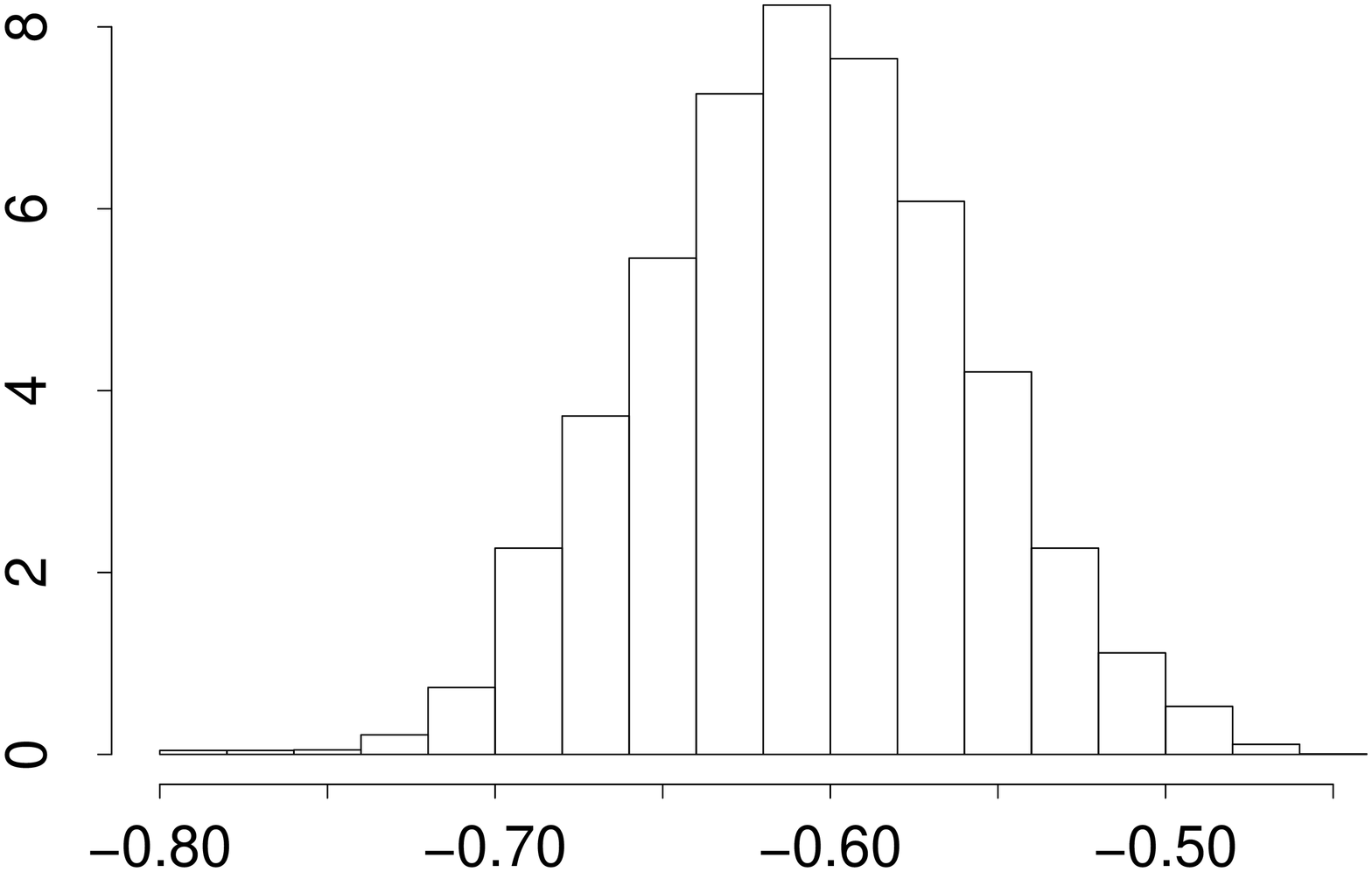} &
\includegraphics[height=4.0cm,width=3.0cm,angle=-90]{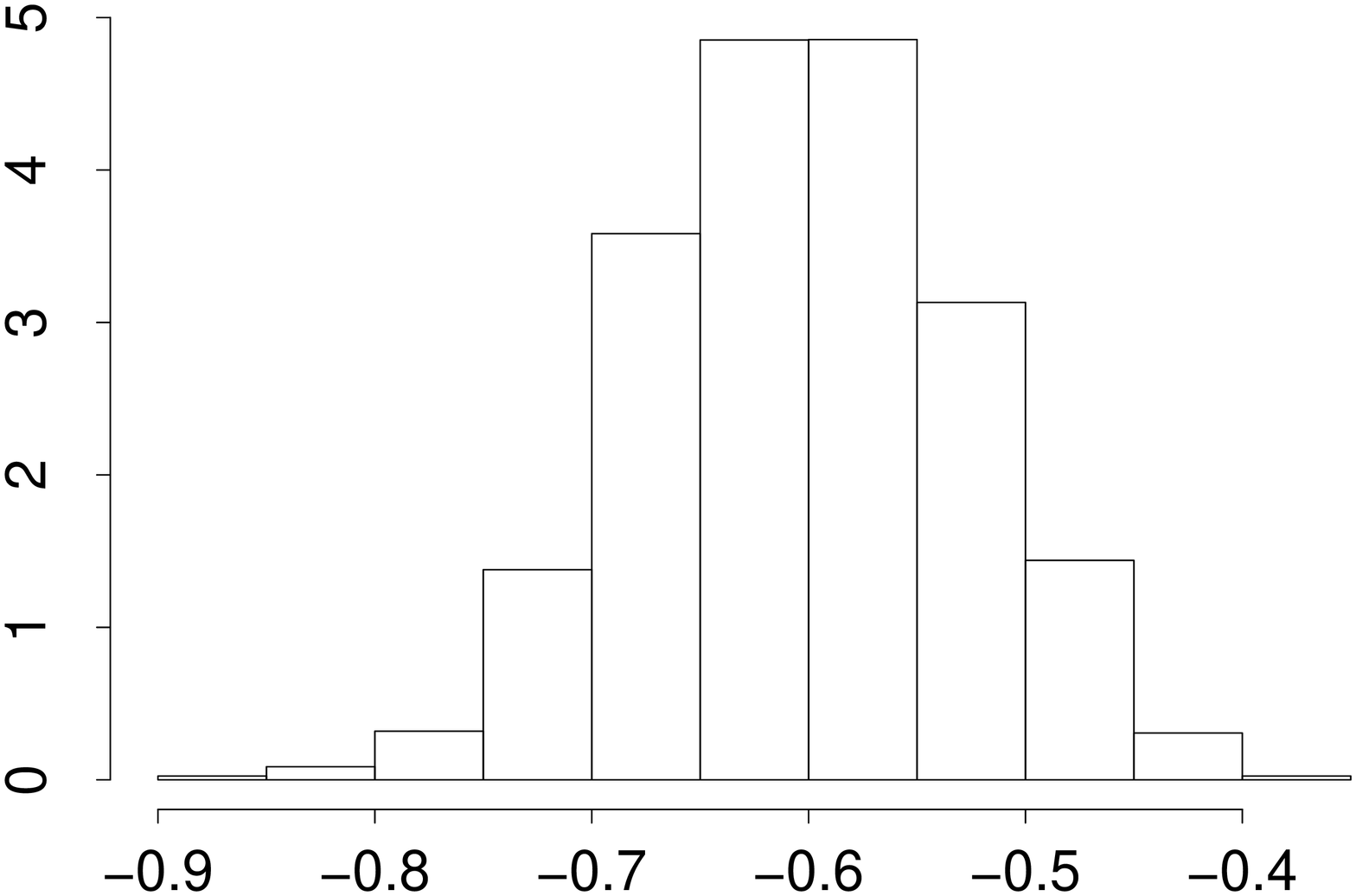} \\
$\theta_0$ & $\theta_1$ & $\theta_2$ \\[-0.4cm]
\includegraphics[height=4.0cm,width=3.0cm,angle=-90]{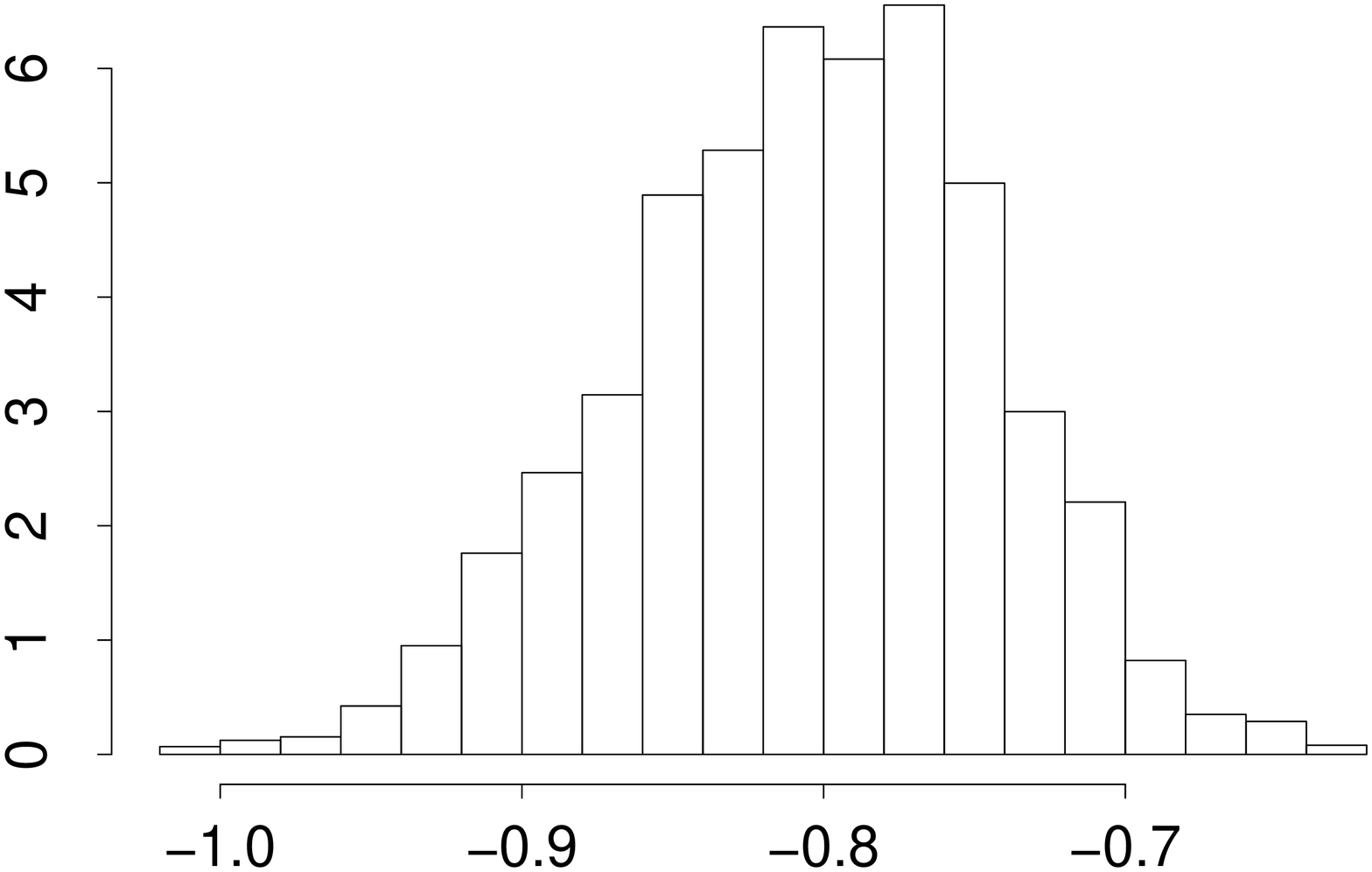} &
\includegraphics[height=4.0cm,width=3.0cm,angle=-90]{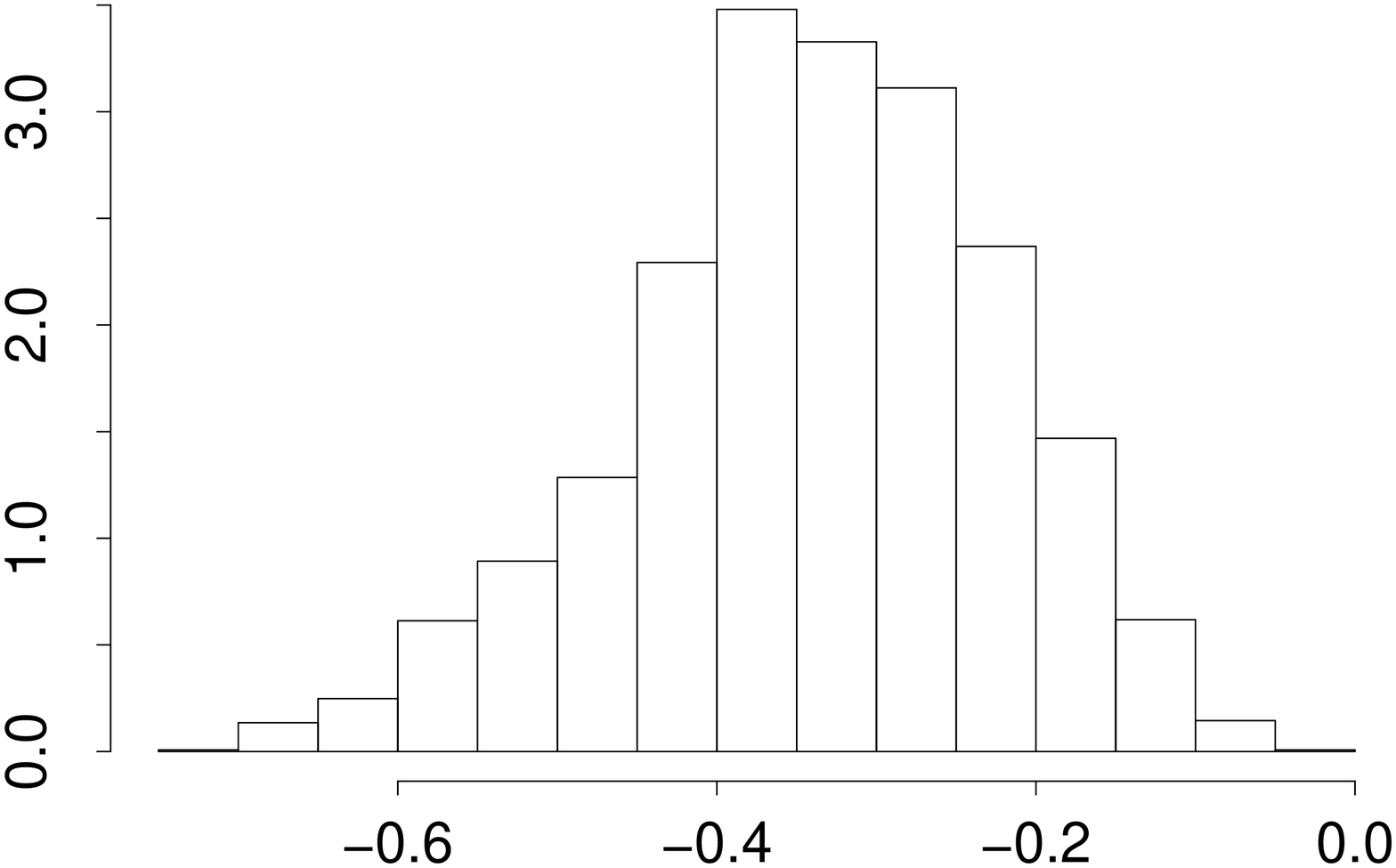} \\
$\theta_3$ & $\theta_4$ \\[-0.6cm]
\end{tabular}
\end{center}
\caption{\label{fig:resultModel3}Real data example: Histograms of the simulated posterior 
values for $\theta_0,\ldots,\theta_4$ when $m=3$.}
\end{figure}

To check the convergence and mixing properties of the reversible 
jump MCMC algorithm we ran three independent runs with different 
starting values. For each of $k=1,2$ and $3$ we had a chain where we
initially set $m=k$ and ran 200 iterations without the transdimensional move.
The idea here is that the chains should essentially converge conditional on
the value of $m$ before we allow the value of $m$ to vary. After these 
$200$ initial iterations we ran the three chains for an additional $4300$ iterations,
now including the transdimensional move. The chain starting with $m=1$ 
then quickly changed to having $m=2$ and $m=3$ and never returned to the state 
$m=1$. The chains starting with $m=2$ and $m=3$ never visited
the state $m=1$. All three chains had several jumps back and fourth 
between $m=2$ and $m=3$ as can be seen from Figure \ref{fig:mchains}.
In the analysis of the Markov chain runs we discard the first $500$ iterations of each run
as burn-in, and use all three runs to estimate the quantities of interest. 
We estimate the posterior probability for model $m=3$ by the fraction of times the chains
are visiting this state, and get $\widehat{p}(m=3|z)=0.680$. To evaluate the uncertainty 
in this number we split the chains into intervals of $100$ iterations and estimate
$p(m=3|z)$ based on each of these intervals. These estimates are close to being
uncorrelated so, considering them as independent, 
we use them to construct an approximate $95\%$ credibility $t$-interval for $p(m=3|z)$.
The interval becomes $[0.619,0.741]$, clearly showing that $m=3$ is the model
with highest a posteriori probability. This demonstrates that even for data sets
where the interactions are quite weak there can be a need for including 
higher-order interactions. Figures \ref{fig:resultModel2} and 
\ref{fig:resultModel3} show estimates of the marginal distributions for the model
parameters for model $m=2$ and $m=3$, respectively. One should note that the 
spread of all of these parameters are well within the variability  of the corresponding standard 
normal prior, so our prior here should not be very influential on these results. We also
observe that all parameters, both for model $m=2$ and $3$, are significantly smaller than zero.
This reflects that zero is the dominating value of the data set and that we in both models have chosen
to set the potential for maximal cliques with only zero values to the reference value of zero.

\section{\label{sec:cr}Closing remarks}
In this report we have shown how we can derive an approximate
forward-backward algorithm by studying how to approximate the
pseudo-Boolean energy function during the summation process. This approximation can then be used to
work with statistical models such as MRFs. It
allows us to produce approximations and bounds of the normalising constant and
likelihood for models that would normally be
too computationally heavy to work with directly. It also 
gives POMM approximations of MRFs that can be used as a surrogate for 
MRFs in more complicated model setups.
We have demonstrated
the accuracy of the approximation and bounds through simple experiments with the
Ising model and a higher-order interaction model, 
and demonstrated some potential applications by simulation experiments.
We have also applied our approximation to a real life data set and here
demonstrated that higher-order interactions may be important even 
for data sets with weak interactions.
We round off now with some possible
future extensions as well as some closing remarks.

The approximation we have defined was inspired by the work in
\citet{art130} and there are many parallels between the two. 
There the energy function was represented as a
binary polynomial and small interactions was dropped while running
the forward-backward algorithm. This worked reasonably well, but
one may worry that dropping many smalls interactions may produce
an approximation no better than when dropping one large interaction.
Moreover, for models with strong interactions the
approximation in \citet{art130} would either include too many terms and thus explode in
run-time, or if the cutoff level was set low enough to run the
algorithm, exclude so many of the interactions that the approximation became
uninteresting. In a sense the work in this report has been an effort to
deal with these issues. We have a clearly defined approximation 
criterion and the construction of the algorithm allows for much
more direct control over the run time. Also, by not just dropping small
terms, but approximating the pseudo-Boolean function in such a way
that we minimise the error sum of squares we manage to get better
approximations of the models with stronger interactions. 

In our setting the sample space of the pseudo-Boolean
function has a probability measure on it, however in our discussion of
approximating pseudo-Boolean functions we have considered each state
in the sample space as equally important. Assuming we are interested
in approximating the normalising constant, this is probably far from
optimal and is reflected in our results. As we can see for the Ising model 
our approximation
works better the smaller the interaction parameter $\theta$. Our initial
approximation attempts to spread the error of removing interactions as
evenly as possible among the states. Ideally, we would like to have small errors in states
with a corresponding high probability, and larger errors in states
with low probability. To get this approximation we should replace
the $\mbox{SSE}$ in (\ref{eq:SSE}) with a 
weighted error sum of squares.
This problem has also been studied in the literature, see
for instance \citet{art151,art150}. However, unlike
the unweighted case an explicit solution is not readily available for
a probability density like the MRF. The iterative method of removing
interactions does not work here, nor can we group the equations like
we do with the SOIR approximation.

In all our examples we consider MRFs defined on a lattice and assume
the potential functions to be translational invariant. Our 
approximations and bounds are, however, also valid for MRFs
defined on an arbitrary graph and applications of the approximation and bounds
in such situations is something we want to explore in the future.

\bibliography{../../../mybib}

\appendix

\section{\label{app:proof3}Proof of Theorem \ref{th:3}}
Expanding $\mbox{SSE}(f,\tilde{\tilde{f}}) = \sum_{x \in \Omega} \left\{f(x)-\tilde{\tilde{f}}(x) \right\}^2$ 
we get,
\begin{align*}
\sum_{x \in \Omega} \left\{f(x)-\tilde{\tilde{f}}(x) \right\}^2 = \sum_{x \in \Omega}
\left\{f(x)-\tilde{f}(x) + \tilde{f}(x)-\tilde{\tilde{f}}(x) \right\}^2 \\
 =  \sum_{x \in \Omega}
\left\{f(x)-\tilde{f}(x) \right\}^2 + 
\sum_{x \in \Omega} \left\{\tilde{f}(x)-\tilde{\tilde{f}}(x) \right\}^2 +\sum_{x \in \Omega}
\{f(x)-\tilde{f}(x)\}\{\tilde{f}(x)-\tilde{\tilde{f}}(x)\}\\
= \mbox{SSE}(f,\tilde{f}) + \mbox{SSE}(\tilde{f},\tilde{\tilde{f}}) + \sum_{x \in \Omega}
\{f(x)-\tilde{f}(x)\}\tilde{f}(x)-\sum_{x \in \Omega}
\{f(x)-\tilde{f}(x)\}\tilde{\tilde{f}}(x).
\end{align*}
To prove the theorem it is thereby sufficient to show that,
\[
\sum_{x \in \Omega}\{f(x)-\tilde{f}(x)\}\tilde{f}(x)-\sum_{x \in
  \Omega}\{f(x)-\tilde{f}(x)\}\tilde{\tilde{f}}(x) = 0.
\]
First recall that we from \eqref{eq:equations} know that,
\begin{equation}\label{eq:Def1}
\sum_{x \in \Omega_\lambda}\left\{f(x) - \tilde{f}(x) \right\} = 0, \text{ } \forall \text{ } \lambda
\in \tilde{S}. 
\end{equation}
Also, since $\tilde{\tilde{S}} \subseteq \tilde{S}$,
\begin{equation}
\sum_{x \in \Omega_\lambda}\left\{f(x) - \tilde{f}(x) \right\} = 0, \text{ } \forall \text{ } \lambda
\in \tilde{\tilde{S}}. 
\label{eq:Def2}
\end{equation}
We study the first term, $\sum_{x \in \Omega}\{f(x)-\tilde{f}(x)\}\tilde{f}(x)$, expand the expression for $\tilde{f}(x)$ outside the parenthesis
and change the order of summation,
\begin{align*}
\sum_{x \in \Omega}\{f(x)-\tilde{f}(x)\}\tilde{f}(x)& = \sum_{x \in \Omega}\left[\{f(x)-\tilde{f}(x)\}
  \sum_{\Lambda \in \tilde{S}} \tilde{\beta}^{\Lambda} \prod_{k \in
  \Lambda}x_k \right] \\
& = \sum_{\Lambda \in \tilde{S}} \left( \tilde{\beta}^{\Lambda}
  \sum_{x \in \Omega} \left[\prod_{k \in
  \Lambda}x_k\{f(x)-\tilde{f}(x)\} \right] \right) \\
& = \sum_{\Lambda \in \tilde{S}} \left[ \tilde{\beta}^{\Lambda}
\sum_{x \in \Omega_\Lambda} \left\{f(x)-\tilde{f}(x) \right\} \right] = 0,
\end{align*}
where the last transition follows from \eqref{eq:Def1}. Using
\eqref{eq:Def2} we can correspondingly show that $\sum_{x \in \Omega}
\{f(x)-\tilde{f}(x)\}\tilde{\tilde{f}}(x) = 0$.

\section{\label{app:proof4}Proof of Theorem \ref{th:4}}
We study the error sum of squares,
\begin{align*}
\sum_{x \in \Omega} \left\{f(x) - \tilde{f}(x) \right\}^2 &=
\sum_{x \in \Omega}\left[\{f(x)-\tilde{f}(x)\}f(x) \right] -
\sum_{x \in \Omega}\left[\{f(x)-\tilde{f}(x)\}\tilde{f}(x) \right] \\
&= \sum_{x \in \Omega} \left[\sum_{\Lambda \in S} \beta^{\Lambda} \{f(x)-\tilde{f}(x)\}\prod_{k \in
  \Lambda}x_k \right] - \sum_{x \in \Omega} \left[\sum_{\Lambda \in \tilde{S}} \tilde{\beta}^{\Lambda} 
\{f(x)-\tilde{f}(x)\}\prod_{k \in
  \Lambda}x_k \right] \\
&= \sum_{\Lambda \in S} \beta^{\Lambda} \left[ \sum_{x \in \Omega_{\Lambda}} \{f(x)-\tilde{f}(x)\} \right] - 
\sum_{\Lambda \in
\tilde{S}} \tilde{\beta}^{\Lambda} \left[ \sum_{x \in \Omega_{\Lambda}} \{f(x)-\tilde{f}(x)\} \right], 
\end{align*}
where the second sum is always zero by \eqref{eq:Def2}. Since $\tilde{S}
\subseteq S$, the first sum can be further split into two parts,
\[
\sum_{\Lambda \in S} \beta^{\Lambda} \left[ \sum_{x \in \Omega_{\Lambda}}
  \{f(x)-\tilde{f}(x)\} \right] = \sum_{\Lambda \in \tilde{S}}
\beta^{\Lambda} \left[ \sum_{x \in \Omega_{\Lambda}}
  \{f(x)-\tilde{f}(x)\} \right] + \sum_{\Lambda \in S\setminus
  \tilde{S}} \beta^{\Lambda} \left[ \sum_{x \in \Omega_{\Lambda}}
  \{f(x)-\tilde{f}(x)\} \right],
\]
where once again the first sum is zero.

\section{\label{app:proof6}Proof of Theorem \ref{th:6}}
From Theorem \ref{th:1} it follows that it is sufficient to 
consider a function $f(x)$ with non-zero interactions $\beta^\Lambda$ only for 
$\Lambda\in S_{\{ i,j\}}$, since we only need to focus on the interactions we want to remove.
Thus we have
\[
f(x) = \sum_{\Lambda\in S_{\{ i,j\}}} \beta^\Lambda \prod_{k\in\Lambda}x_k
\mbox{~~~and~~~}
\widetilde{f}(x) = \sum_{\Lambda\in \widetilde{S}} \widetilde{\beta}^\Lambda \prod_{k\in \Lambda} x_k,
\]
and we need to show that then
\begin{equation}\label{eq:resultProof}
\tilde{\beta}^\Lambda = \left\{ \begin{array}{@{}ll}
- \frac{1}{4}\beta^{\Lambda\cup\{ i,j\}} & \mbox{~~if $\Lambda \cup \{i,j\} \in S$}, \\
\hspace*{0.26cm} \frac{1}{2}\beta^{\Lambda\cup\{ i\}} & \mbox{~~if $\Lambda\cup\{ i\}\in S$ 
  and $\Lambda\cup\{ j\}\not\in S$}, \\
\hspace*{0.26cm} \frac{1}{2}\beta^{\Lambda\cup \{ j\}} & \mbox{~~if $\Lambda\cup\{ i\}\not\in S$ 
  and $\Lambda\cup\{ j\}\in S$}, \\
\hspace*{0.26cm}0 & \mbox{~~otherwise.}
\end{array}\right.
\end{equation}
We start by defining the sets
\[
R_\Lambda = \{ \Lambda\setminus\{ i\},\Lambda\setminus\{ j\},
\Lambda\setminus\{ i,j\}\} \mbox{~~for $\Lambda\in S_{\{i,j\}}$},
\]
and note that these sets are disjoint, 
and, since we have assumed $S$ to be dense, $R_\Lambda\subseteq \widetilde{S}$.
Defining also the residue set 
\[
T = \widetilde{S} \setminus \left( \bigcup_{\Lambda\in S_{\{ i,j\}}}
R_\Lambda\right)
\]
we may write the approximation error $f(x)-\widetilde{f}(x)$ in 
the following form,
\[
f(x)-\widetilde{f}(x) = 
\sum_{\Lambda\in S_{\{ i,j\}}} \left\{
\left( \beta^\Lambda x_ix_j -
\sum_{\lambda\in R_\Lambda} \widetilde{\beta}^\lambda
\prod_{k\in \Lambda\setminus \lambda} x_k
\right) \prod_{k\in \Lambda\setminus\{ i,j\}} x_k\right\}
- \sum_{\Lambda \in T} \widetilde{\beta}^\Lambda 
\prod_{k\in \Lambda} x_k.
\]
Defining
\[
\Delta f^\Lambda(x_i,x_j) = \beta^\Lambda x_i x_j - \sum_{\lambda\in R_\Lambda} 
\widetilde{\beta}^\lambda \prod_{k\in \Lambda\setminus\lambda}x_k
= \beta^\Lambda x_ix_j - \left( \widetilde{\beta}^{\Lambda\setminus\{ i,j\}} + 
\widetilde{\beta}^{\Lambda\setminus\{ j\}}x_i +
\widetilde{\beta}^{\Lambda\setminus\{ i\}}x_j\right)
\]
we have
\begin{equation}\label{eq:errorProof}
f(x)-\widetilde{f}(x) = \sum_{\Lambda\in S_{\{ i,j\}}} \Delta f^\Lambda(x_i,x_j) 
\prod_{k\in \Lambda\setminus\{ i,j\}} x_k
- \sum_{\Lambda\in T} \widetilde{\beta}^\Lambda
\prod_{k\in\Lambda}x_k.
\end{equation}
Inserting this into (\ref{eq:equations}) and switching the order of summation we get
\[
\sum_{x\in\Omega_\lambda}\left\{ f(x)-\widetilde{f}(x)\right\} = 
\sum_{x\in\Omega_\lambda}\left(
\sum_{\Lambda\in S_{\{ i,j\}}} \Delta f^\Lambda(x_i,x_j) \prod_{k\in \Lambda\setminus\{ i,j\}} x_k
- \sum_{\Lambda\in T} \widetilde{\beta}^\Lambda
\prod_{k\in \Lambda}x_k\right)
\]
\[
= \sum_{\Lambda\in S_{\{ i,j\}}} \left( \sum_{x\in \Omega_\lambda} \Delta f^\Lambda(x_i,x_j) 
\prod_{k\in \Lambda\setminus \{ i,j\}} x_k\right) - \sum_{\Lambda\in T}\left(
\sum_{x\in\Omega_\lambda} \widetilde{\beta}^\Lambda
\prod_{k\in \Lambda} x_k\right) 
\]
\begin{equation}\label{eq:equationsProof}
= 
\sum_{\Lambda\in S_{\{ i,j\}}} \left(\sum_{x\in \Omega_{\lambda\cup (\Lambda\setminus\{ i,j\}}} 
\Delta f^\Lambda(x_i,x_j)\right) 
- \sum_{\Lambda\in T}\left( \sum_{x\in\Omega_{\lambda\cup T}} \widetilde{\beta}^\Lambda\right)
= 0 \mbox{~~~for all $\lambda\in \widetilde{S}$}
\end{equation}
We now proceed to show that this system of equations has a solution where
$\widetilde{\beta}^\Lambda = 0$ for $\Lambda\in T$ and 
$\sum_{x\in \Omega_{\lambda\cup (\Lambda\setminus\{ i,j\}}} \Delta f^\Lambda(x_i,x_j) = 0$
for each $\Lambda\in S_{\{ i,j\}}$. Obviously for each $\Lambda$ the function 
$\Delta f^\Lambda(x_i,x_j)$ has only our possible values, namely
$\Delta f^\Lambda(0,0)$, $\Delta f^\Lambda(1,0)$, $\Delta^\Lambda (0,1)$ and 
$\Delta f^\Lambda (1,1)$. Thus the sum 
$\sum_{x\in \Omega_{\lambda\cup (\Lambda\setminus \{ i,j\}}} \Delta f^\Lambda(x_i,x_j)$ is simply 
given as a sum over these four values multiplied by the number of times they 
occur. Consider first the case where $\lambda$, and thereby also $\lambda\cup (\Lambda\setminus
\{ i,j\})$ does not contain $i$ or $j$. Then the four values 
$\Delta f^\Lambda(0,0)$, $\Delta f^\Lambda(1,0)$, $\Delta^\Lambda (0,1)$ and 
$\Delta f^\Lambda (1,1)$ will occur the same number of times, so
\[
\sum_{x\in \Omega_{\lambda\cup (\Lambda\setminus\{ i,j\})}} \Delta f^\Lambda (x_i,x_j) = 
\frac{|\Omega_{\lambda\cup (\Lambda\setminus\{ i,j\}}|}{4} \left(
\Delta f^\Lambda(0,0)+\Delta f^\Lambda(1,0)+\Delta f^\Lambda(0,1) + 
\Delta f^\Lambda(1,1)\right).
\]
Next consider the case when $\lambda$, and thereby also 
$\lambda\cup (\Lambda\setminus\{ i,j\})$, contains $i$, but not $j$.
Then $x_i=1$ in all terms in the sum, so the values $\Delta f^\Lambda(0,0)$
and $\Delta f^\Lambda(0,1)$ will not occur, whereas the values $\Delta f^\Lambda
(1,0)$ and $\Delta f^\Lambda(1,1)$ will occur the same number of times. Thus,
\[
\sum_{x\in \Omega_{\lambda\cup (\Lambda\setminus\{ i,j\})}} \Delta f^\Lambda (x_i,x_j) = 
\frac{|\Omega_{\lambda\cup (\Lambda\setminus\{ i,j\}}|}{2} \left\{
\Delta f^\Lambda(1,0)+ 
\Delta f^\Lambda(1,1)\right\}.
\]
When $\lambda$ contains $j$, but not $i$ we correspondingly get
\[
\sum_{x\in \Omega_{\lambda\cup (\Lambda\setminus\{ i,j\})}} \Delta f^\Lambda (x_i,x_j) = 
\frac{|\Omega_{\lambda\cup (\Lambda\setminus\{ i,j\}}|}{2} \left\{
\Delta f^\Lambda(0,1)+ 
\Delta f^\Lambda(1,1)\right\}.
\]
The final case, that $\lambda$ contains both $i$ and $j$, will never occur since
$\lambda\in \widetilde{S}$ and 
all interaction involving both $i$ and $j$ have been removed from 
$\widetilde{S}$.
We can now reach the conclusion that if we can find a solution for
\begin{align*}
\Delta f^\Lambda(0,0)+\Delta f^\Lambda(1,0)+\Delta f^\Lambda(0,1)+\Delta (1,1) &=
0,\\
\Delta f^\Lambda(1,0)+\Delta f^\Lambda(1,1) & = 0, \\
\Delta f^\Lambda(0,1)+\Delta f^\Lambda(1,1) &= 0,
\end{align*}
for all $\Lambda\in S_{\{ i,j\}}$ we also have a solution for (\ref{eq:equationsProof})
as discussed above. Using our expression for $\Delta f^\Lambda(x_i,x_j)$, the above three
equations become
\begin{align*}
\beta^\Lambda - \left( 4\widetilde{\beta}^{\Lambda\setminus\{ i,j\}} + 
2\widetilde{\beta}^{\Lambda\setminus\{ i\}} + 2\widetilde{\beta}^{\Lambda\setminus\{ j\}}\right) &=0,\\
\beta^\Lambda - \left( 2\widetilde{\beta}^{\Lambda\setminus\{ i,j\}} + 
\widetilde{\beta}^{\Lambda\setminus\{ i\}} + 2\widetilde{\beta}^{\Lambda\setminus\{ j\}}\right) &=0,\\
\beta^\Lambda - \left( 2\widetilde{\beta}^{\Lambda\setminus\{ i,j\}} + 
2\widetilde{\beta}^{\Lambda\setminus\{ i\}} + \widetilde{\beta}^{\Lambda\setminus\{ j\}}\right) &=0.
\end{align*}
Since the sets $R_\Lambda$ are disjoint, the three equations above can be solved separately for 
each $\Lambda$, and the solution is 
$\widetilde{\beta}^{\Lambda\setminus\{ i,j\}} = -\frac{1}{4}\beta^\Lambda$ and
$\widetilde{\beta}^{\Lambda\setminus\{ i\}}=\widetilde{\beta}^{\Lambda\setminus\{ j\}}=
\frac{1}{2}\beta^\Lambda$. Together with $\widetilde{\beta}^\Lambda=0$ for $\Lambda\in T$ this
is equivalent to (\ref{eq:soir}) in the theorem. Inserting the values we have 
found for $\widetilde{\beta}^\Lambda$ in (\ref{eq:errorProof}) we get
\[
\Delta f^\Lambda(x_i,x_j) = \left(x_ix_j + \frac{1}{4} - \frac{1}{2}x_i - \frac{1}{2}x_j\right)\beta^\Lambda.
\]
Inserting this into the above expression for $f(x)-\widetilde{f}(x)$, and using that we know
$\widetilde{\beta}^\Lambda=0$ for $\Lambda\in T$ we get (\ref{eq:soirerror}) given in the 
theorem.

\end{document}